\documentclass[12pt,preprint]{aastex}

\usepackage{epsfig}

\let\gsim=\ga

\let\lsim=\la

\begin{document}

%\newcommand{\vdag}{(v)^\dagger}
%\newcommand{\myemail}{tsuruta@physics.montana.edu}
%\shorttitle{Evolution of Population III Stars}
%\shortauthors{Ohkubo et al.}]

\title{Core-Collapse Very Massive Stars: Evolution, Explosion, 
and Nucleosynthesis of Population III 500 -- 1000 $M_{\odot}$ Stars}

\author{Takuya Ohkubo\altaffilmark{1}, Hideyuki Umeda\altaffilmark{1}, Keiichi Maeda\altaffilmark{2}, Ken'ichi Nomoto\altaffilmark{1,3}, Tomoharu Suzuki\altaffilmark{1}, Sachiko Tsuruta\altaffilmark{4}, and Martin J. Rees\altaffilmark{5}}

%\affil{Department of Astronomy, Graduate School of Science, University of Tokyo, 7-3-1 Hongo, Bunkyo-ku, Tokyo 113-0033, Japan; ohkubo@astron.s.u-tokyo.ac.jp, umeda@astron.s.u-tokyo.ac.jp, nomoto@astron.s.u-tokyo.ac.jp, suzuki@astron.s.u-tokyo.ac.jp}
%\author{Sachiko Tsuruta}
%\affil{Department of Physics, Montana State University,\\
%       Bozeman, MT},
%\and
%\author{Martin J. Rees}
%\affil{Institute of Astronomy, Cambridge University,\\
%       Cambridge, UK}

\altaffiltext{1}{Department of Astronomy, School of Science, University of Tokyo, 7-3-1 Hongo, Bunkyo-ku, Tokyo 113-0033, Japan; ohkubo@astron.s.u-tokyo.ac.jp, umeda@astron.s.u-tokyo.ac.jp, nomoto@astron.s.u-tokyo.ac.jp, suzuki@astron.s.u-tokyo.ac.jp}
\altaffiltext{2}{Department of Earth Science and Astronomy, graduate school of Arts and Sciences, University of Tokyo, 3-8-1, Komaba, Meguro-ku, Tokyo 153-8902, Japan; maeda@esa.c.u-tokyo.ac.jp}
\altaffiltext{3}{Research Center for the Early Universe, School of Science, University of Tokyo, 7-3-1 Hongo, Bunkyo-ku, Tokyo 113-0033, Japan}
\altaffiltext{4}{Department of Physics, Montana State University, Bozeman, MT 59717-3840; uphst@gemini.msu.montana.edu}
\altaffiltext{5}{Institute of Astronomy, Cambridge University, Madingley Road, Cambridge CB3 0HA, UK; mjr@ast.cam.ac.uk}

\begin{abstract}

We calculate evolution, collapse, explosion, and nucleosynthesis
of Population III very-massive stars with 500$M_{\odot}$ and
1000$M_{\odot}$. Presupernova evolution is calculated in spherical symmetry.
Collapse and explosion are calculated by a
two-dimensional code, based on the bipolar jet models. 
We compare the results of
nucleosynthesis with the abundance patterns of intracluster
matter, hot gases in M82, and extremely metal-poor stars in the
Galactic halo. It was found that both 500$M_{\odot}$ and 1000$M_{\odot}$ models
enter the region of pair-instability but continue to undergo 
core collapse. In the presupernova stage, silicon burning regions occupy a
large fraction, more than 20\% of the total mass. For moderately aspherical explosions, the
patterns of nucleosynthesis match the observational data of both
intracluster medium and M82. Our results
suggest that explosions of Population III core-collapse very-massive stars
contribute significantly to the chemical evolution of gases in
clusters of galaxies. For Galactic halo stars, our [O/Fe] ratios
are smaller than the observational abundances. However, our
proposed scenario is naturally consistent with this outcome. The
final black hole masses are $\sim 230M_{\odot}$ and $\sim 500M_{\odot}$ for the 
$500M_{\odot}$ and 1000$M_{\odot}$ models, respectively. This result may support the view
that Population III very massive stars are responsible for the
origin of intermediate mass black holes which were recently
reported to be discovered.

\end{abstract}

\keywords{nuclear reactions, nucleosynthesis, abundances -- stars: evolution -- stars: supernovae: general -- stars: abundances -- galaxies: starburst -- galaxies: intergalactic medium}

\section{Introduction}

One of the most interesting challenges in astronomy is to
investigate the mass and properties of first generation
"Population III (Pop III)" stars, and how various elements have
been synthesized in the early universe. Just after the Big Bang
these elements were mostly only H, He and a small amount of light
elements (Li, Be, B, etc). Heavier elements, such as C, O, Ne, Mg,
Si and Fe, were synthesized during the evolution of later
generation stars, and massive stars exploded as supernovae (SNe),
releasing heavy elements into space.

Stars end their lives differently depending on their initial masses $M$. 
Here the Pop III stars are assumed to undergo too little mass loss 
to affect the later core evolution. Then the 
fates of Pop III stars are summarized as follows. 
Those stars lighter than 8$M_\odot$ form white dwarfs. Those with
8$M_\odot$ -- 130$M_\odot$ undergo ONe-Fe core collapse at a last
stage of their evolution leaving neutron stars or black holes.
Some of these stars explode as the core-collapse supernovae. Stars
with 130$M_\odot$ -- 300$M_\odot$ undergo electron-positron pair
creation instability during oxygen burning, releasing more energy
by nuclear burning than the gravitational binding energy of the
whole star, and hence these stars disrupt completely as the
pair-instability supernovae (PISN). Stars with 300$M_\odot$ - $10^5M_\odot$ 
also enter into the pair-instability region but continue to collapse. 
Fryer, Woosley, \& Heger (2001) calculated evolution of 260 and 300 $M_\odot$ stars 
and obtained the result that a 260 $M_\odot$ star ends up as a PISN and 
a 300 $M_\odot$ star collapsed. 
Stars over $\sim 10^5M_\odot$ collapse owing to general relativistic 
instability before reaching the main-sequence. 
The core collapse SNe (Type
II, Ib and Ic SNe) release mainly $\alpha$-elements such as O, Mg,
Si and Ca and some Fe-peak elements as well.

It has been suggested that the initial mass function (IMF) of Pop
III first stars may be different from the present one - that more
massive stars existed in the early universe (e.g., Nakamura \&
Umemura 1999; Abel, Bryan, \& Norman 2000; Bromm, Coppi, \& Larson
2002; Omukai \& Palla 2003).  Some authors (e.g., Wasserburg and
Qian 2000; Qian, Sargent, Wasserburg 2002; Qian and Wasserburg
2002; \& Yoshida et al. 2004) argued that existence of very massive stars (VMSs) in the
early universe is consistent with abundance data of Ly$\alpha$
systems. Numerical simulations by, e.g., Bromm \& Loeb (2004),
indicate that the maximum mass of Pop III stars to be formed will
be $\sim$ 300$M_\odot$ -- 500$M_\odot$.  Omukai \& Palla (2003),
however, point out that under certain conditions VMSs much heavier
than 300$M_\odot$ can be formed in the zero-metallicity
environment. Tan \& McKee (2004) calculated star formation by 
taking rotation and disk structure and concluded that first stars should 
be much more massive than $30M_\odot$.  
Another scenario for the formation of VMSs for any metallicity 
has been
presented by Ebisuzaki et al. (2001) and Portegies Zwart et al.
(1999, 2004a, 2004b), where VMSs are formed by merging of less
massive stars in the environment of very dense star clusters.

%Note that the definition 'VMS' is not identical for different authors. For 
%example, Tumlinson et al. 2004 uses 'VMS' as stars ${\rm M > 140M_{\odot}}$. 
%$140{\rm M_{\odot}}$ is the boundary between core-collapse and PISN, so 
%this definition is physically plausible. Venkatesan \& Truran 2003 refers 
%to 100-1000${\rm M_{\odot}}$ as 'VMS'. These definitions both include stars 
%which end up as PISNe. 
In the present paper, we call the stars with $M \gsim 10^5 M_{\odot}$
"Super-Massive 
Stars (SMSs)", and the stars with $M = 130M_{\odot} - 10^5M_{\odot}$ "Very-Massive Stars (VMSs)". Among "VMSs" we define $M > 300M_{\odot}$ 
stars as "Core-Collapse Very-Massive Stars (CVMSs)", in order to clarify 
the distinction between the PISN mass range and the core-collapse range. 
Here we focus on CVMSs, and deal with 500 and 1000 
$M_{\odot}$ models.  

Such CVMSs might have released a large amount of
heavy elements into space by mass loss and/or supernova
explosions, and they might have significantly contributed to
the early galactic chemical evolution, if they were the
source of reionization of intergalactic H and He (e.g., Gnedin
\& Ostriker 1997; Venkatesan, Tumlinson, \& Shull 2003). The reionization of
intergalactic He has traditionally been attributed to quasars.
However, according to the results of the {\it the Wilkinson
Microwave Anisotropy Probe} ({\sl WMAP}) observation in 2003,
reionization in the universe took place as early as 0.2-0.3
billion years after the Big Bang (redshift $z \gsim 20$)(Kogut et
al. 2003). Then these Pop III CVMSs might provide a better
alternative channel which could operate at redshifts higher than
what is assumed for quasars (Bromm, Kudritzki, \& Loeb 2001).

The question of whether CVMSs ($\sim 300M_{\odot} - 10^5M_{\odot}$) actually
existed is of great importance, for instance, to understand the
origin of intermediate mass black holes (IMBHs)($\sim 5 \times
10^{(2 - 4)} M_\odot$). Stellar mass black holes ($\sim
10M_{\odot}$) are formed as the central compact remnants of
ordinary massive (25 - 130$M_\odot$) stars at the end of their
evolution, while supermassive black holes(SMBHs) ($\sim {10}^5$ -
${10}^9 M_{\odot}$) are now known to exist in the center of almost
all galaxies (e.g., Kormendy \& Richstone 1995; Bender 2004).
IMBHs have not been found until recently. However, there is a
strong possibility that some IMBHs have been, indeed, found (e.g.,
Barth, Green, \& Ho 2005 for most recent review).  Matsumoto et al. (2001)
reported possible identification of a $\gsim$ 700$M_{\odot}$ black
hole in M82, by using {\sl Chandra} data.  As to formation of
SMBHs there are several scenarios (e.g., Rees 2002, 2003). SMBHs
may be formed directly from supermassive halos of dark matter
(e.g., Marchant \& Shapiro 1980; Bromm and Loeb 2003). Ebisuzaki
et al. (2001) suggested a scenario where IMBHs grow to a SMBH by
merging and swallowing of many of these objects. If CVMSs actually
existed, they could be considered as natural progenitors of IMBHs.

Motivated by these backgrounds, here we calculate the evolution,
collapse, explosion, and nucleosynthesis of Pop III CVMSs (over
300$M_\odot$). These stars are expected to form black holes
directly at the end of evolution. It has not been known yet if they
will explode as SNe. However, if the star is rotating the whole
star will not become a black hole at once, but it is expected to
form an accretion disk around the central remnant (e.g. Shibata \&
Shapiro 2002).  After forming an accretion disk, jet-like
explosions may occur by extracting energy from the accretion disk
and/or the black hole itself (Fryer et al. 2001;
MacFadyen et al. 2001; Maeda \& Nomoto 2003). Therefore, in our
current explosion and collapse calculations we adopt a
two-dimensional approach including accretion along the equatorial
direction and jets toward the polar direction. 

We compare our results of nucleosynthesis with the observed
abundance data of intracluster medium (ICM), intergalactic medium (IGM), 
gases in the central part of M82, and
extremely metal-poor (EMP) stars in the Galactic halo. Since it
is very difficult to observe directly the explosions of the first
generation stars due to the large distance (redshift $z \gsim
20$), currently {\sl comparison of the kind carried out in this
study will offer a powerful method to support the existence of
such very massive stars.}

After describing the basic methods adopted and the assumptions
made for our models in section 2, the results are presented
and discussed in section 3, and they are compared with
observations in section 4. Further discussion and concluding
remarks are given in the last section, section 5.

\section{Methods, Assumptions \& Models}

We calculate evolution, core collapse, explosion, and
nucleosynthesis of very-massive stars with 500$M_\odot$ and
1000$M_\odot$. As mentioned in Introduction, such massive stars
may be formed in a metal-free environment. We start our
evolutionary calculations by assuming that the stars have 500 and
1000$M_\odot$, with zero-metallicity, on the pre-main sequence. As
our starting approximation we neglect radiative mass loss due to
zero metallicity (Kudritzki 2000).  Ibrahim et al. (1981), Baraffe
et al. (2001), and Nomoto et al. (2003) showed that pulsational
mass loss is not so effective for metal-free stars, so we also
neglect the pulsational mass loss. To calculate presupernova
evolution we adopt the stellar evolution code constructed by Umeda
\& Nomoto (2002) based on the Henyey method. This code is
developed from the codes constructed by Nomoto \& Hashimoto
(1988), and Umeda et al. (1999). The nuclear reaction network for calculating
nucleosynthesis and energy generation at each stage of the
evolution is developed by Hix \& Thielemann (1996). We include 51
isotopes up to Si until He burning ends, and 240 up to Ge
afterwards. Our evolutionary calculations are carried out from the
pre-main sequence up to the iron-core collapse where the central
density reaches as high as $2\times 10^{10} {\rm g cm^{-3}}$. When
the temperature reaches $5 \times {10}^9$ K, where "nuclear
statistical equilibrium" (NSE hereafter) is realized, the
abundance of each isotope is determined for a given set of
density, temperature, and $Y_{\rm e}$. Here $Y_{\rm e}$ is the
number of electrons per nucleon, defined as: \\ \begin{equation}
    Y_{\rm e} = \sum_{i} \frac{Z_{i}}{A_{i}} X_{i} .  \label{eqn:Yedef} \end{equation} where $Z_{i}$ is the atomic number, $A_{i}$ is the mass number, and $X_{i}$ is the mass fraction of species i. $Y_{\rm e}$, as well as density and temperature, is a key quantity to determine the abundance of each element. We assume NSE at $\log T$ (K) $ \geq 9.7$.

The explosion mechanism of core-collapse supernovae is not well
understood.  Moreover, we do not know beforehand how very massive
stars over 300$M_\odot$ explode due to strong gravitation even
when they are rotating. In this study, therefore, instead of going
into the problem of whether such massive stars actually explode,
we investigate the conditions required for these stars to explode,
by exploring several situations with various models. For the
explosion in the hydrodynamical simulation, we adopt the
two-dimensional (2D) Newtonian hydrodynamical code constructed by
Maeda \& Nomoto (2003) and Maeda (2004).  This code adopts the
Eulerian coordinate and solves Euler equations based on Roe's
scheme (Hachisu et al. 1992, 1994). Previously 2D simulations of
jet-induced supernova explosions have been carried out by many
authors for ordinary massive stars with $\sim$ 25$M_\odot$ to
40$M_\odot$ (Khokhlov et al. 1999; MacFadyen et
al. 2001; Nagataki 2000; Maeda et al. 2002; Maeda \& Nomoto 2003; Maeda et al. 2006).
However, there have been no such detailed calculations for
CVMSs with $M \gsim 500 M_{\odot}$. 
Because temperatures are so high at the
explosion, the pressure is radiation-dominated, and hence we use
the equation of state for the radiation and electron-positron pairs with the adiabatic index
of 4/3.

The explosion models to be explored are summarized in Tables~\ref{tab:ModelsH} and ~\ref{tab:ModelsL}.  Jets, 
which are supposed to be injected from the accretion disk, are considered to be 
the energy source of the explosion.  We first choose the initial black hole mass, 
and the outer matter accretes toward the central object. Because our hydrodynamical 
code includes gravitational force, the final black hole mass and the ejected mass are 
determined as the results of the calculations for a set of given parameters 
(Maeda \& Nomoto 2003).  One of the purposes of this study is to explore the 
condition for the stars to explode when jets are injected. As typical cases we 
choose the initial black hole mass, $M_{\rm BH0}$, at 100$M_{\odot}$ for the 
1000$M_{\odot}$ star and 50$M_{\odot}$ for the 500$M_{\odot}$ star. However, in order 
to investigate the dependence of results on this parameter, we also explore 
larger or smaller values of $M_{\rm BH0}$.

We adopt 
the spherical polar coordinate with the number of meshes set to
$150 \times 60$ for five models (higher resolution; A-1, A-2, B-1, B-2, and B-3), and $100 \times 30$ for the rest (lower resoluton). The latter models are chosen so that we can search for 
parameter dependence quickly. We describe our results 
mainly based on those of higher resolution models. However, we 
include lower resolution models to explore the detailed dependence 
on parameters. We also calculate Model B-2 with lower resolution, in order 
to numerically compare the two resolutions. The difference 
of the quantities we give in Tables~\ref{tab:EMtableH}, ~\ref{tab:EMtableL}, 
~\ref{tab:EjectStellarH}, and ~\ref{tab:EjectStellarL} between the two 
resolutons is around 10 \%.
 
At the beginning of hydrodynamical simulations, the region $M_{\rm BH0} < 
M_r$ is mapped onto the computational domain. 
The central part ($M_r \le M_{\rm BH0}$) is displaced by a point mass. 
The inner boundary of the simulations is set at the radius $R_0$ 
(see Tables~\ref{tab:ModelsH} and ~\ref{tab:ModelsL}).  
In the computational domain, we assume that the effect of rotation is negligible. 
This assumption applies if the specific angular momentum 
$j_{17}=j/10^{17}$ cm$^{2}$ s$^{-1}$ in the progenitor star is in 
the range $6.3 \le j_{17} <<45$, where the lower and upper limits 
correspond to the conditions that the disk forms beyond the schwarzshild radius and well below 
the inner boundary of our computational domain. If $j_{17} \sim 6.3$, which 
is favorable in order to make an efficiently accreting disk 
(Narayan, Piran, \& Kumer 2001), then the rotational force is 
at most a few percent of the gravitational force at the inner boundary.

For the jet injection, we choose various values for the parameter, 
$\theta_{\rm jet}$, the angle from the polar axis.  
The jet is injected into the direction of $0 \leq \theta \leq \theta_{\rm jet}$. 
At the direction $\theta > \theta_{\rm jet}$,   
the inner boundary is treated as follows: 
it is set to be transmitted (absorbed; i.e., vanishing radial gradient of all variables) 
or reflected if the material just above the boundary 
has negative (accreting) or positive sign. 
These boundary conditions are used in previous studies on jet-induced 
supernova explosions (Khokhlov et al. 1999; Macfadyen, Woosley, \& Heger 2001, 
Maeda \& Nomoto 2003). 
It should be noted that by using the transmitted boundary condition 
for the accretion case and neglecting pressure and rotational support below the boundary, 
we may overestimate the accretion rate.

The energy and mass injected by the jets per unit time are connected with the properties 
of accreting matter as (Maeda \& Nomoto 2003):
\begin{equation}
    \dot{E}_{\rm jet} = \epsilon \dot{M}_{\rm acc}c^2 =
    \dot{E}_{\rm thermal} + (\frac{1}{2} \rho_{\rm jet} v_{\rm jet}^2)
        {v_{\rm jet}} A_{\rm jet}
\label{eqn:dotE}
\end{equation}
\begin{equation}
    \dot{M}_{\rm jet} = \mu \dot{M}_{\rm acc}
      = \rho_{\rm jet} {v_{\rm jet}} A_{\rm jet} \label{eqn:dotM} \end{equation} where $\dot{M}_{\rm acc}$ is the accretion rate, $\dot{E}_{\rm jet}$ the injected energy per unit time, $\dot{M}_{\rm jet}$ the mass spouted per unit time, $\epsilon$ the energy transformation efficiency, $\mu$ the mass fraction of the jets to the accreted matter, $\rho_{\rm jet}$ the jet density, $v_{\rm jet}$ the jet velocity, and $A_{\rm jet}$ the area 
over which the jet is spouted, respectively. We treat $\epsilon$ and $\mu$ as free parameters to be varied to explore the explosion energy.

We consider two cases for the form of the injected energy. One is that almost all the energy of the jet is given as kinetic energy (Case A). The other is that almost all the energy is given as thermal energy (Case B).  We introduce a parameter $F_{\rm thermal}$ defined as: \begin{equation}
    F_{\rm thermal} = \frac{\dot{E}_{\rm thermal}}{\dot{E}_{\rm jet}}  \label{eqn:Fth} \end{equation} i.e., the ratio of thermal energy in the jet to the total jet energy per unit time. This parameter is set to 0.01 for Case A, and to 0.9 or 0.95 for Case B. By using equations 2, 3 and 4 we obtain the jet velocity: \begin{equation}
      {v_{\rm jet}} = {\left[ \frac{2 \epsilon
      (1 - F_{\rm thermal})}{\mu} \right]}^{1/2} c. \label{Vjet} \end{equation}

Equations 2,3, and 4 give a complete set of jet properties at 
the inner boundary.
In Case A the jet carries most energy toward the polar
(jet-injected) direction, and hence the model is highly
non-spherical. For this case, we set the ratio between the two
parameters, $\epsilon$/$\mu$, to 0.1. Because we perform a
Newtonian calculation, the larger we set $\epsilon$ the larger we
need to set $\mu$, so that the jet material does not exceed or
approach the speed of light. For Case B, on the other hand, the
models, due to the dominant thermal energy, become more spherical,
because thermal motion is random and non-directional. For this
case, we set larger $\epsilon$ values for the same $\mu$ compared
with Case A. This means that the jet is something like a hot
bubble.

One of our primary purposes is to investigate how much heavy
elements are synthesized and ejected by the explosion. Therefore
we stop the calculations when all of the following conditions are
satisfied: 
\begin{itemize}
  \item[1] $\dot{M}_{\rm acc}$ decreases enough below $0.1
M_{\odot}$
   s$^{-1}$, i.e., $\lsim 0.02 M_{\odot}$ s$^{-1}$. 
  \item[2] Total explosion energy $E_{\rm tot}$ becomes much larger than the
   absolute gravitational binding energy $|E_{\rm grav}|$, $E_{\rm
   tot} \gsim 10 |E_{\rm grav}|$.  \\
   These criteria mean that accretion has almost stopped.  
  \item[3] The maximum temperature of the matter decreases below $8 \times
   {10}^8$ K.  \\
           This means that explosive nucleosynthesis no longer occurs at such low temperatures.
\end{itemize}
Under these criteria, calculations sometimes end before jets reach the stellar surface.

Maeda \& Nomoto (2003) used helium stars (the hydrogen envelope is
removed by mass loss) as the initial models, and carried out the
calculations for about 100 s until the jet reaches the stellar
surface and the expansion becomes homologous.  In contrast the
radii of the stars we use here are in the order of ${10}^2
R_\odot$ ($\sim {10}^{13} {\rm cm}$) because they have the
hydrogen-rich envelope. Because we investigate the first
generation stars, the mass loss will not be effective due to the
metal-free environment. Therefore, it is reasonable to examine a
hydrogen-rich star rather than a He star.

We calculated explosive nucleosynthesis by using temporal histories of density and temperature stored during hydrodynamical calculations. The reaction network we use includes 280 isotopes up to $^{79}{\rm Br}$. At high temperatures, $T_{9} = T/10^9$ K $> 5 $, NSE is realized. We use the "NSE" code (Hix \& Thielemann 1996) for $T_{9} > 6$.

The jet matter should be included in the ejected matter and we
need to calculate its nucleosynthesis.  We do not know which of
the accreted matter is injected as jets, and so the final chemical
composition is uncertain. However, Pruet et al. (2004) carried out
nucleosynthesis of disk wind for various $Y_{\rm e}$ values.
MacFadyen \& Woosley (1999) and MacFadyen (2003) also considered
disk wind.  Based on these works in this study we make the
following assumptions: 
\begin{itemize}
   \item[1] Because the jet matter is injected
through the inner region (from the accretion disk), it should have
experienced high temperatures at which NSE is realized ($T_{9} >
5$).
   \item[2] The jet matter expands adiabatically after it is
injected (i.e., entropy is conserved). 
   \item[3] The accreted matter
is mostly accreted while the accretion rate $\dot{M}_{\rm acc}$ is
of the order of 10 - ${10}^2 M_{\odot}$ s$^{-1}$.  It is likely that $\rho$ 
varies depending on when the jet material is injected. Therefore $s$ (entropy density, $\propto \frac{T^3}{\rho}$) is likely to vary as well.  
   \item[4] For the value of $Y_{\rm e}$ we assume $0.48 \leq Y_{\rm e} \leq 0.52$.
\end{itemize}

Based on these assumptions we start the calculation of
nucleosynthesis of the jet matter from $T_{9} =6$, using the
historical temperature and density data of the first test particle
of the jet (injected at the first stage of the explosion) in Model
A-2 (history A), and the changed entropy data (double (history B) 
or triple (history C) the density at the same temperature). In other 
words, we use three $\rho - T$ histories 
(history A:($\rho (t),T(t)$), 
history B:($2\rho (t),T(t)$), and 
history C:($3\rho (t),T(t)$)), where the set ($\rho (t), T (t)$ ) 
is given by the hydrodynamic simulations. 
$Y_{\rm e}$ is parameterized at 0.48, 0.49, 0.50, 0.51, and 0.52.
We calculate 15 patterns and average these results to the first
approximation. $Y_{\rm e}$ and entropy of the jet material can
change when it is ejected. Therefore here we consider combination
of jets with different values of these parameters. These
assumptions still include large uncertainty, but our aim is just
to roughly estimate the amount of $^{56}{\rm Ni}$. The larger the
mass of the jet is, the larger we expect the uncertainties to be.

\section{Results}

\subsection{Presupernova Evolution}

\subsubsection{Evolutionary Tracks}

\begin{figure}
%\epsfxsize=10cm   %width of figure - will enlarge/reduce the figures
%\epsfbox{fig3.eps}
%\figurebox{2cm}{3cm}{} %to have a box alone
\plotone{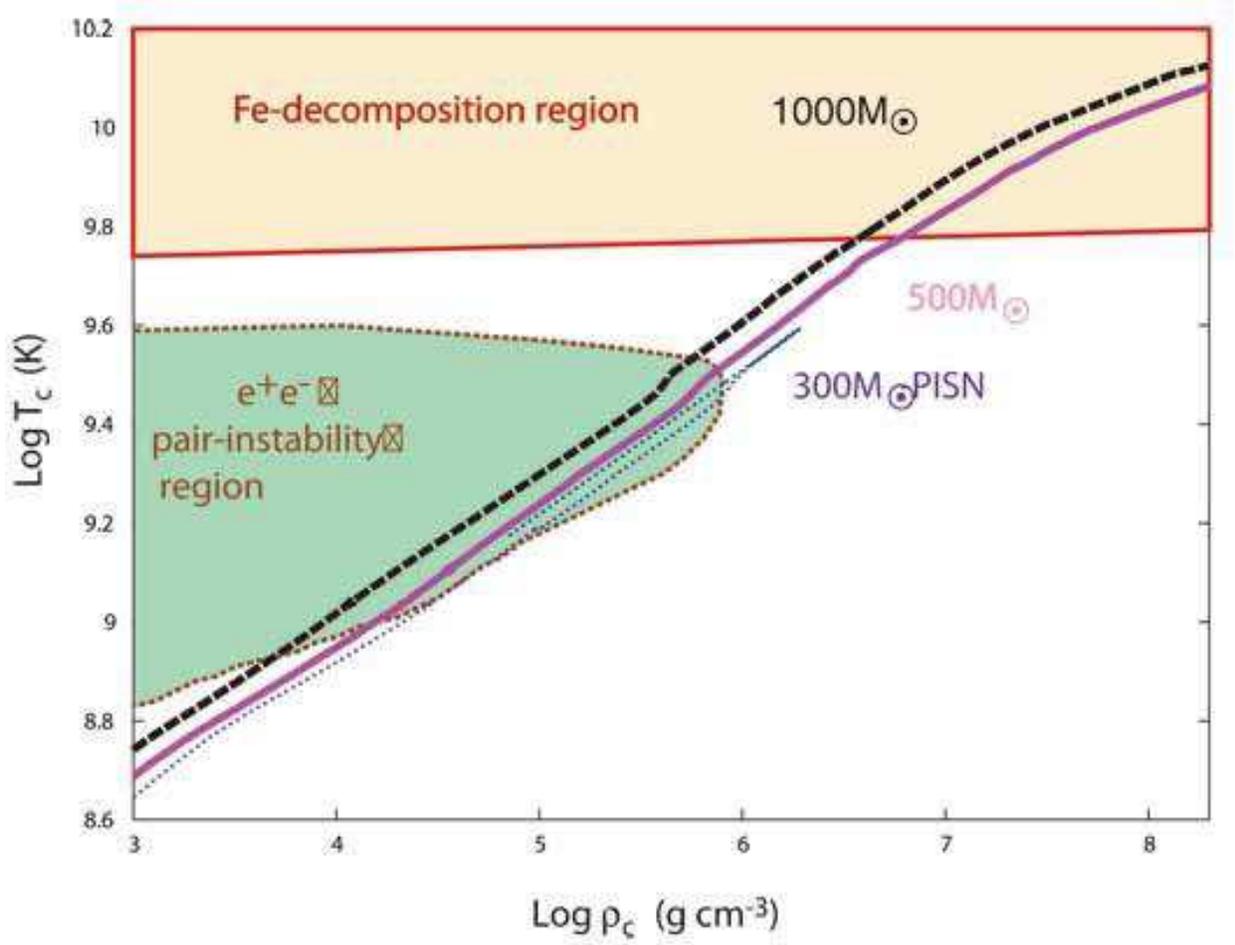}
\caption{Evolutionary tracks of central temperature and density of the stars with 300$M_\odot$ (thin dotted line), 500$M_\odot$ (thick solid line),
and 1000$M_\odot$ (thick dashed line). \label{track}}
\end{figure}

Figure~\ref{track} shows the evolutionary tracks of the central
density - temperature relation for 500$M_\odot$ and
1000$M_\odot$ stars.  We also plot, for comparison, the track of
a 300$M_\odot$ star, which results in the pair-instability
supernova.  Generally, more massive stars have higher entropies
(lower densities) at the same temperatures (i.e., at the same
burning stage).  Although each star passes through the region of
electron-positron pair-instability, both 500$M_\odot$ and
1000$M_\odot$ stars proceed to iron-core collapse
(Fe-decomposition region in Figure~\ref{track}), unlike the 300$M_\odot$
star. The 500$M_\odot$ and 1000$M_\odot$ stars do not become
pair-instability supernovae though they pass through the
pair-instability region, because the energy released at this stage
is less than the gravitational binding energy of the star (Rakavy
et al. 1967; Bond et al. 1984; Glatzel et al. 1985; Woosley 1986).

\subsubsection{Presupernova Model}

\begin{figure}
%\epsfxsize=10cm   %width of figure - will enlarge/reduce the figures
%\epsfbox{fig3.eps}
%\figurebox{2cm}{3cm}{} %to have a box alone
\epsscale{.70}
\plotone{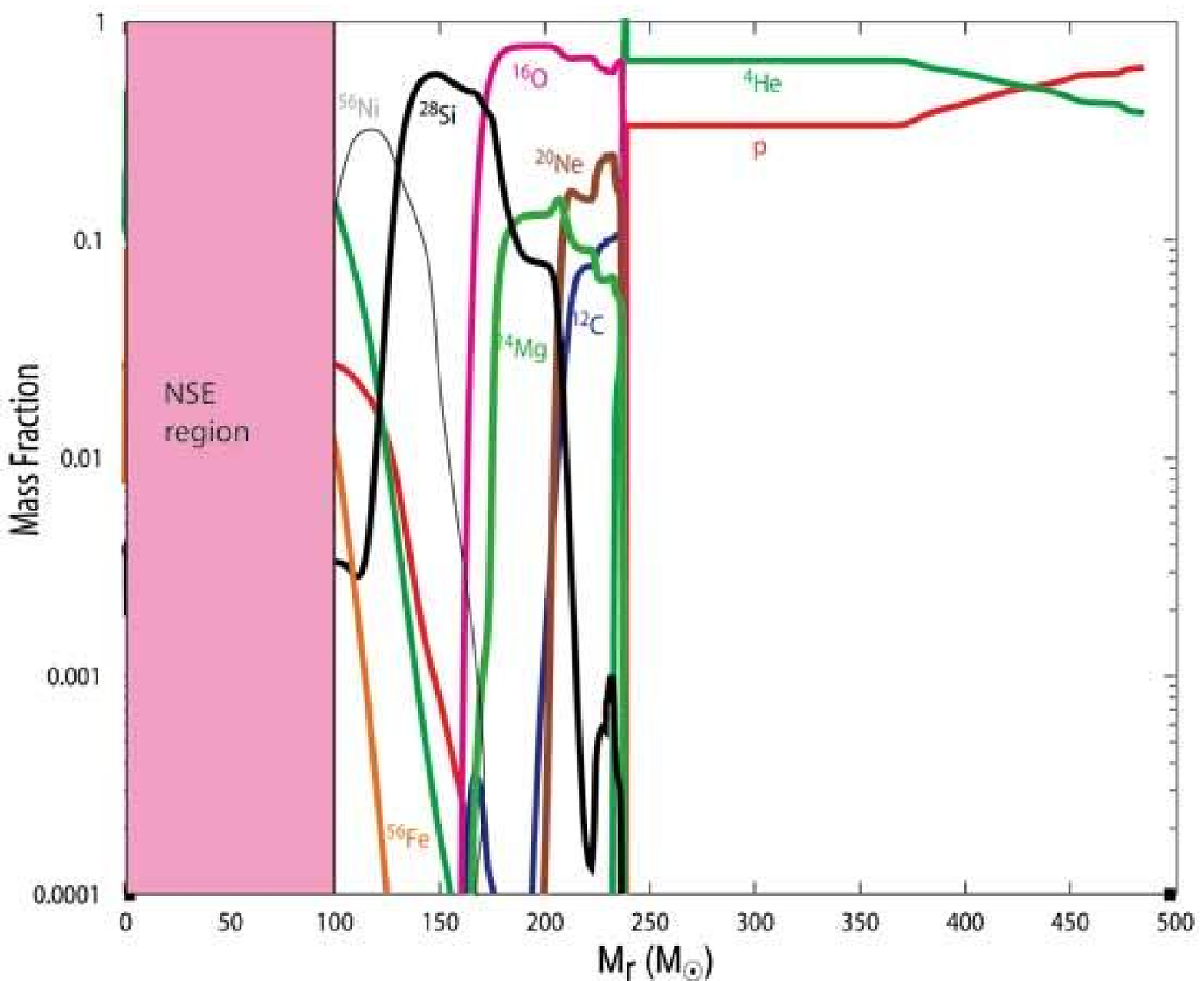}
\plotone{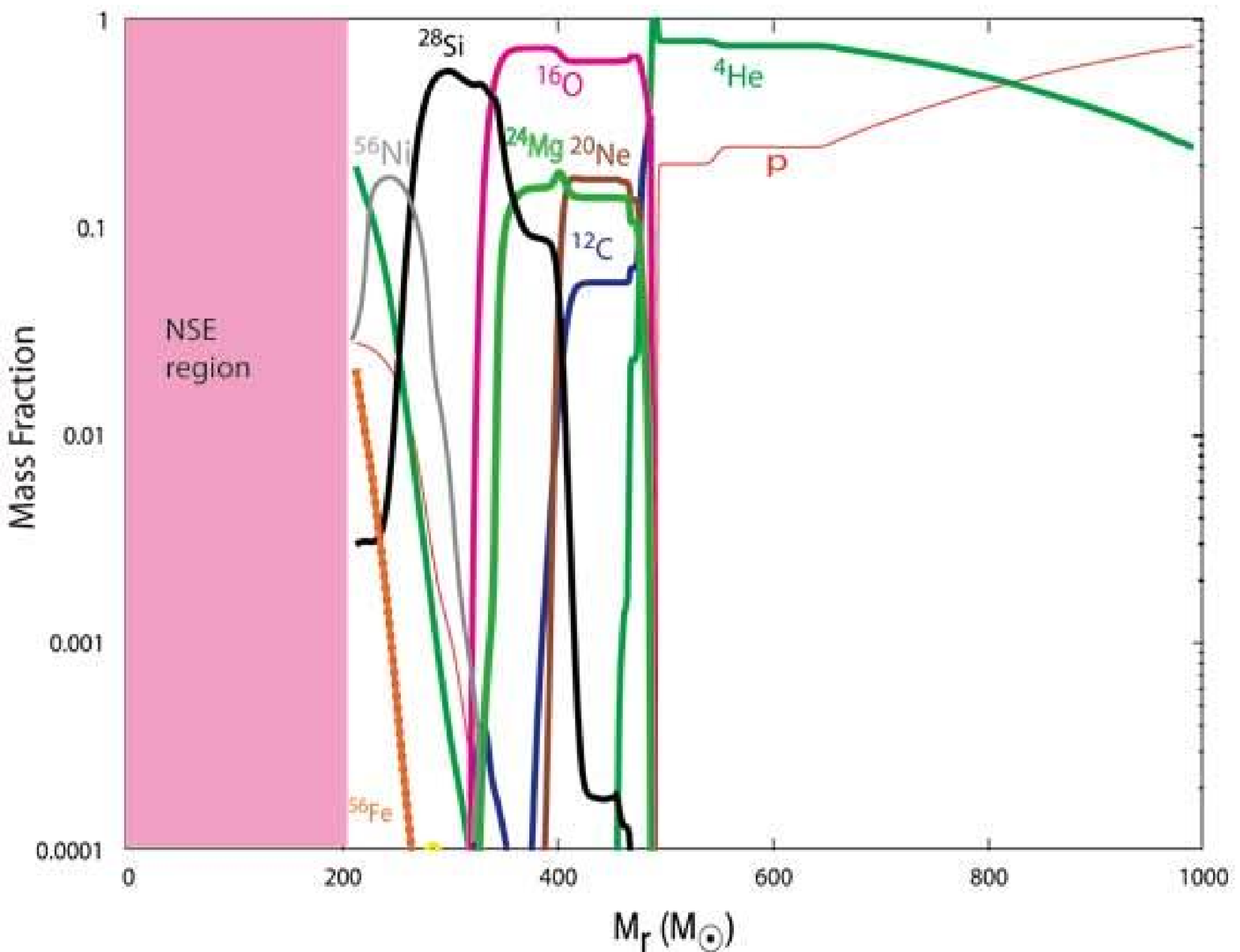}
\caption{Chemical composition just before the explosion (when the central density reaches 
$10^{10}$ g cm$^{-3}$) of the 500$M_\odot$ star (upper panel), and
1000$M_\odot$ star (lower panel). The iron core occupies more than 20\% of the total mass for both
cases.
\label{fig:ChemiCon}}
\end{figure}

Figure~\ref{fig:ChemiCon} shows the presupernova chemical
composition for the 500$M_\odot$ and 1000$M_\odot$ stars. In the
region labeled as "NSE region", NSE is realized.  For this
region, we calculate the evolutional changes in terms of ($Y_{\rm
e}, \rho, T$) to obtain the NSE abundances.  One can see the
onion-like structure from the center to the surface - the
iron-core, silicon layer, oxygen layer, helium layer, and
hydrogen-rich layer. Here we define the iron-core as the region
where the mass fraction of Si is less than 10\%. The iron-core
occupies up to 130$M_\odot$ of mass from the center for the
500$M_\odot$ star, and 250$M_\odot$ for the 1000$M_\odot$ star.
For both cases, they occupy a quarter of the total mass.  This
fraction is much larger than that in ordinary massive stars. For
example, in a 25$M_\odot$ star, the iron-core is about
1.6$M_\odot$(Umeda et al. 1999), less than 10\% of the total mass.
The reason is the difference of the density and temperature
structure.

\begin{figure}
%\epsfxsize=10cm   %width of figure - will enlarge/reduce the figures
%\epsfbox{fig3.eps}
%\figurebox{2cm}{3cm}{} %to have a box alone
\plotone{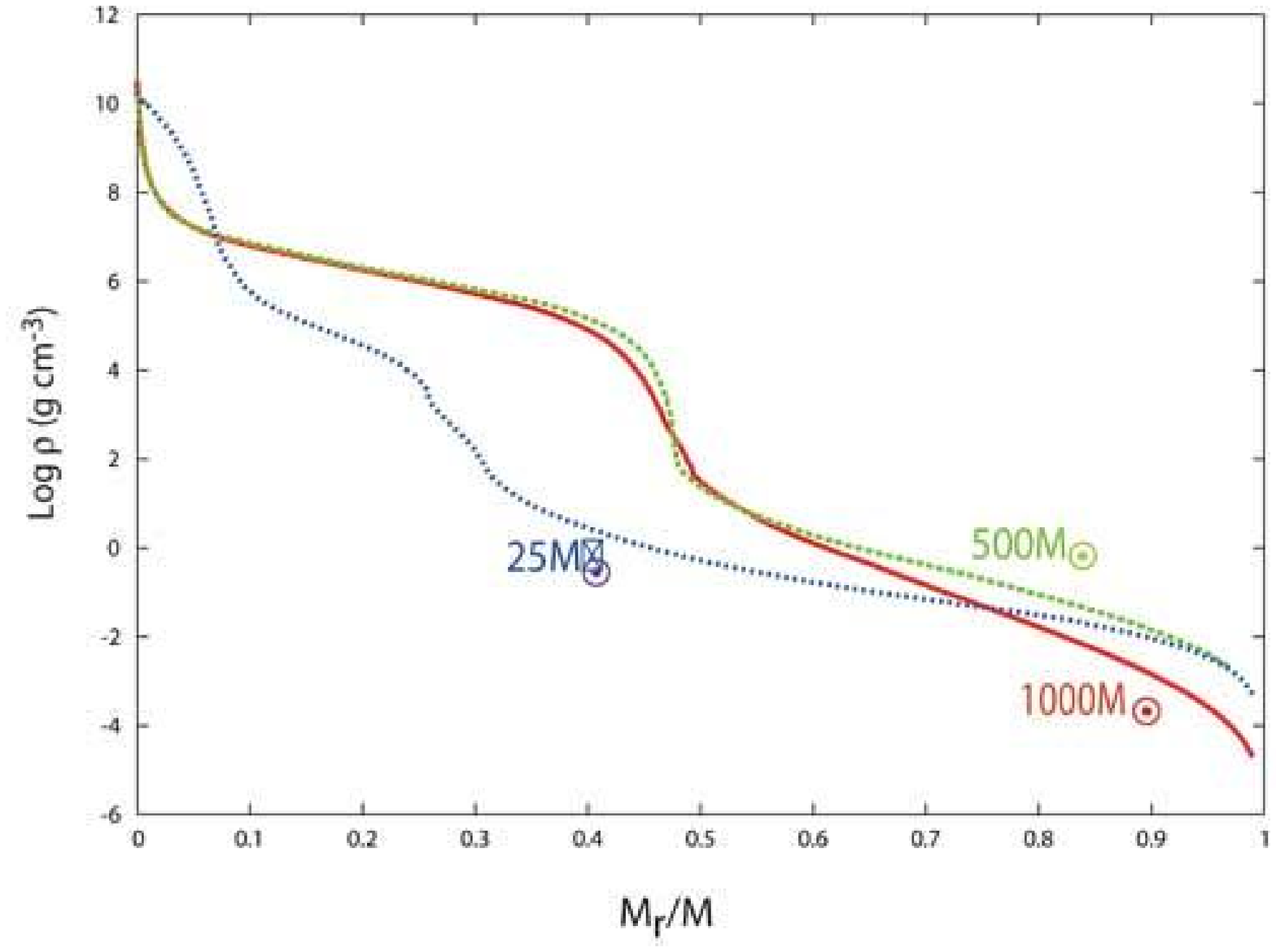}
\plotone{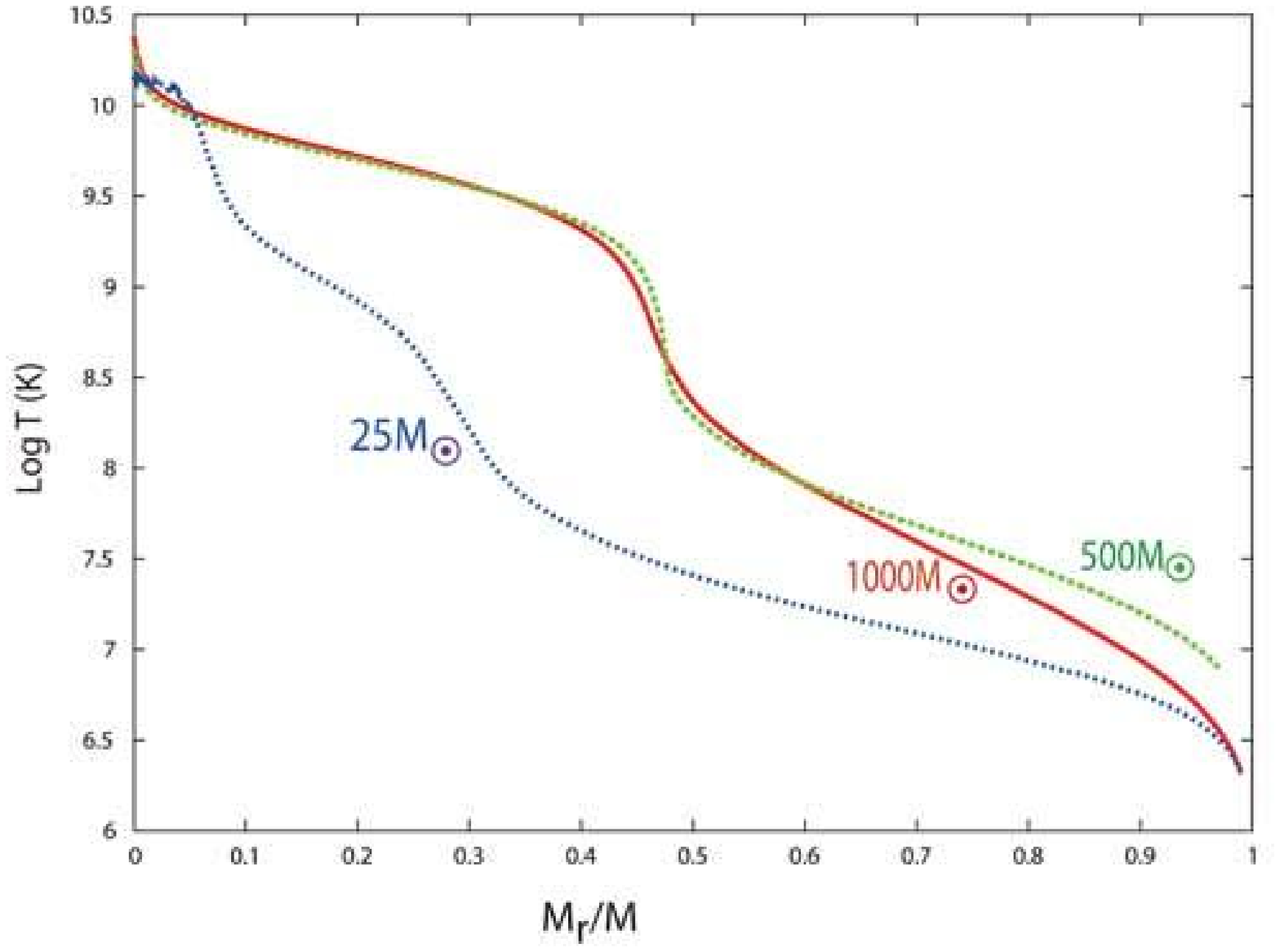}
\caption{Density structure (top panel) and temperature structure (bottom
panel) of 25$M_\odot$, 500$M_\odot$,
1000$M_\odot$ models. The horizontal axis is
the mass fractions $M_{\rm r}/M_{\rm total}$. The vertical axis shows
the density (top), and the temperature (bottom),
respectively.
The data of 25$M_\odot$ is from Umeda \& Nomoto (2003).
\label{qMr-Rho}}
\end{figure}

Figure~\ref{qMr-Rho} shows the density and temperature structure 
of the two stars just before the explosion (when the central density 
reaches $10^{10}$ g cm$^{-3}$), which is compared with 
the 25$M_\odot$ model. The density and temperature gradients for the 
500$M_\odot$ and 1000$M_\odot$ stars are smaller than those of the 
25$M_\odot$ star, and hence the regions with high temperature and high 
density are larger. Then the fraction of the iron-core is larger. 
The large drop of density at $M_{\rm r}/M_{\rm total} \sim 0.5$ for 
the 500$M_\odot$ and 1000$M_\odot$ stars in Figure~\ref{qMr-Rho} 
corresponds to the boundary between the oxygen and helium layer.

\subsection{Explosion Hydrodynamics}

\begin{figure}
\plotone{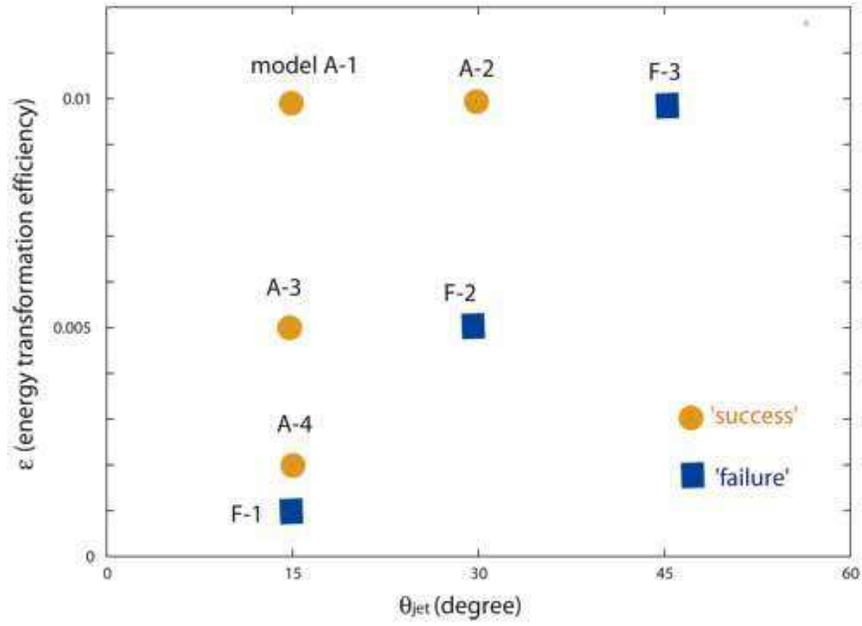}
\caption{Models in which explosion occurs (filled circles) or not (filled squares),
depending on two parameters $\theta_{\rm jet}$ and $\epsilon$
for 1000$M_{\odot}$ models.
The other parameters are set at $\mu = 10\epsilon$, $M_{\rm BH0} = 100M_{\odot}$, $f = 0.01$
(see Table~\ref{tab:ModelsH}, ~\ref{tab:ModelsL}, and ~\ref{tab:Failure}).  
\label{fig:Success}}
\end{figure}

We describe the results of explosion hydrodynamics in this subsection 
and nucleosynthesis in the next subsection, showing several figures. All 
figures are based on the results of high resolution models, except for 
Figure~\ref{fig:Success}.

\subsubsection{Explosion Energy and Ejected Mass}

In Tables~\ref{tab:EMtableH} and ~\ref{tab:EMtableL} we summarize for each model the total explosion energy,
final black hole mass, and mass of the jets. The total explosion
energy is of the order of ${10}^{54}$ erg for most cases, except
that in Model A-4 it is of the order of ${10}^{53}$ erg.  Model
A-4 is almost at the border between the 'successful' and 'failed'
explosions. Actually, we also try to calculate the case which has
the same parameters as Model A-4 except that $\epsilon$ and $\mu$
are half the values of A-4 (see Model F-1 in Table~\ref{tab:Failure}), 
but in this model the jet promptly
falls back to the central remnant after it is injected, and hence
the explosion fails. In this model, the total energy is still
negative and the stellar matter moves toward the central remnant
more than 200 s after the beginning of the accretion. The absolute
value of gravitational binding energy over the region outside of
the central 100$M_{\odot}$ core is as high as ${10}^{55}$ ergs for
the 1000$M_{\odot}$ model. In this case energy injection is too
weak for the jet to proceed outward.

Table~\ref{tab:Failure} summarizes the models in which explosion ends up as
'failure'. Figure~\ref{fig:Success} shows the models in which the
explosion either occurs or not, depending on the two parameters
$\theta_{\rm jet}$ and $\epsilon$. One can see that the minimum
$\epsilon$ needed for the successful explosion becomes higher if
$\theta_{\rm jet}$ is larger, as in Model A-2 and Case B (most of
the injected energy is given as thermal energy). Actually,
explosion energies tend to be lower in such models than those in
Case A models with $\theta_{\rm jet} = {15}^o$.  In Model A-2 and
Case B models, the injected energy tends to diffuse into the
direction apart from the polar direction, and so the jet is weak
even for the polar direction.

\begin{figure}
\epsscale{1.0}
\plottwo{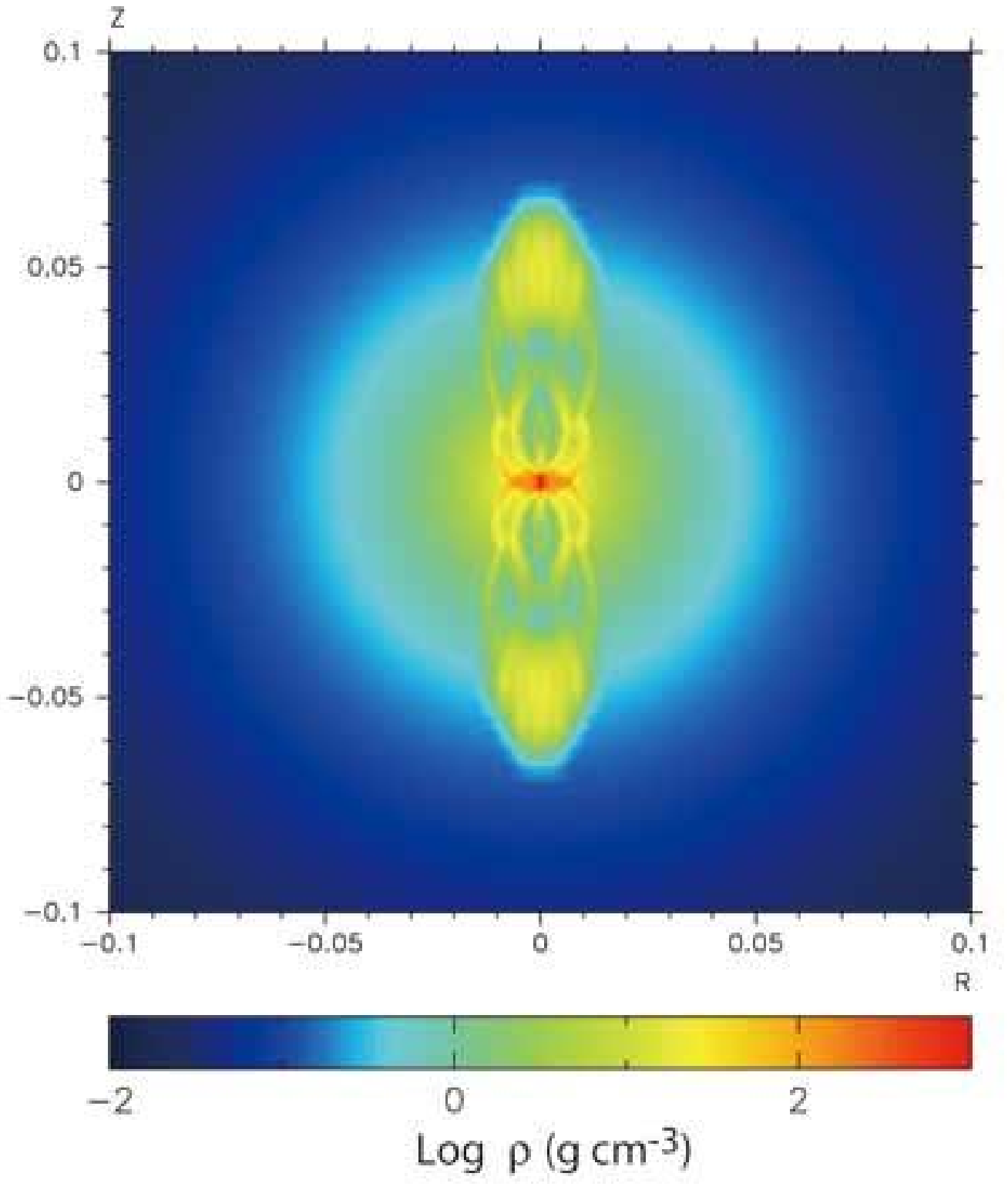}{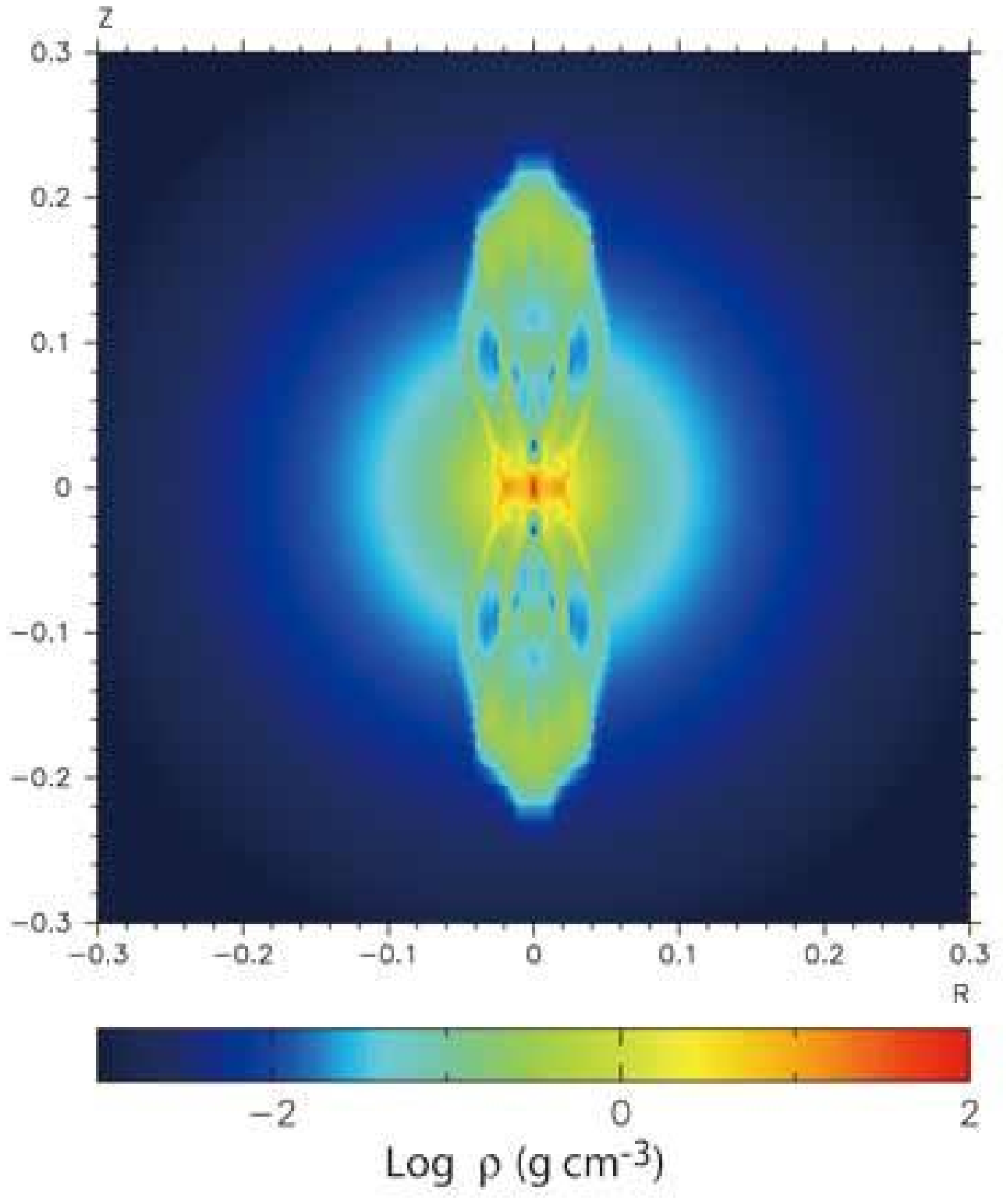}
\plottwo{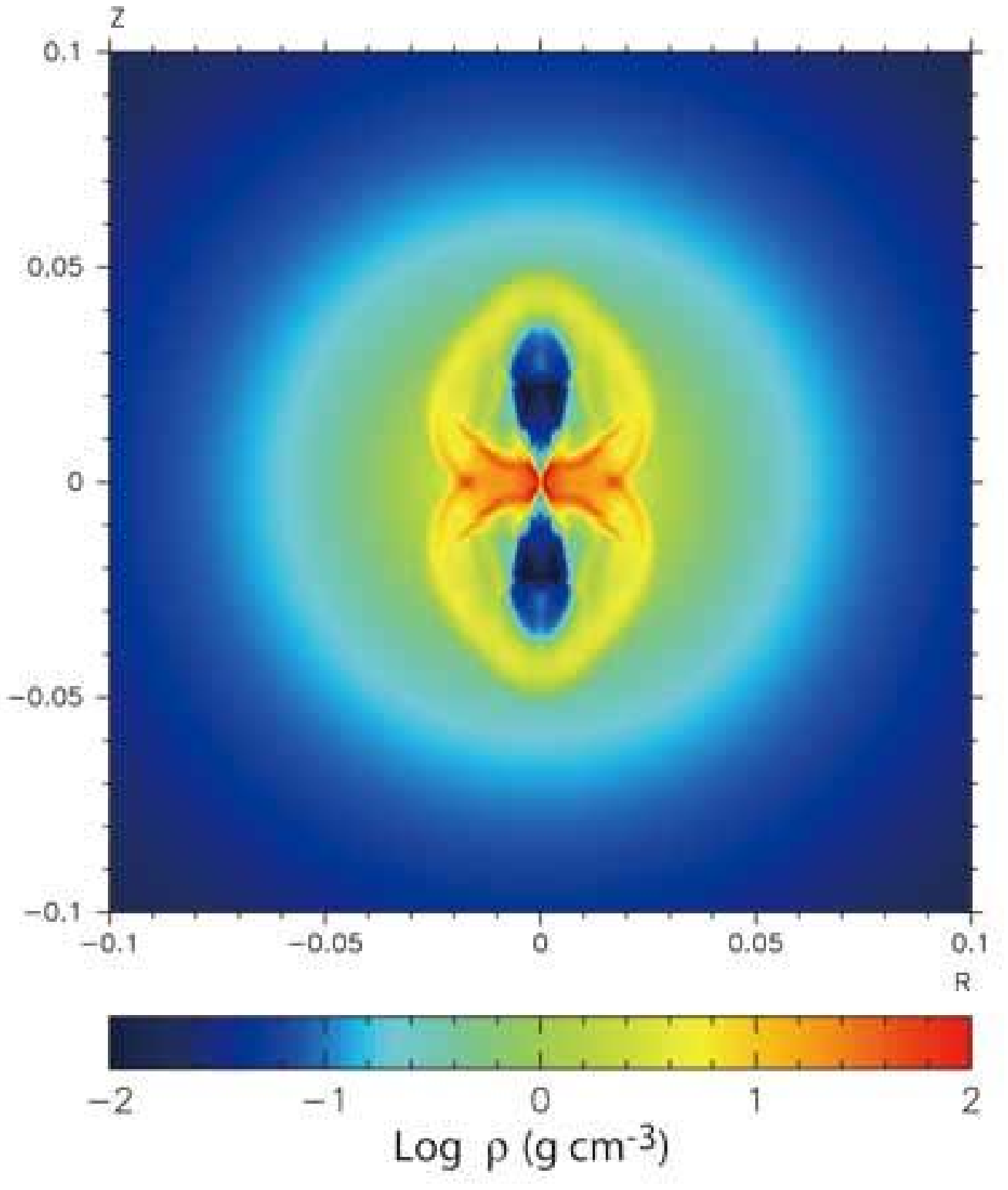}{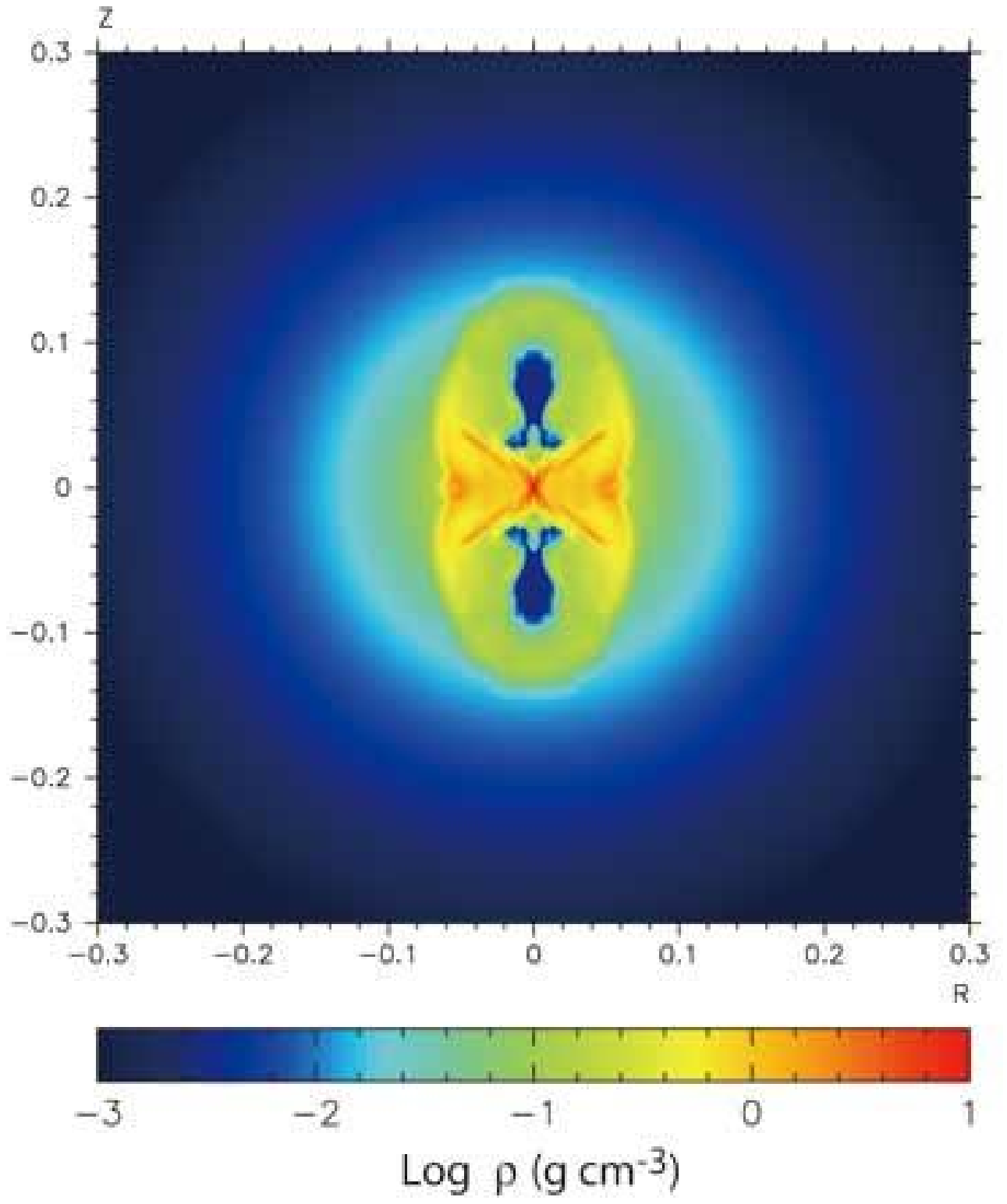}
\caption{Snapshots of density structure showing how the jet is propagating
at 30 second and 100 second after we started the calculation. The dial is normalized by the star's radius ($\sim 7.7 \times
 10^{12} $ cm). Note that 
${\theta}_{\rm jet}$ is set at ${15}^o$ for both models. top left: 30s (A-1); top right: 100s (A-1); bottom left: 30s (B-1); bottom right: 100s (B-1). \label{fig:Snapshot}}
\end{figure}

Figure~\ref{fig:Snapshot} shows how the jet is propagating through the star 
by plotting the density structures. One can see that the jet is strongly 
propagating in the polar direction for A-1, while for B-1 the jet is broadened 
toward the side directions due to the random heat motion.

The final black hole mass and ejected mass are also important. For the 500 $M_{\odot}$ and the 1000$M_{\odot}$ models, these values are $\sim 230M_{\odot}$ and $500M_{\odot}$, respectively.

\subsubsection{Direction-Dependent Features}

\begin{figure}
\epsscale{.30}
\plotone{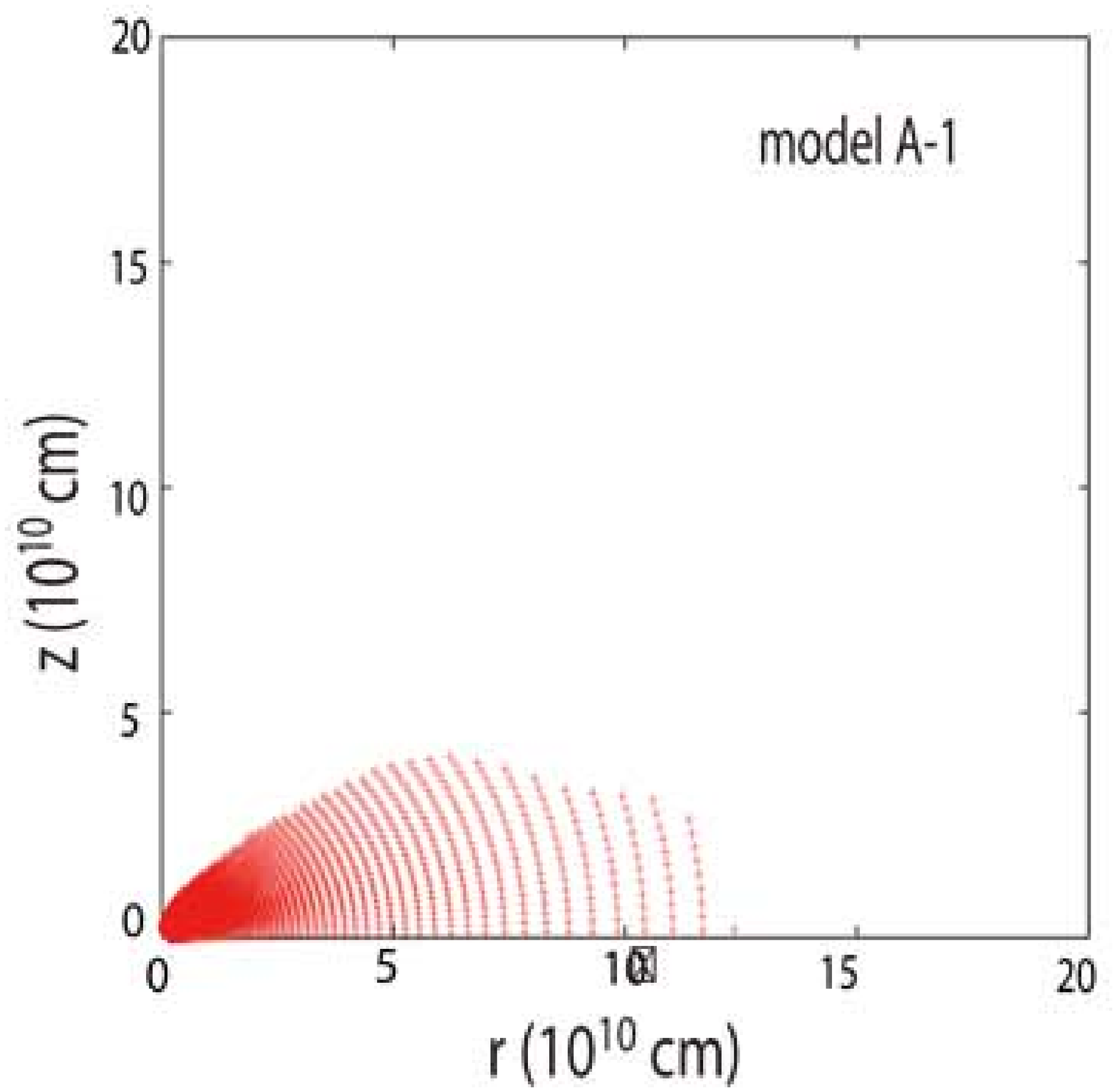}
\plotone{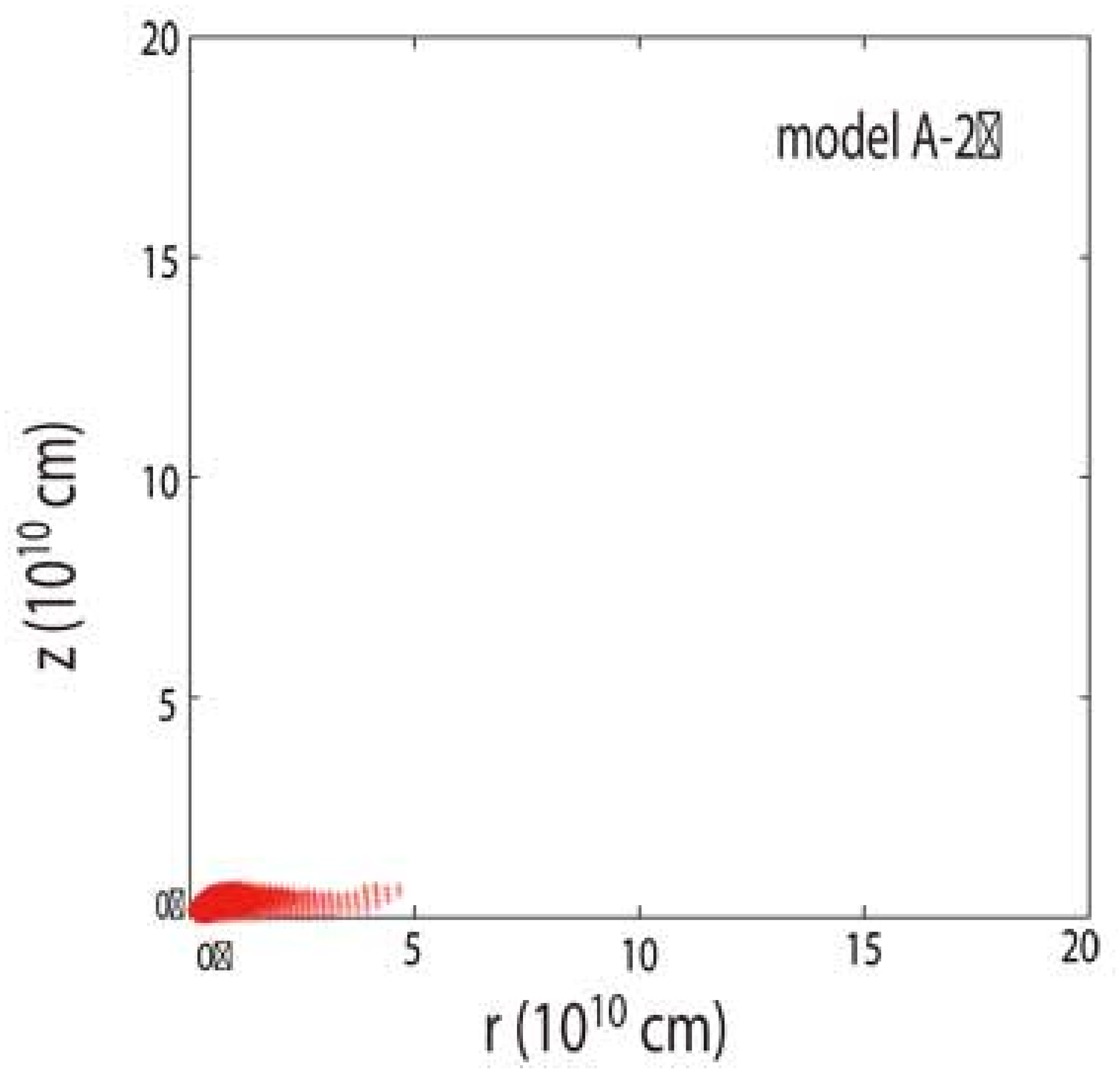}
\plotone{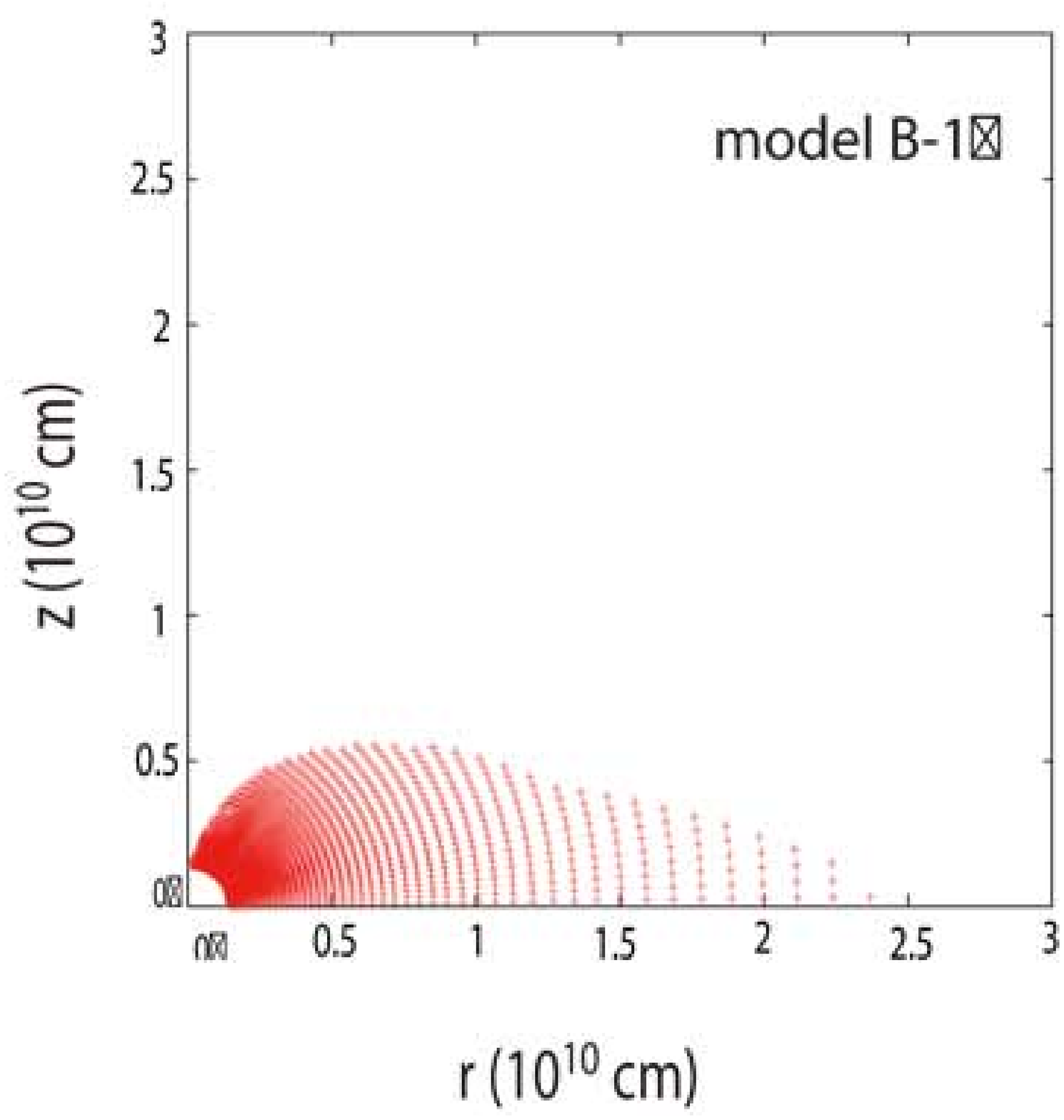}
\caption{Initial radial positions of the matter that will be accreted into the central black hole for models A-1 (left), A-2 (middle),
and B-1 (right). The blank region near the origin corresponds to
the region of black hole initially formed.
\label{fig:masscutMR1}}
\end{figure}

Figure~\ref{fig:masscutMR1} shows the regions where the matter
will be accreted onto the central black hole.  In these figures we
can see the extent of asphericity and the amount of accreting
matter. These panels show the initial positions (just before the
explosion) of the accreted matter.  In Model A-2 in which
$\theta_{\rm jet}$ is twice (${30}^o$) the other models (${15}^o$)
and Case B, asphericity is weakened to some extent and the amount
of the accreting matter is less compared with the models with
$\theta_{\rm jet} = {15}^o$ and the same energy transformation
efficiency $\epsilon$. For models with $\theta_{\rm jet} =
{15}^o$, the stellar matter toward the direction $\theta >
\theta_{\rm jet}$ almost accretes up to 500$M_{\odot}$ (even a
part of the helium layer) on the mass coordinate. On the other
hand, in Model A-2 and Case B models a large amount of matter
inner than the 500$M_{\odot}$ core is ejected. At the same time,
if asphericity becomes weaker the threshold efficiency for the
successful explosion becomes more strict-- that is, larger
$\epsilon$ is needed. This is because more energy diffuses toward
the equatorial direction.

\begin{figure}
\epsscale{.30}
\plotone{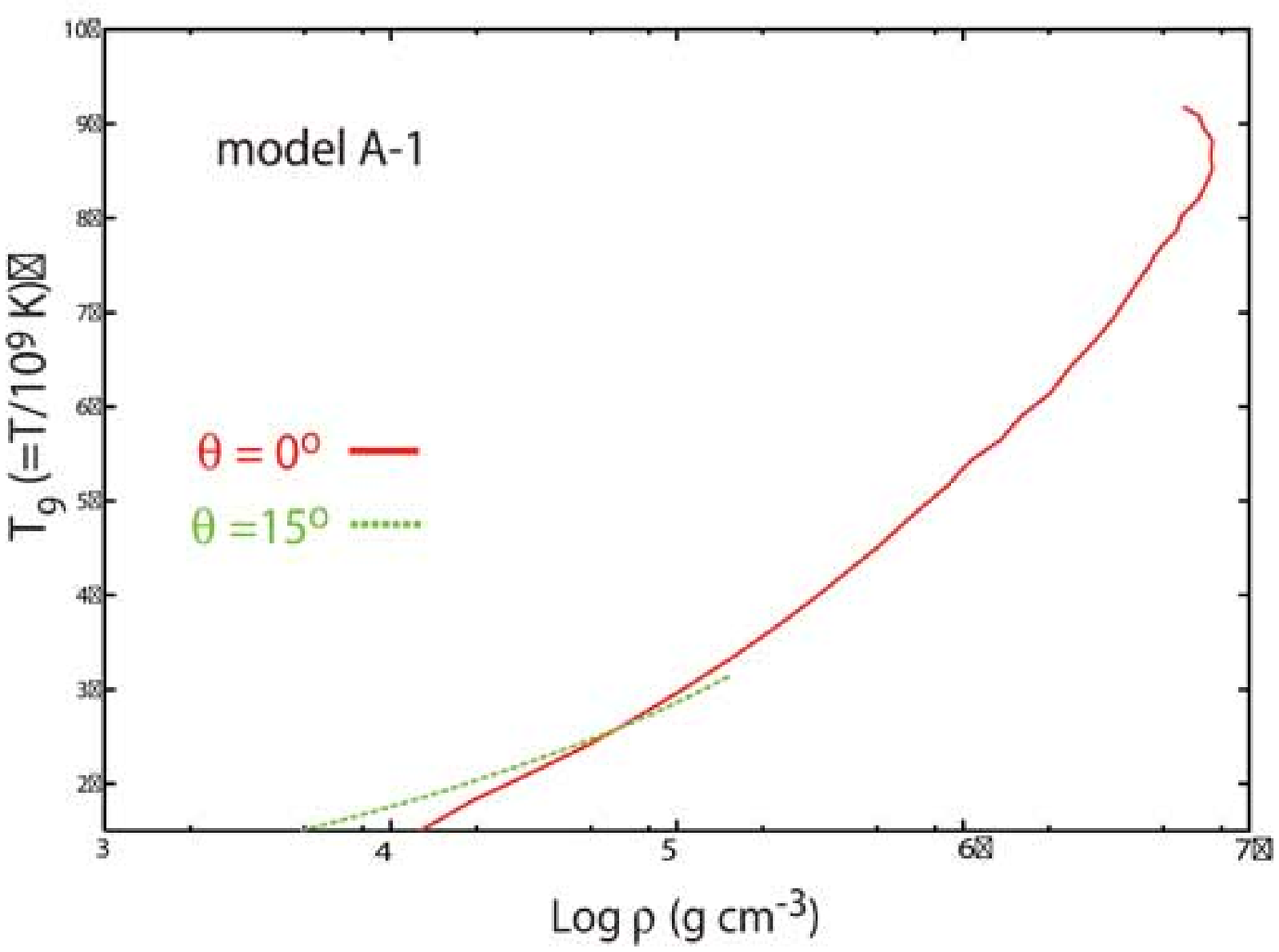}
\plotone{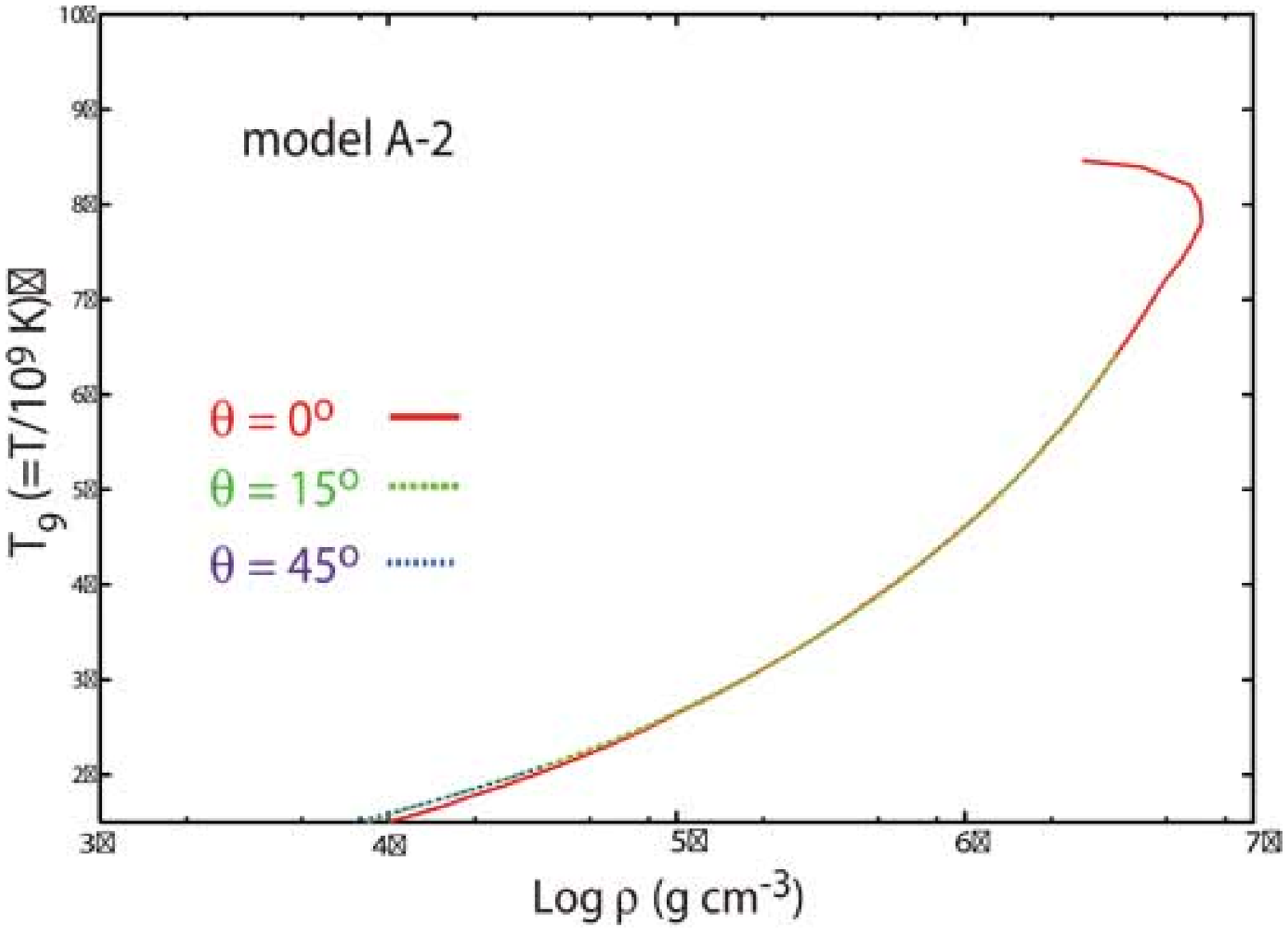}
\plotone{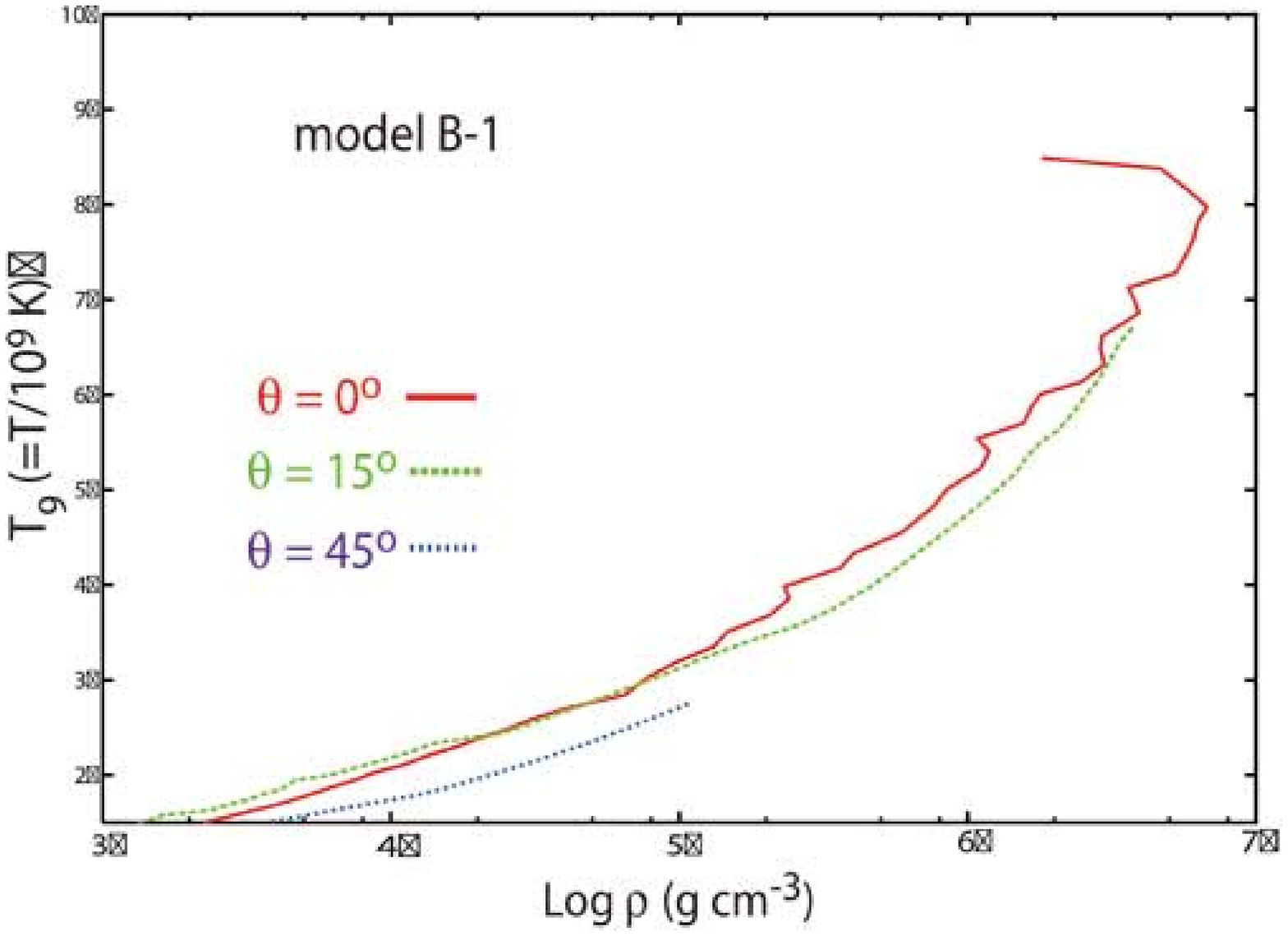}
\caption{Maximum temperatures and densities of each mesh point for the directions of $\theta = {0}^o$,
$\theta = {15}^o$, $\theta = {45}^o$ for models A-1 (left), A-2 (middle),
and B-1 (right).
\label{fig:maxTR1}}
\end{figure}

In Figure~\ref{fig:maxTR1} we show the maximum temperatures which
each mesh reaches and the densities at the maximum temperatures
for the $z$ (polar)-direction, $\theta = {15}^o, \theta = {45}^o$.
We may pay special attention to the maximum $T_{9}$ for the
$\theta = {15}^o$ and $\theta = {45}^o$ direction. For some models
(e.g., A-1 in the left panel) with $\theta_{\rm jet} = {45}^o$,
the maximum $T_{9}$ does not appear within the range of these
graphs, because the matter which can experience such high
temperatures is in the inner region, and hence such matter all
accretes in these models. However, for the other cases, the inner
matter is ejected and the explosive nucleosynthesis occurs even
for the $\theta = {45}^o$ direction.

\subsection{Explosive Nucleosynthesis}

When a shock arrives the shocked region is compressed and heated, drastically raising the density and temperature, and then the explosive nucleosynthesis occurs. The products of this event are characterized by the peak temperature. We first summarize the main products at different peak temperatures and then describe the results of the calculations.

\subsubsection{Explosive Burning and Products}

If the peak temperature $T_{\rm peak}$ exceeds $5 \times {10}^9$ K, NSE is realized.  In such regions 'complete silicon burning' occurs and then Fe-group elements (such as Mn, Co, Fe, Ni) are produced. The main product is $^{56}{\rm Ni}$, which eventually decays into $^{56}{\rm Fe}$.

In the complete silicon burning region, at lower density for a given temperature the reaction rate decreases, and the number of free-particles may exceed the NSE value. Or, if the initial temperature is higher, free-particles become more abundant because in NSE the number of these particles is a high-powered function of temperature. This situation is called '$\alpha$-rich freezeout', and it tends to produce the Fe-group elements and nuclei to the high-Z side of the peak (e.g. Thielemann, Nomoto, \& Hashimoto 1996; Arnett 1996).

If $4 \times {10}^9 {\rm K} < T_{\rm peak} < 5 \times {10}^9 {\rm K}$, {\sl incomplete Si burning} occurs. In such regions, Si is not all converted into the Fe-group elements but remains or is converted to the elements such as $^{32}{\rm S}$, $^{36}{\rm Ar}$, and $^{40}{\rm Ca}$.

If $3 \times {10}^9 {\rm K} < T_{\rm peak} < 4 \times {10}^9 {\rm K}$, {\sl explosive oxygen burning} occurs, which produces $^{28}{\rm Si}$ and $^{32}{\rm S}$, while the original $^{16}{\rm O}$ composition stays the same.

If $2 \times {10}^9 {\rm K} < T_{\rm peak} < 3 \times {10}^9 {\rm K}$, {\sl explosive carbon burning} occurs, which produces $^{20}{\rm Ne}$ and $^{24}{\rm Mg}$. The original $^{12}{\rm C}$ remains because the burning does not proceed during such a short time scale.

If $T_{\rm peak} < 2 \times {10}^9 {\rm K}$, almost no explosive burning occurs, and so the original chemical composition realized during the hydrostatical burning phase is conserved.

\subsubsection{Direction-Dependent Features}

\begin{figure}
\epsscale{1.2}
\plottwo{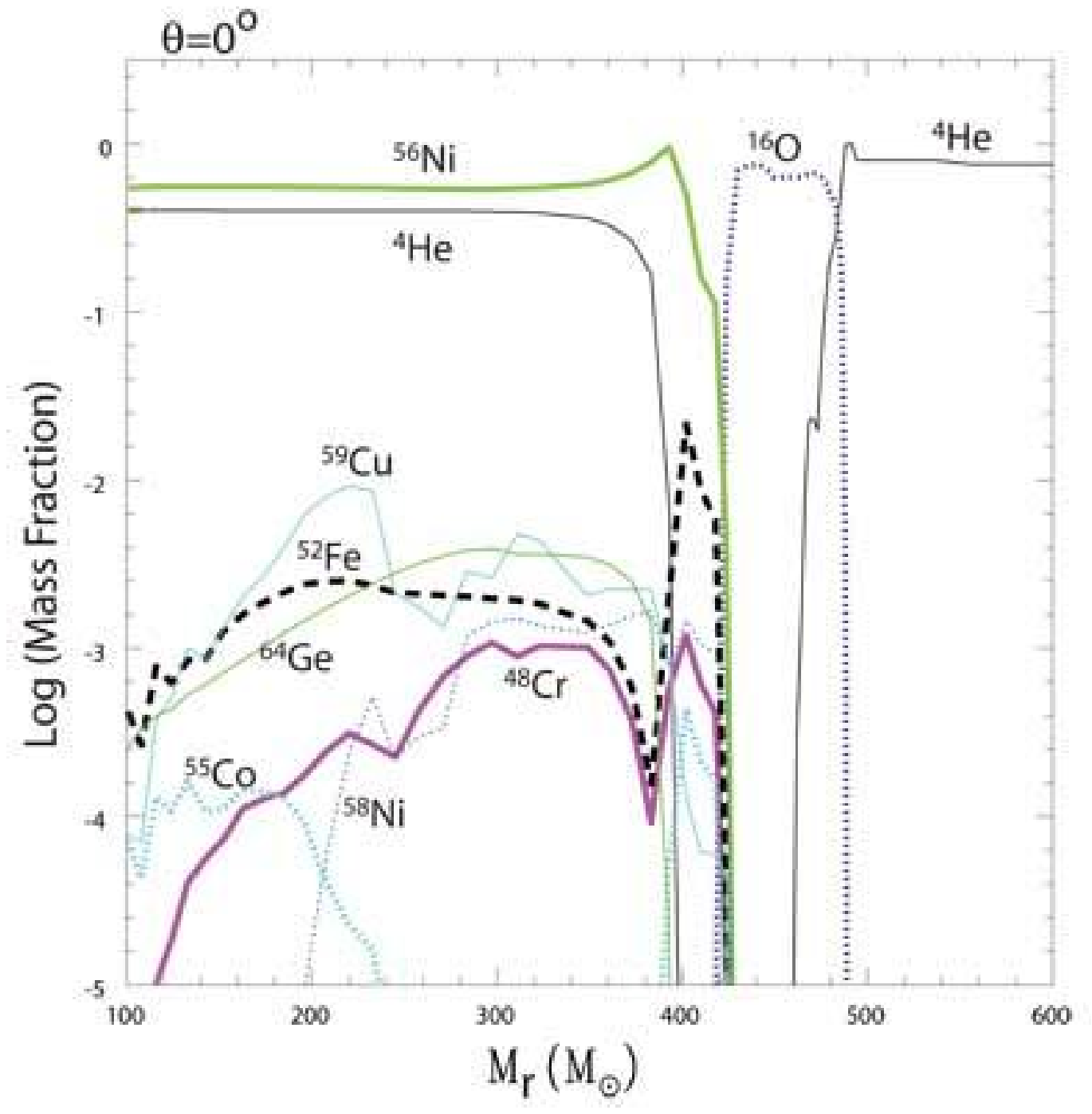}{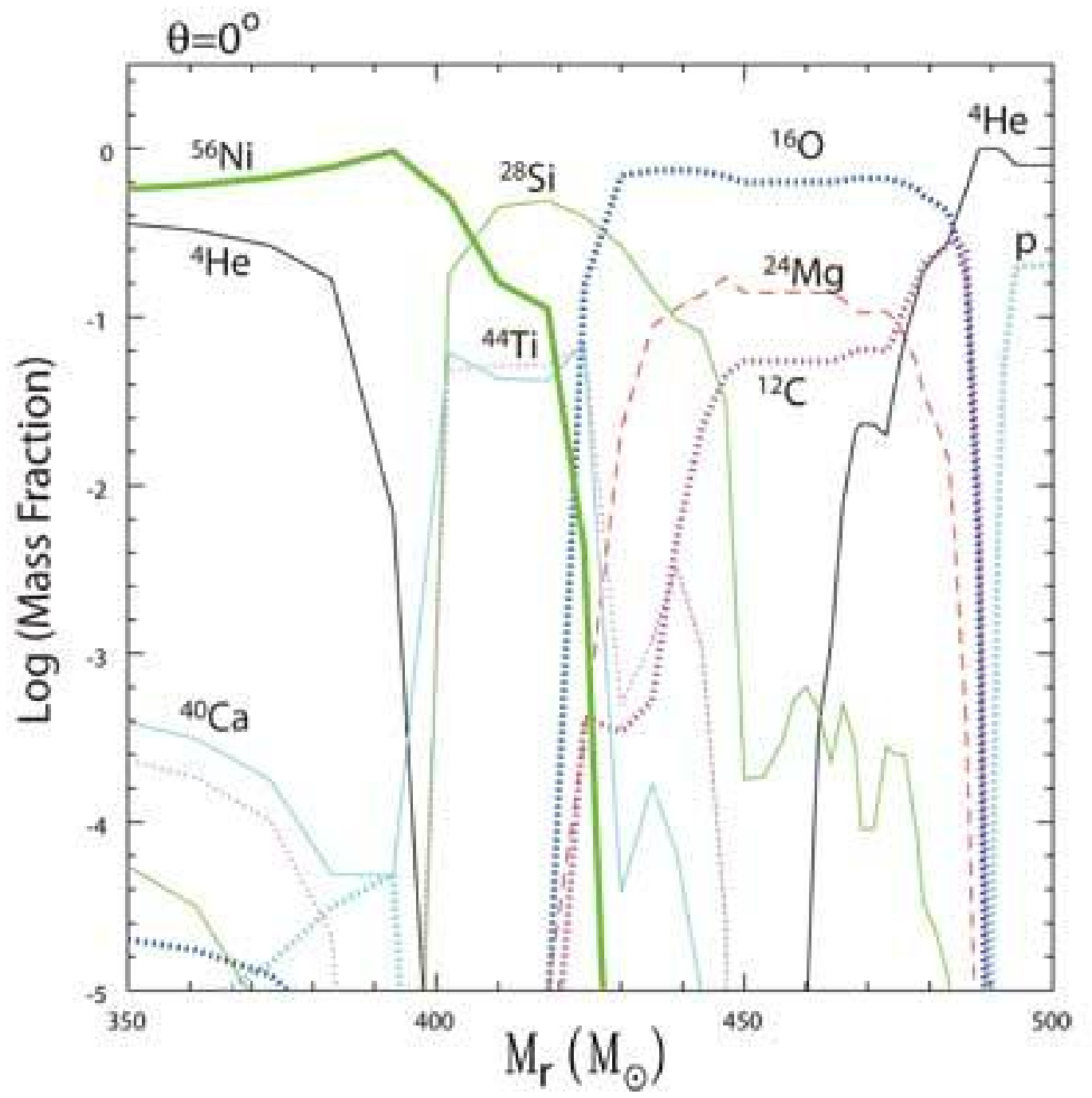}
\plottwo{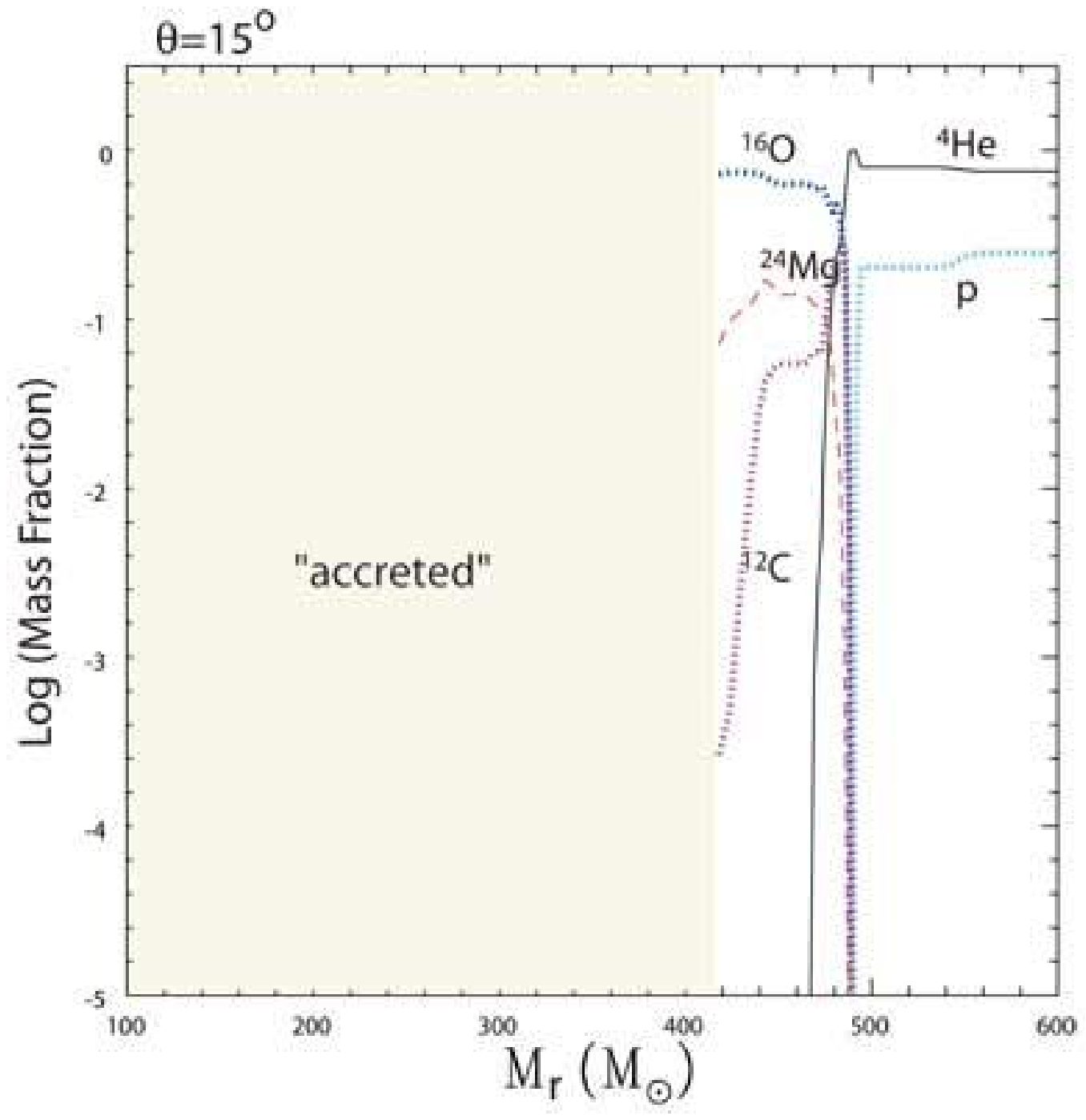}{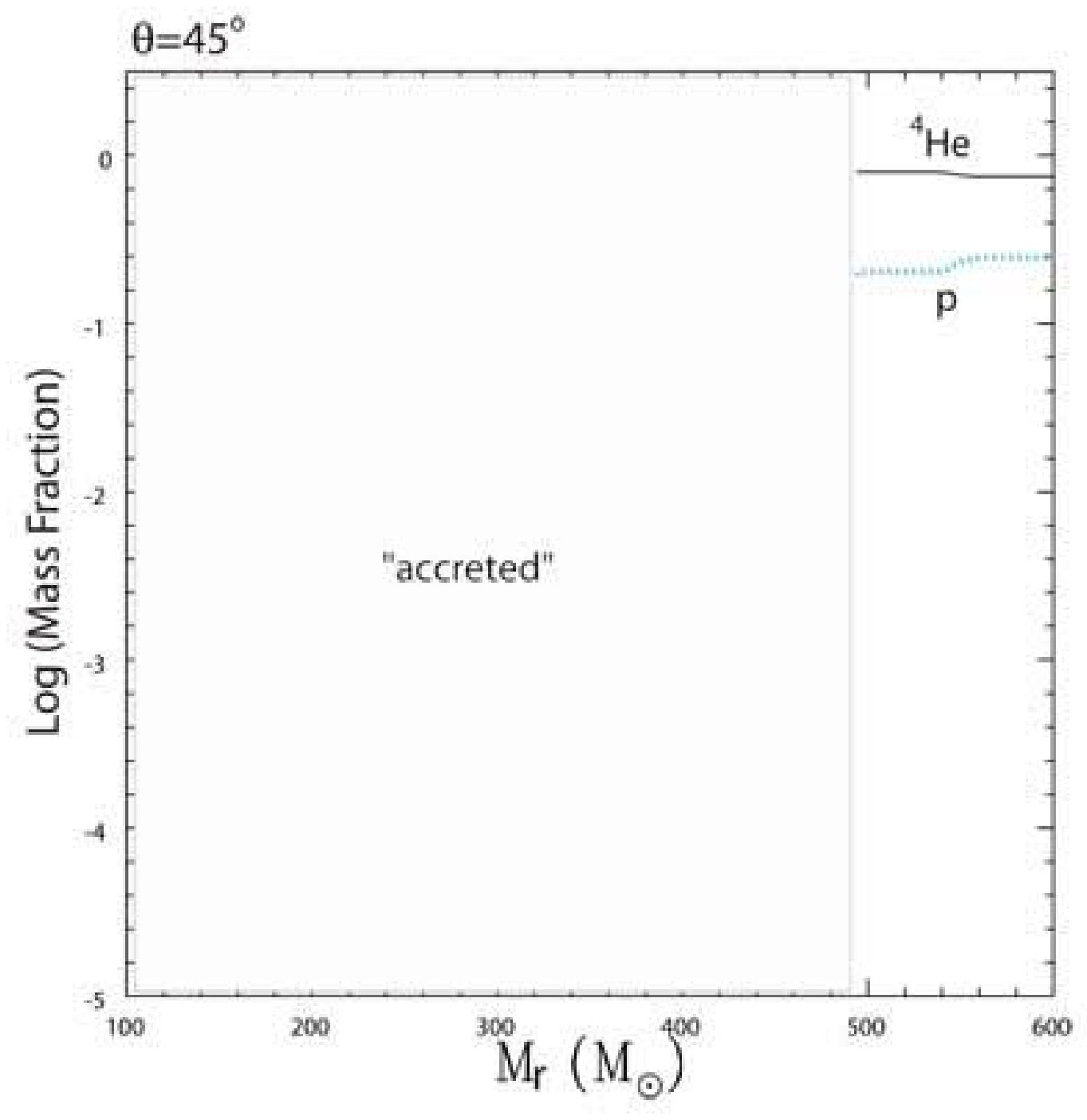}
\caption{Distributions of elements: Fe-group elements for $\theta = 0^o$ (top left), $\alpha$-elements for $\theta = 0^o$ (top right),
$\alpha$-elements for $\theta = {15}^o$ (bottom left), $\alpha$-elements for $\theta = {45}^o$ (bottom right), for Model A-1. Note that the 
mass range is set to $350 - 500 M_{\odot}$ in the top right panel to see clearly the distributions of $\alpha$-elements, while in others it is set to 
$100 - 600 M_{\odot}$
\label{fig:DegBution1}}
\end{figure}

\begin{figure}
\epsscale{1.2}
\plottwo{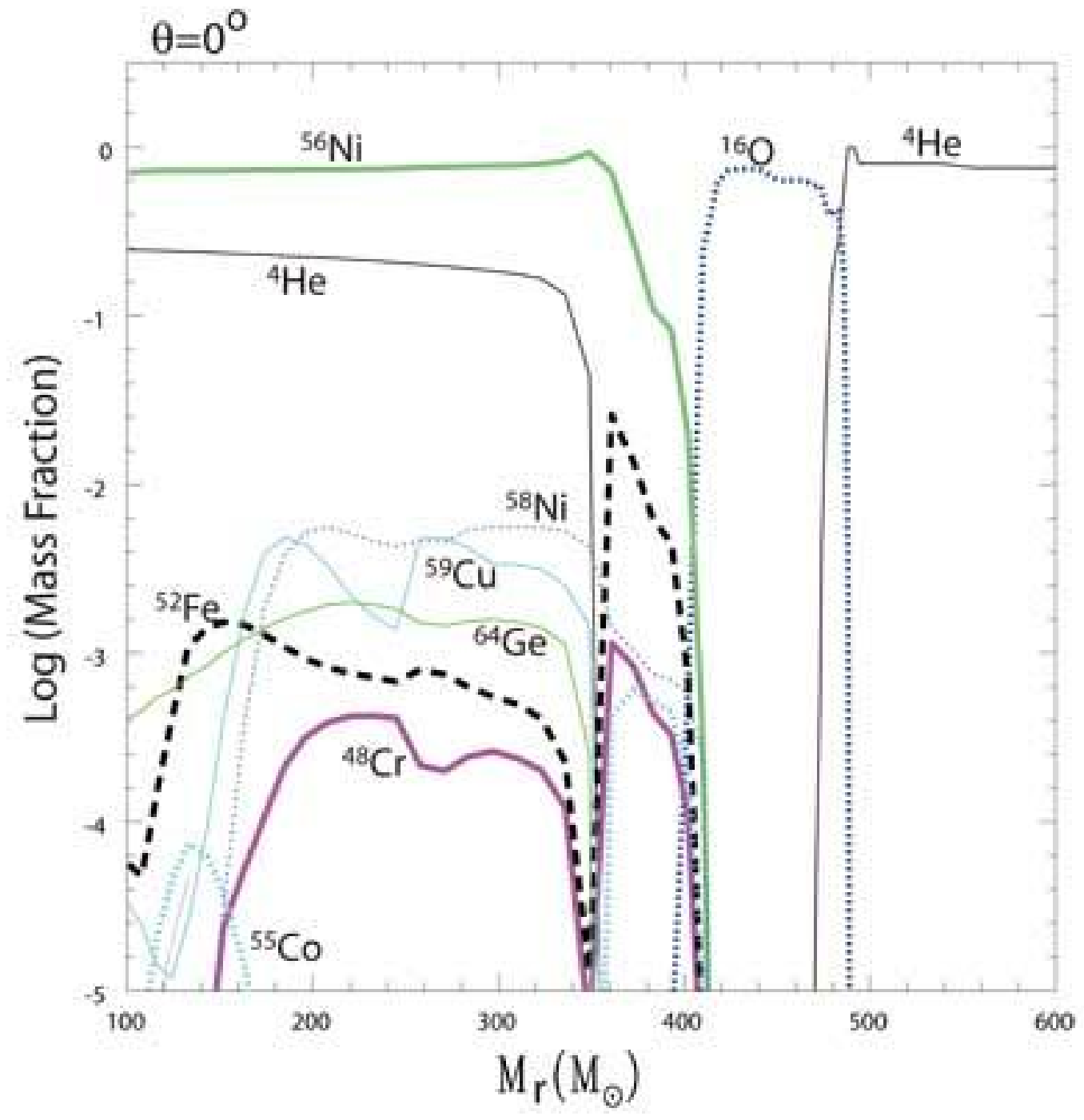}{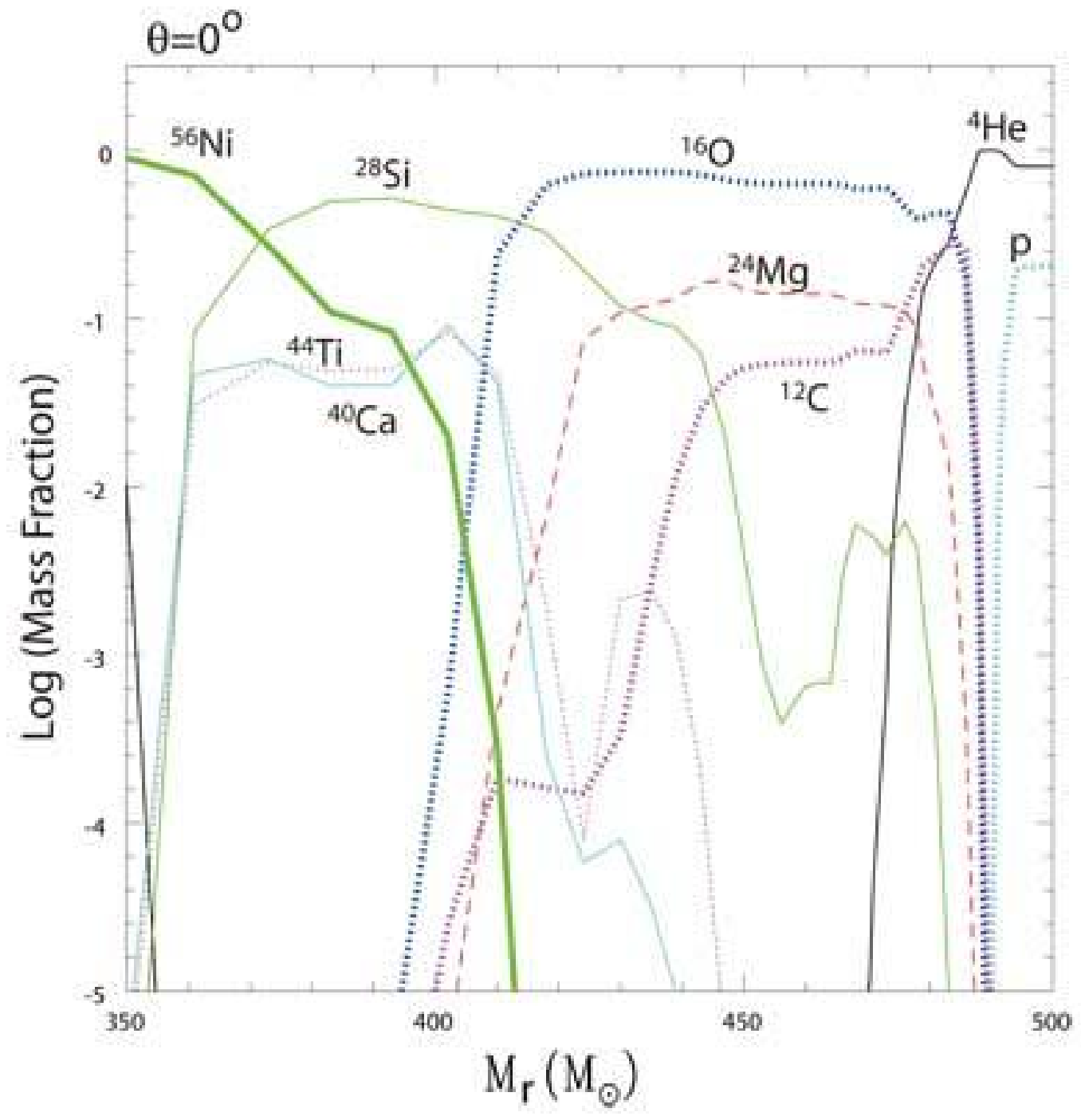}
\plottwo{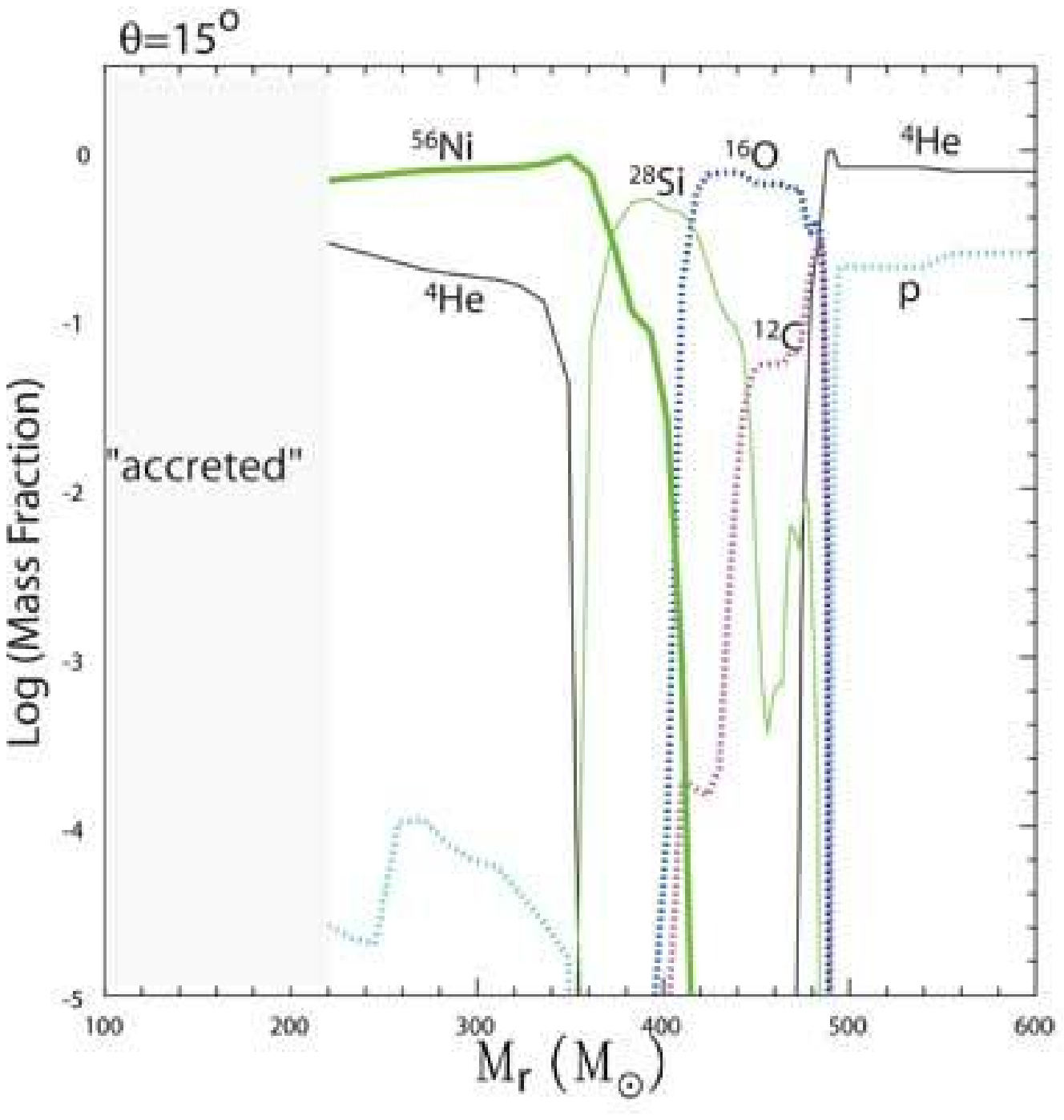}{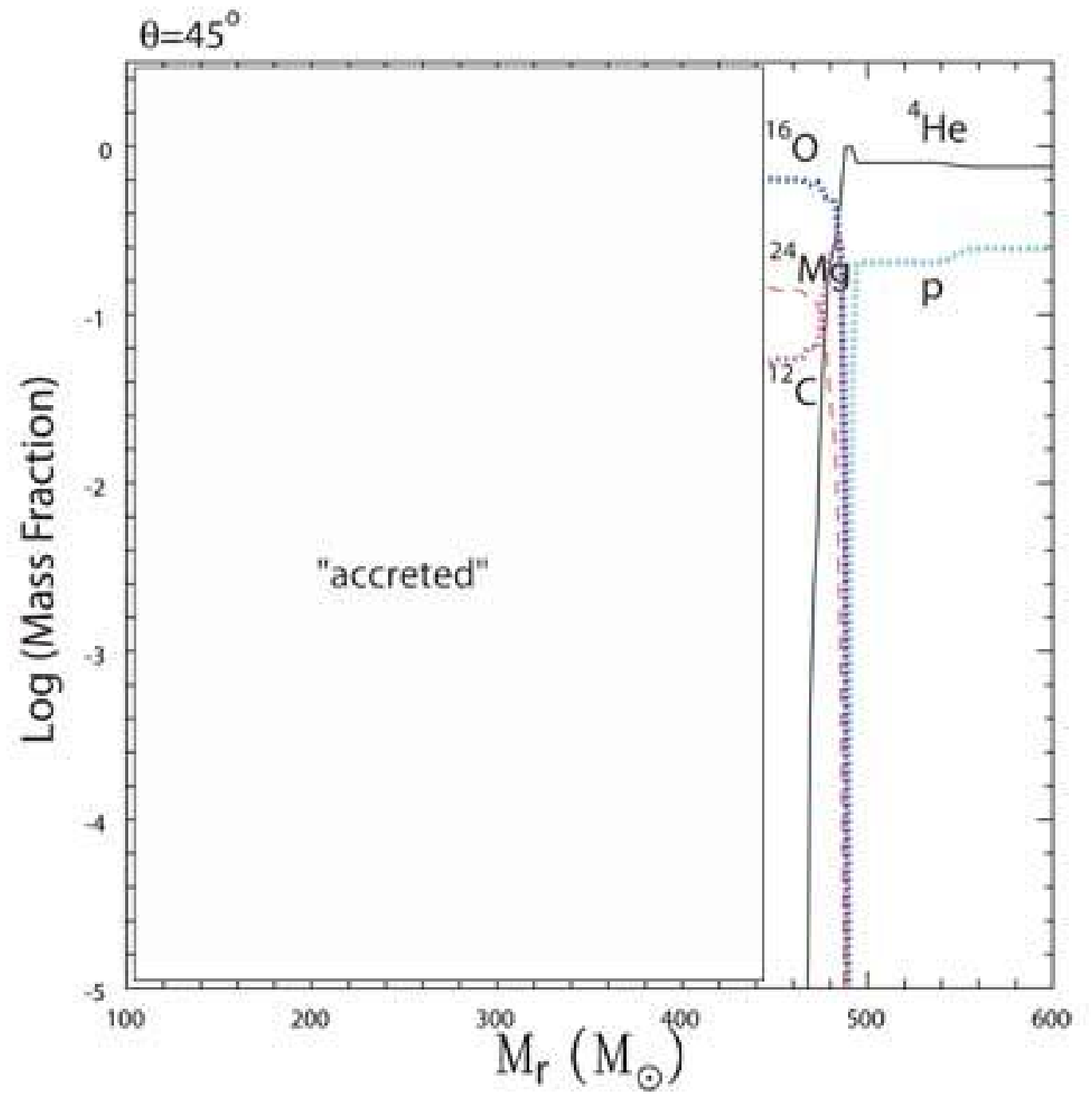}
\caption{Same as Figure~\ref{fig:DegBution1}, but for Model A-2.
\label{fig:DegBution2}}
\end{figure}

\begin{figure}
\epsscale{1.2}
\plottwo{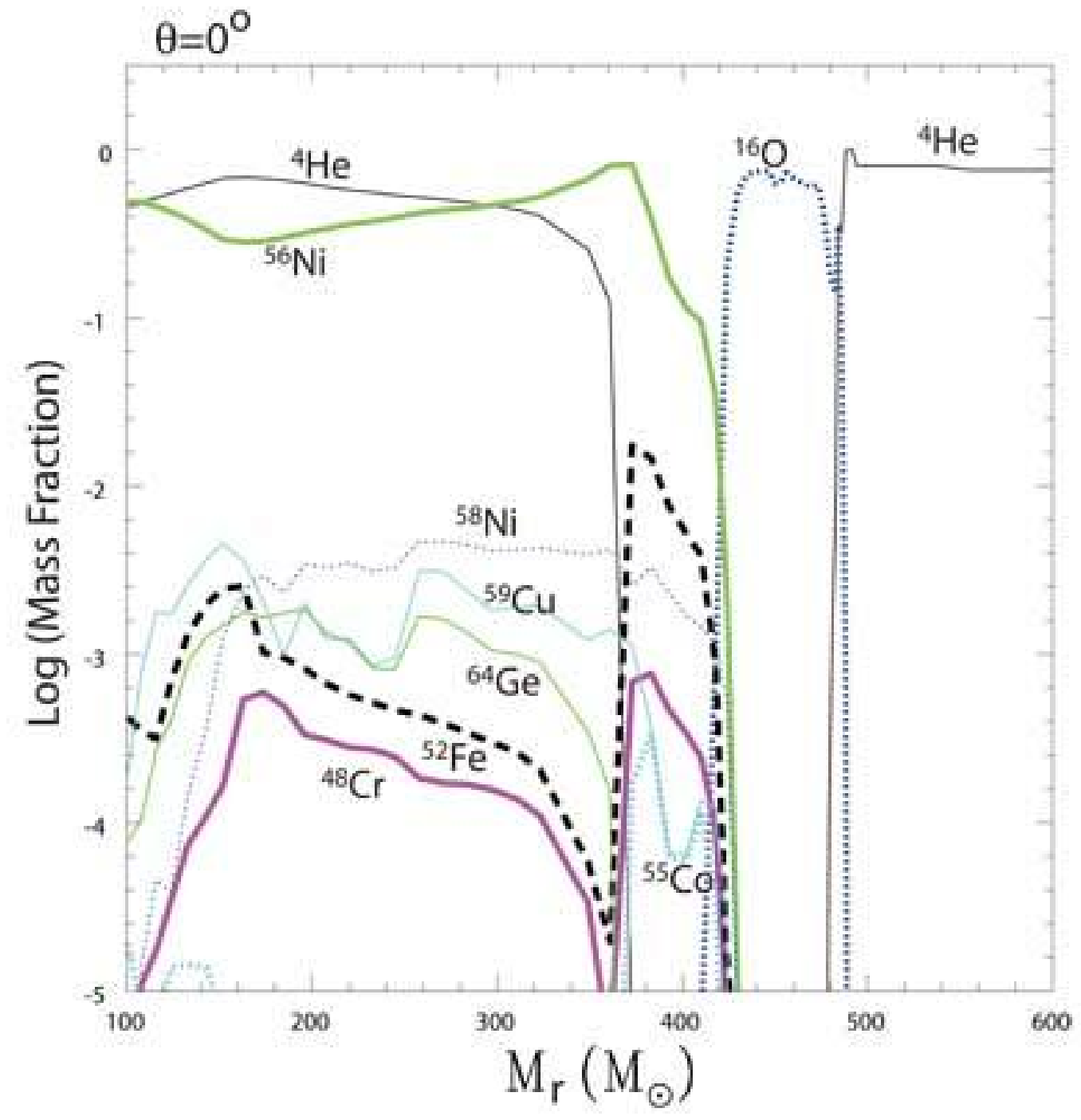}{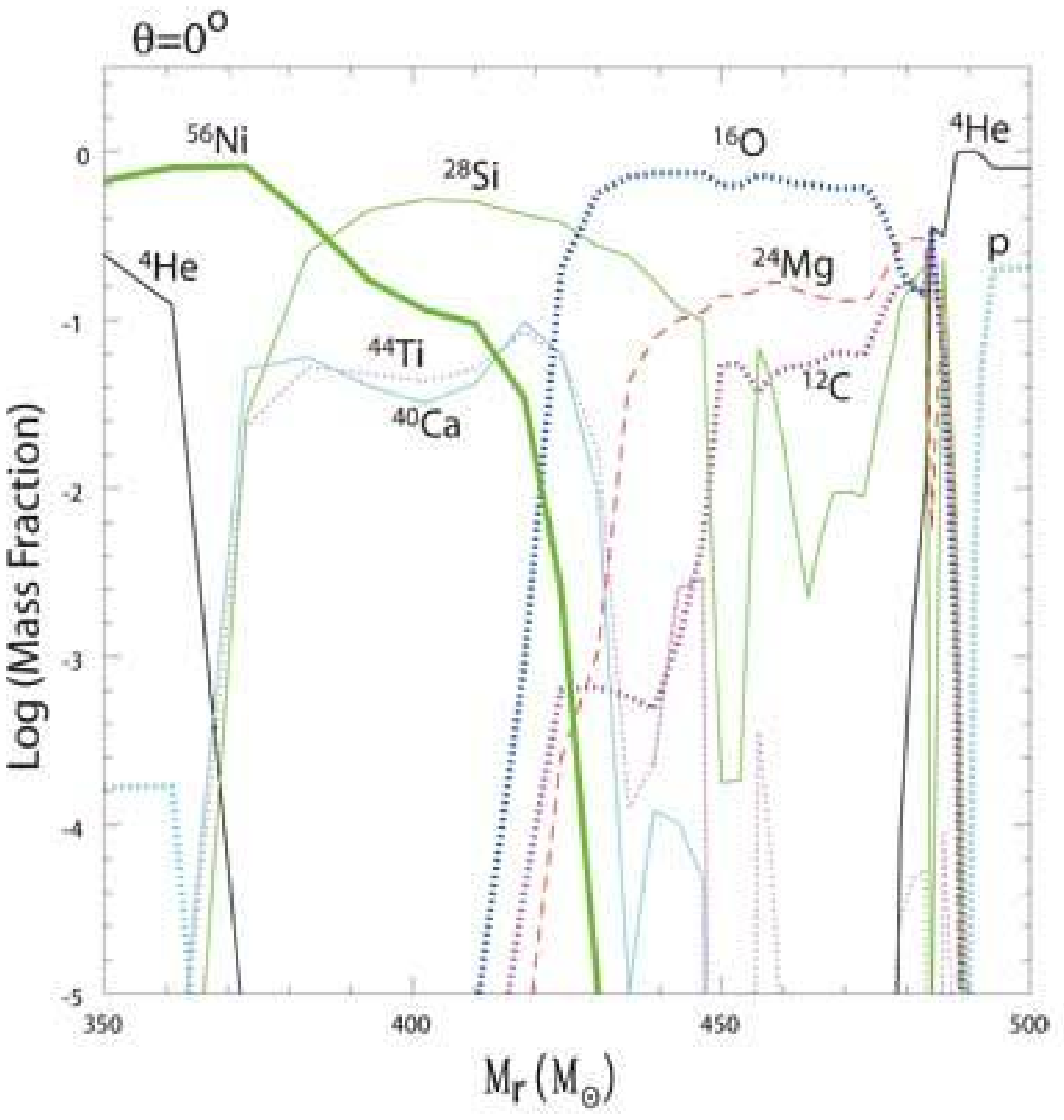}
\plottwo{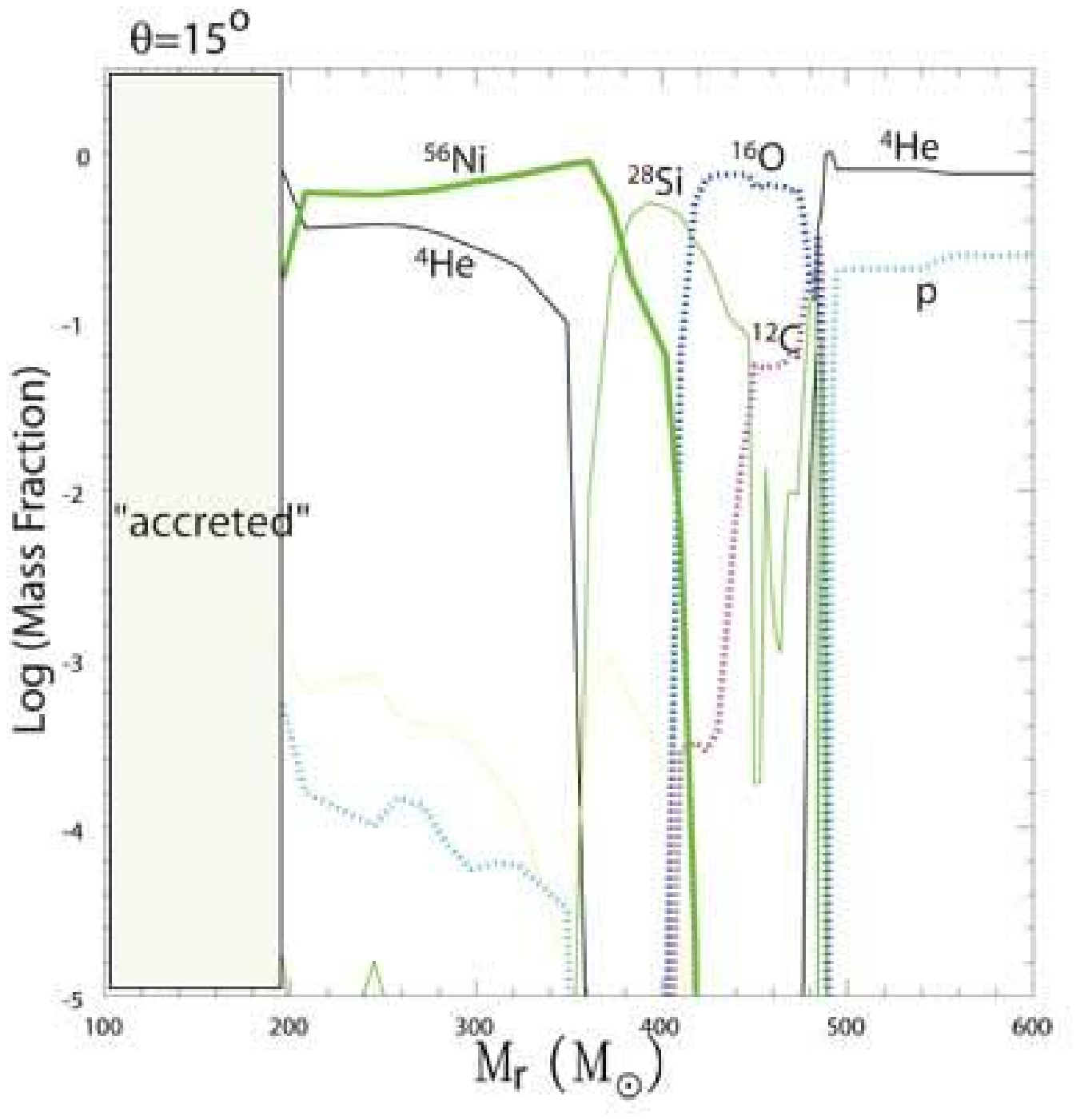}{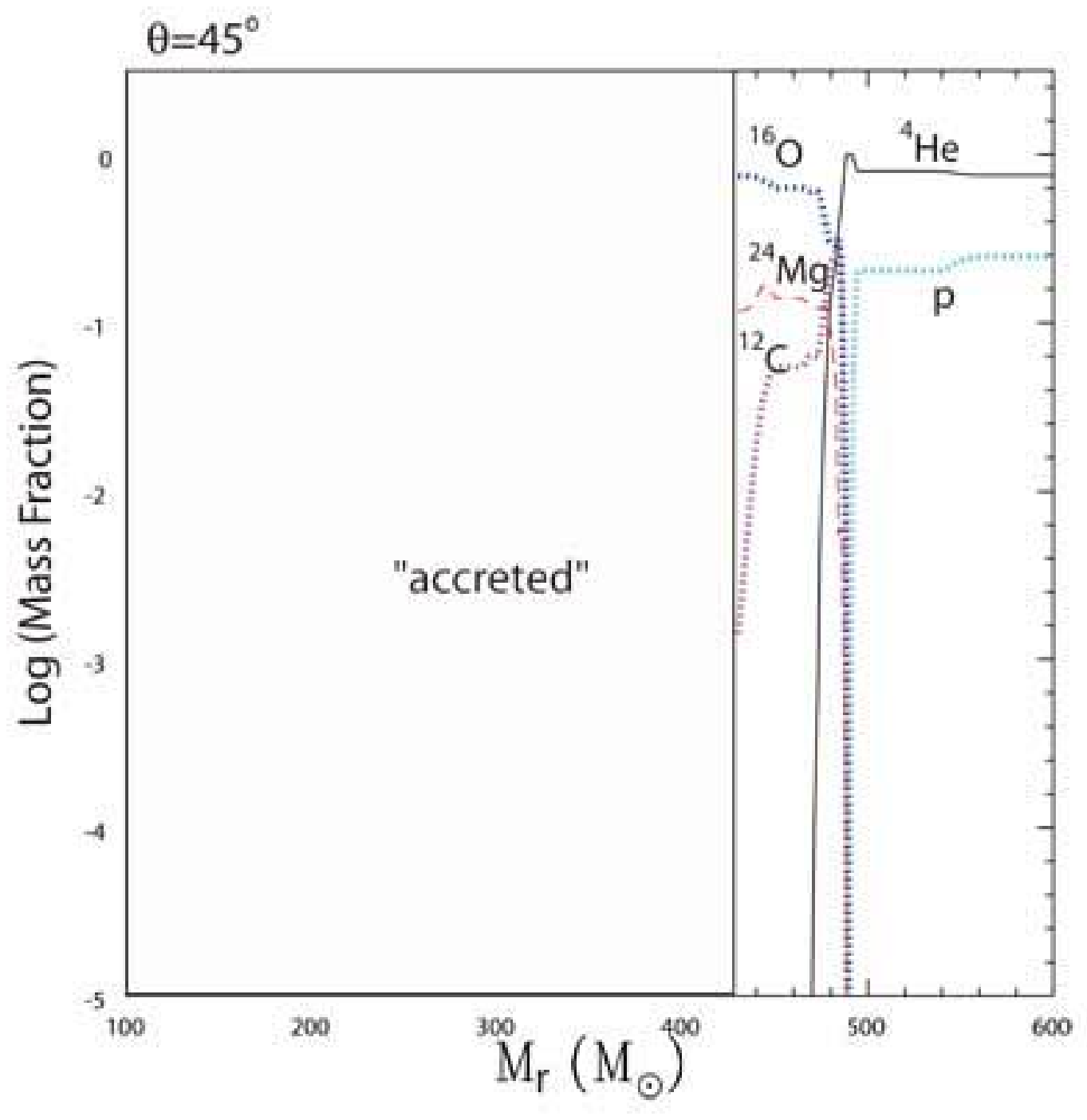}
\caption{Same as Figure~\ref{fig:DegBution1}, but for Model B-1.
\label{fig:DegBution3}}
\end{figure}

As typical interesting cases, Figures~\ref{fig:DegBution1} -- ~\ref{fig:DegBution3} show 
the distribution of elements after the explosive nucleosynthesis for Models A-1, A-2 and B-1, 
respectively. In each figure the upper left panel shows the Fe-group elements in the polar 
direction. The upper right panel shows the $\alpha$ - elements in the polar direction, 
the lower left panel shows the $\alpha$ - elements at $\theta = {15}^o$, and the lower 
right panel shows the $\alpha$ - elements at $\theta = {45}^o$. In each model, complete 
silicon burning region shows strong $\alpha$-rich freezeout. The upper panel (polar direction) 
in each figure shows that $^{56}{\rm Ni}$ is synthesized dominantly up to 400$M_{\odot}$ from 
the center.  
Compared with Figure 2 which shows the chemical composition just before the explosion, 
one can see that oxygen is consumed in the region with 350 - 400 $M_{\odot}$.

For directions $\theta = {15}^o$ and $\theta = {45}^o$ the silicon
and oxygen layers considerably accrete, and even a part of the
helium layer accretes for $\theta = {45}^o$ for Model A-1. On the
other hand, in Models A-2 and B-1 the complete silicon burning
region still remains for $\theta = {15}^o$ and the oxygen layer
remains for $\theta = {45}^o$. This is because the shock is
diffused to the equatorial directions more than Model A-1.

\begin{figure}
\epsscale{0.5}
\plotone{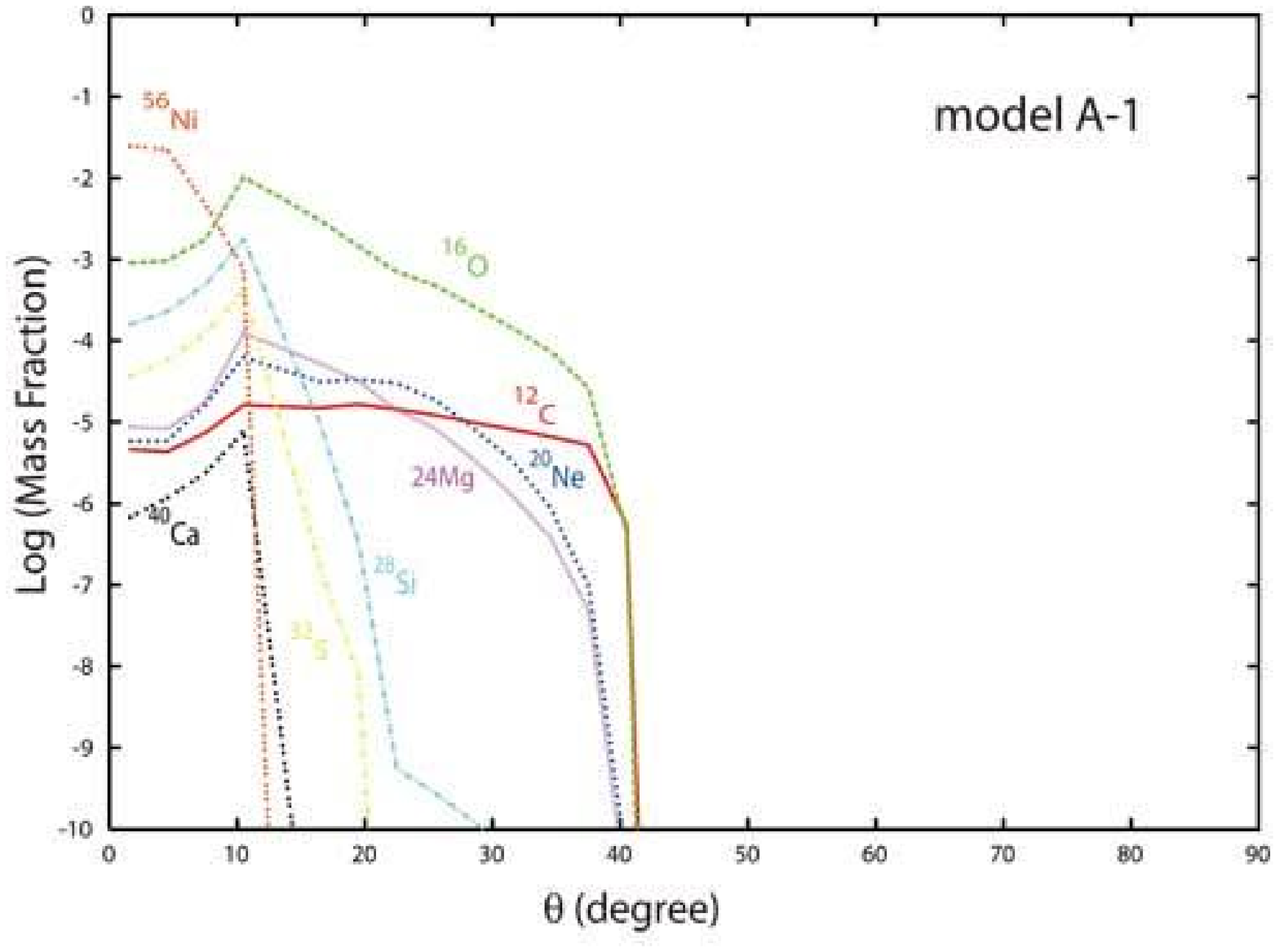}
\plotone{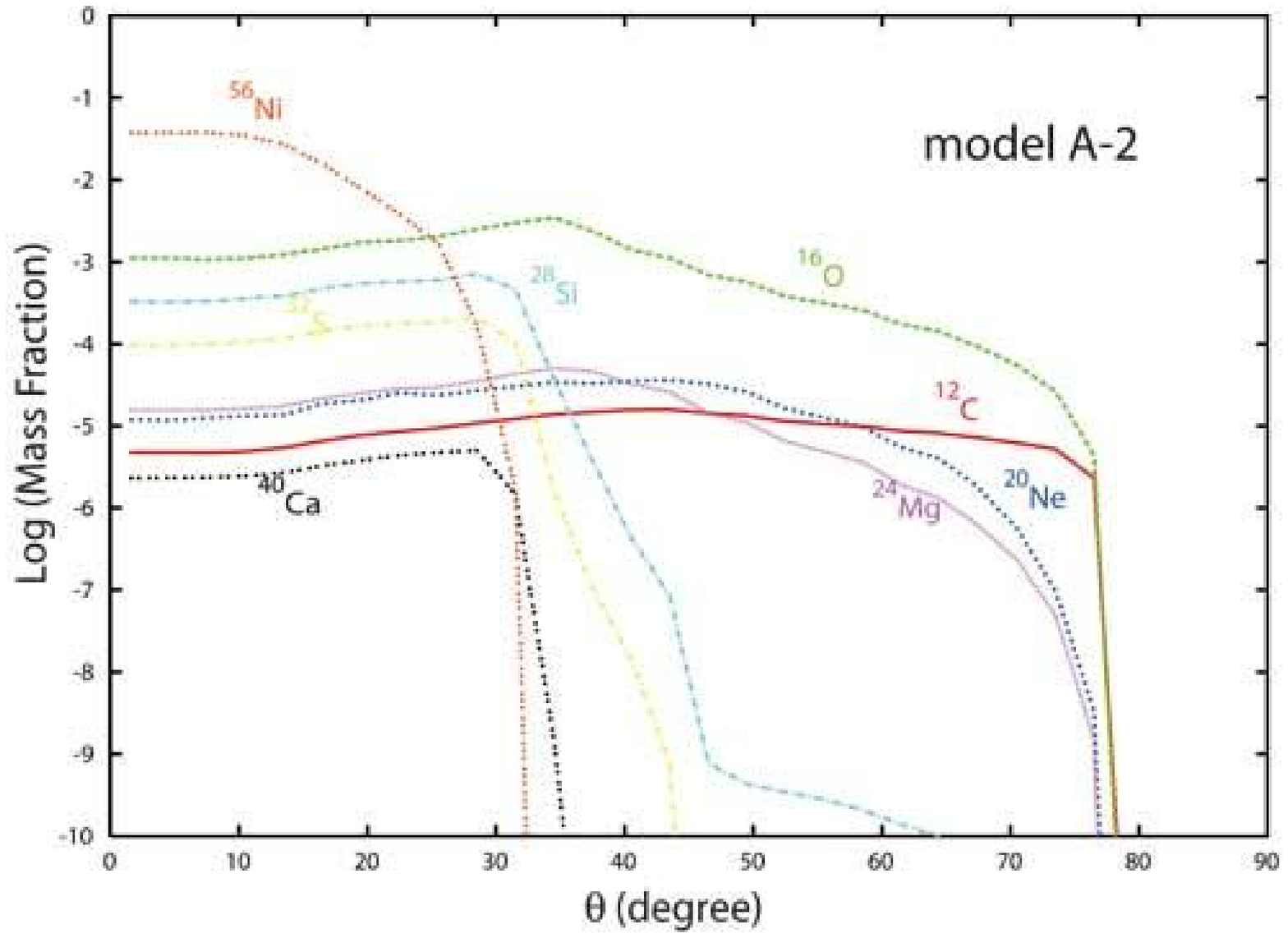}
\plotone{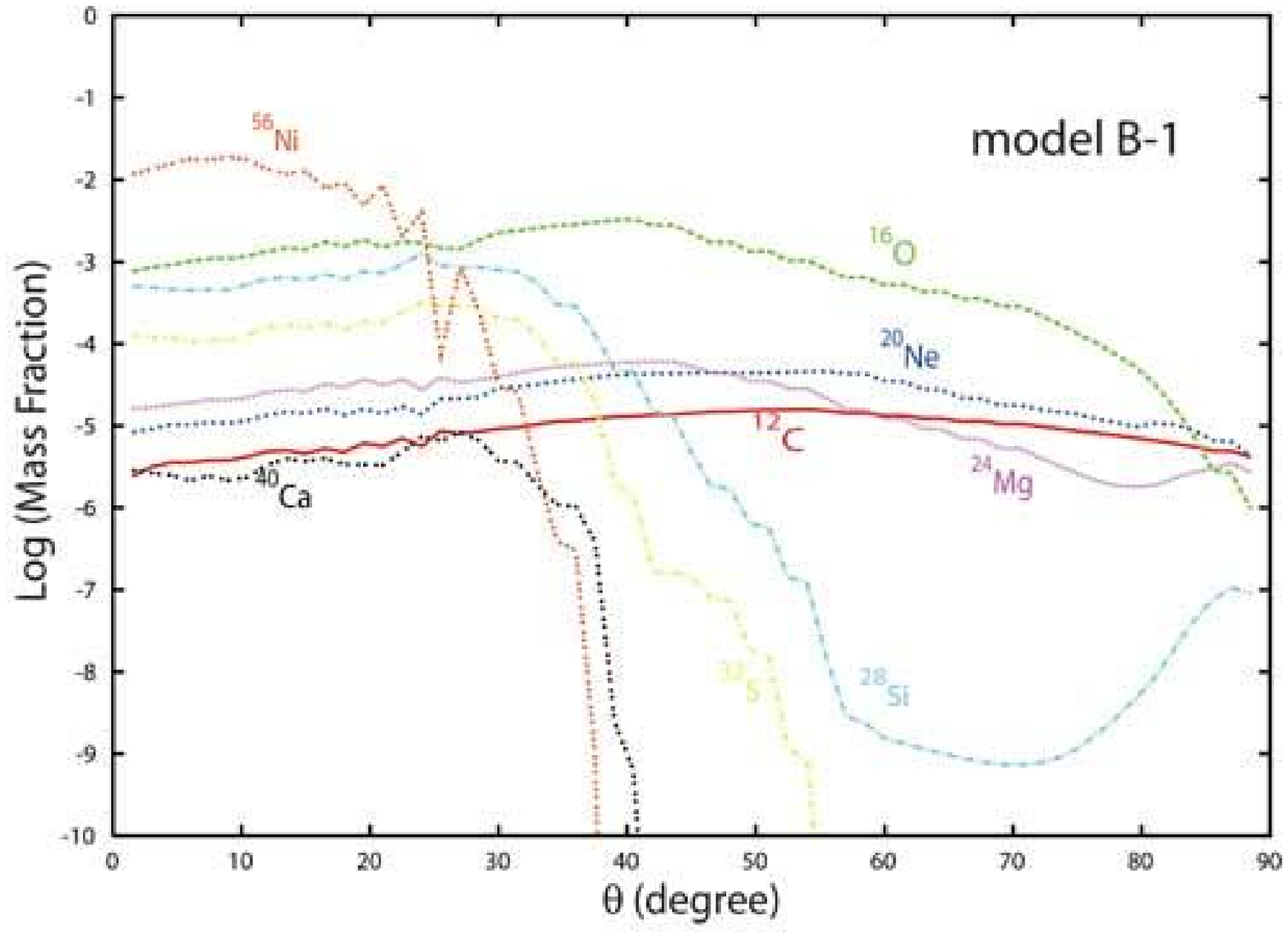}
\caption{Mass fractions of ejected $\alpha$ - elements and $^{56}{\rm Ni}$ as a function of the direction $\theta$.
These figures are for models A-1, A-2, and B-1.
These values are obtained by integrating over radial direction for each $\theta$.
\label{fig:DegFrac1}}
\end{figure}

Figure~\ref{fig:DegFrac1} shows the mass fractions of $^{56}{\rm Ni}$ 
and $\alpha$-elements for each $\theta$ integrated over the radial direction. 
These figures clearly show how much matter is ejected - for example, 
the ejected mass of the oxygen layer by seeing the mass fraction of 
$^{16}{\rm O}$. For the $\theta_{\rm jet} = {15}^o$ models in Case A (e.g., A-1) 
there is no ejected matter except helium and hydrogen in the directions 
$\theta > {45}^o$ (see also Figure~\ref{fig:DegBution1}). 
However, for Case B and the $\theta_{\rm jet} = {30}^o$ models of Case A {\ (e.g., A-2 or B-1) the  
$^{56}{\rm Ni}$ synthesized region and $\alpha$-element rich 
region are broadened to around $\theta = {30}^o$ and $80^{o}$.}

\subsubsection{Composition of Jet Material}

\begin{figure}
\epsscale{0.5}
\plotone{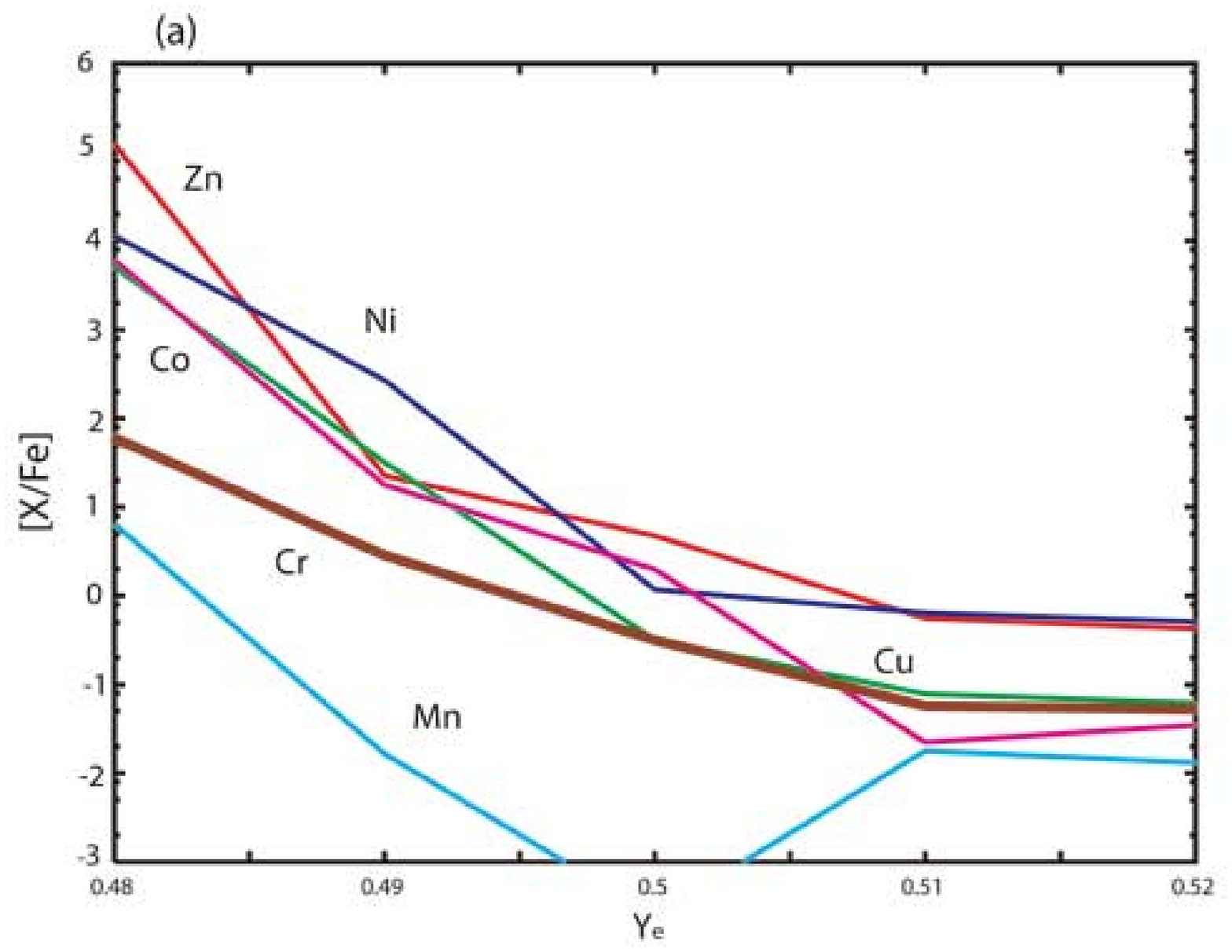}
\epsscale{1.1}
\plottwo{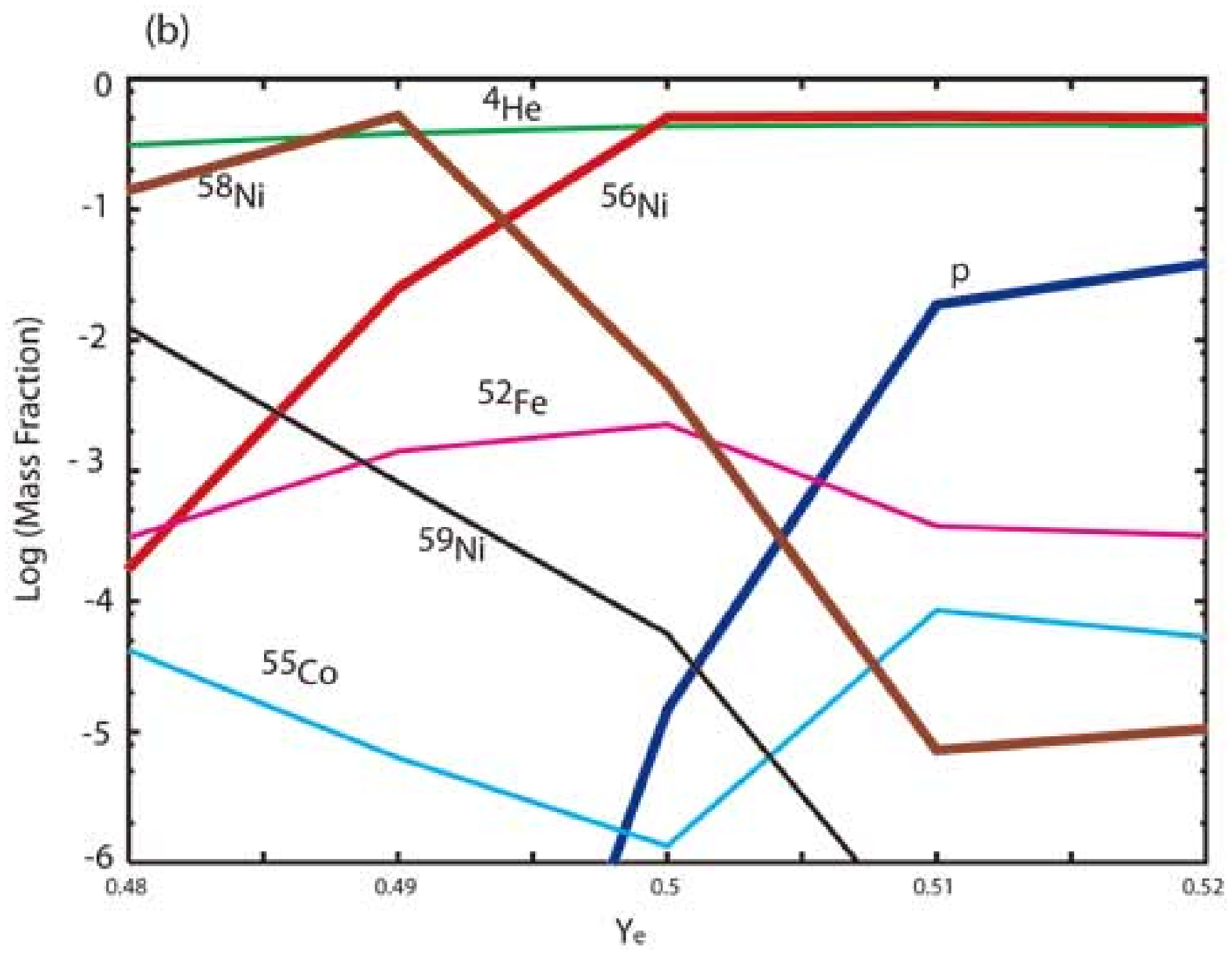}{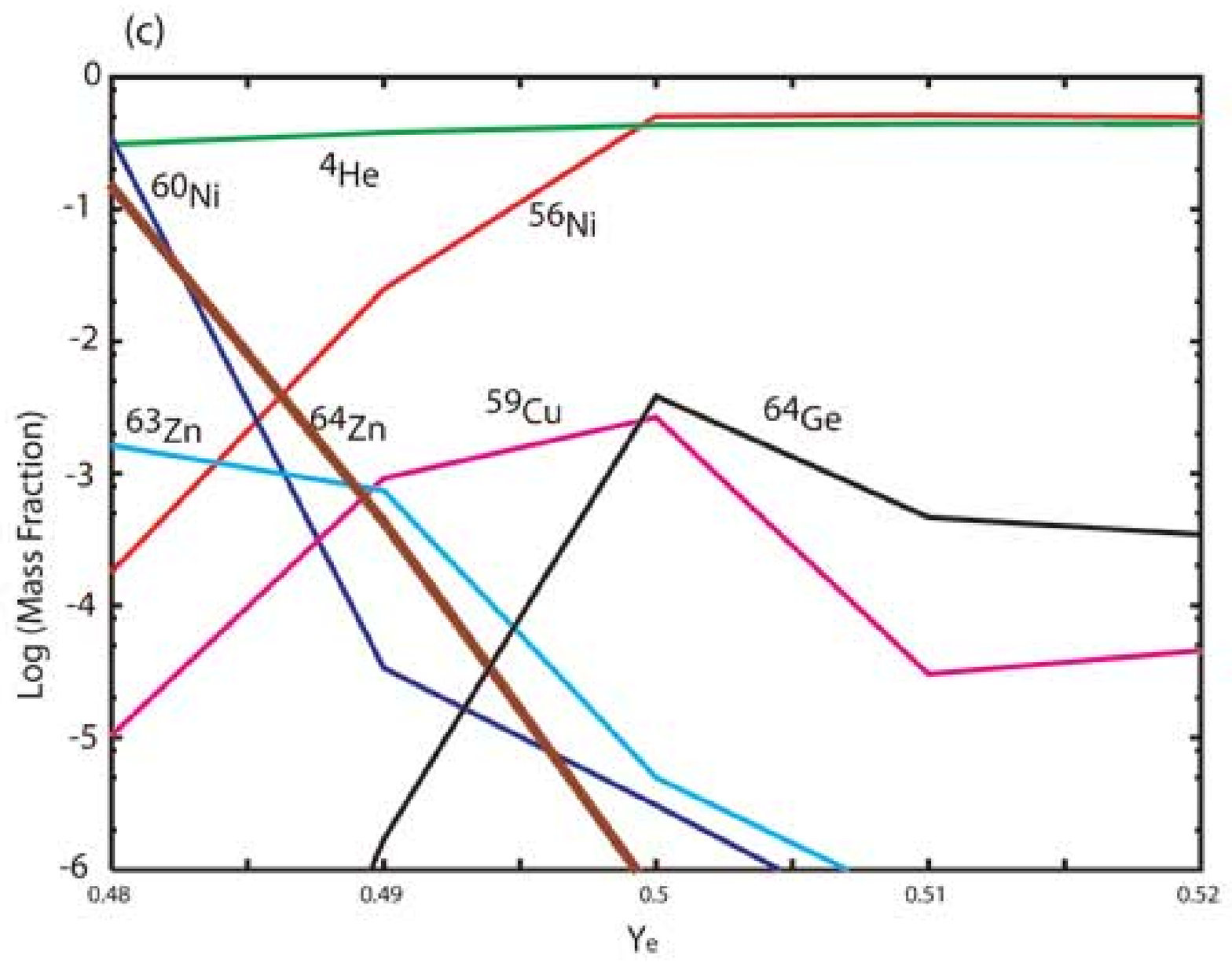}
\caption{(a): [X/Fe] for Fe-group elements as a function of  $Y_{\rm e}$ for history A.
(b), (c): Mass fractions of Fe-group elements, proton and $^{4}{\rm He}$ as a function of $Y_{\rm e}$ for history A. Both (b) and (c) are for history A. Some different elements are shown separately in (b) and (c) for clarity.
\label{fig:ABPjet1}}
\end{figure}

Figure~\ref{fig:ABPjet1} shows [X/Fe] (top panel) and mass
fractions (bottom panels) for the Fe-peak elements as a function
of $Y_{\rm e}$. Note that we assume the temperature of the jet
material reaches higher than $5 \times {10}^9$ K, and therefore it
consists mostly of the Fe-group elements and $^{4}{\rm He}$.

Peculiar features are seen particularly when $Y_{\rm e} < 0.5$. Co, Cu, Ni and Zn are dramatically abundant relative to Fe (500-1000 times larger than the solar values) as shown in Figure~\ref{fig:ABPjet1}. When $Y_{\rm e} < 0.5$ the mass fraction of synthesized $^{56}{\rm Ni}$ is very small, less than 10\% for $Y_{\rm e} = 0.49$ and less than 0.1\% for $Y_{\rm e} = 0.48$. Then a large amount of neutron-rich nuclei, such as $^{58}{\rm Ni}$, $^{60}{\rm Ni}$, and $^{59}{\rm Ni}$ (which decays into $^{59}{\rm Co}$), $^{63}{\rm Zn}$ (which decays into $^{63}{\rm Cu}$), and $^{64}{\rm Zn}$, are synthesized. In these situations neutron-rich $^{64}{\rm Zn}$ is directly synthesized rather than $^{64}{\rm Ge}$, which decays into $^{64}{\rm Zn}$. The rise of [Cr/Fe] and [Mn/Fe] for $Y_{\rm e} < 0.5$ is mainly due to the small fraction of $^{56}{\rm Ni}$ rather than the increase of the fractions of Cr and Mn.

On the other hand, for $Y_{\rm e} > 0.5$ most of the products are $^{56}{\rm Ni}$ and $^{4}{\rm He}$, similar to the case where $Y_{\rm e} = 0.5$. The main effect of $Y_{\rm e}$ larger than 0.5 is the existence of free protons.

\begin{figure}

\plotone{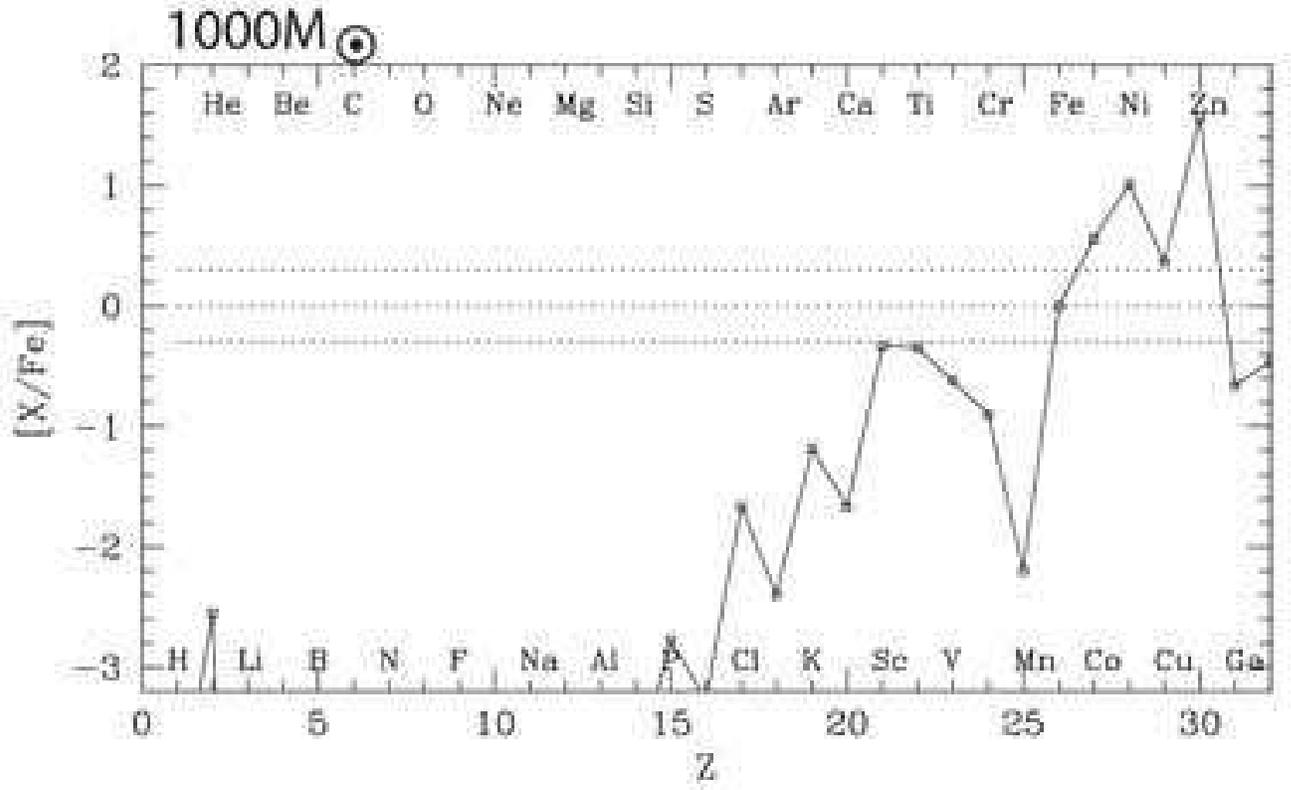}
\caption{Abundance pattern of jet material averaged for 15 cases (5 $Y_{\rm e}$ values times 3 density-temperature
histories).
\label{fig:ABPjet2}}
\end{figure}

As our first step for the treatment of the jet material, Figure~\ref{fig:ABPjet2} shows the abundance pattern of jet material averaged over 15 patterns (5 $Y_{\rm e}$ values $\times$ 3 entropy values). [Zn/Fe] and [Ni/Fe] are larger than the solar values due to the effects of small $Y_{\rm e}$ regions, while [Cr/Fe] and [Mn/Fe] are smaller due to the effects of large $Y_{\rm e}$ regions. The averaged mass fraction of $^{56}{\rm Ni}$ is about 40\%.  We multiply the mass fraction of each nucleus by the jet mass and add it to the total abundance pattern.

\subsection{Ionization Rates, Heavy Element Yield, and Ionization Efficiency}

The suggestion that VMSs are responsible for the reionization of
HI and HeI is not a new one (e.g., Gnedin \& Ostriker 1997).
Bromm, Kudritzki, \& Loeb (2001) calculated the stellar atmosphere
models for Pop III main-sequence CVMSs of 300 - 1000 $M_\odot$ and
obtained the effective temperatures of log $T_{\rm eff}$ (K)
$\sim$ 5.05, which are higher than log $T_{\rm eff}$ (K) $\sim$
4.81 of Pop I stars with the same mass and slightly higher than
log $T_{\rm eff}$ (K) = 4.85 - 5.0 for Pop III 15 - 90 $M_\odot$
stars (Tumlinson \& Shull 2000). Thanks to the high effective
temperature, Pop III CVMSs give high production rates of ionizing
radiations $\sim$ $1.6 \times 10^{48}$ photons s$^{-1}$
${M_{\odot}}^{-1}$ for H I ionization, $1.1 \times 10^{48}$
photons s$^{-1}$ ${M_{\odot}}^{-1}$ for He I ionization, and $3.8
\times 10^{47}$ photons s$^{-1}$ ${M_{\odot}}^{-1}$ for He II
ionization. These numbers correspond to $\sim$ 16, 14 and 75 times
higher than the corresponding values with a Salpeter IMF (see
Bromm et al. 2001), and therefore, they are sufficient for
completely reionizing IGM.

Daigne et al. (2004) estimate the efficiency of supplying UV 
photons and chemical enrichment of IGM simultaneously. These authors suggest 
that the IMF that essentially formes less than 100 $M_{\odot}$ is favorable. 
However, this conclusion is due to the assumption that all CVMSs collapse entirely to a black hole.
Venkatesan \& Truran (2003) considered the relation between the
reionizing radiation and metal enrichment of IGM, 
using stellar atmosphere models and model yields available
at that time. For the model yields, they assumed no metal ejection
by stars of $\sim 30 - 130 M_{\odot}$ and also $M \gsim 300 M_{\odot}$.
Following their argument, here we compute the reionization
efficiency for our CVMSs using model yields in the present work.

Adopting the mass of heavy elements ejected by our 1000 $M_{\odot}$
star model, $M_{\rm Z} \sim 50 M_{\odot}$, the conversion
efficiency ($\eta_{\rm Lyc}$) of energy produced in the HI
ionizing radiation divided by the energy produced in the rest mass
of metals ($M_{\rm Z} c^2$) is $\eta_{\rm Lyc} \sim 0.05$. (We
used Eq. 1 of Venkatesan \& Truran 2003). Here we use the
timescale of $t_{\rm ms} = 2 \times 10^6$ years for the 1000
$M_{\odot}$ Pop III star. With these values, the number of
ionizing photons per baryon in the universe generated in
association with the IGM metallicity $Z_{\rm IGM} \sim 10^{-4}$,
obtained for our model, is $N_{\rm Lyc}/N_{\rm b} \sim 150$.  (We
used Eq. 2 of Venkatesan \& Truran 2003). Note that this value
well exceeds the value required for reionization of inter galactic
hydrogen, $1 < N_{\rm Lyc}/N_{\rm b} \lsim 10$ (see Somerville et
al. 2003). Therefore, contrary to the earlier results, our 
conclusion is that CVMSs can
contribute significantly to reionization of IGM in the early
epochs.

\section{Integrated Abundance Patterns and Comparison with Observations}

The abundance pattern, the mass ratio of each element to be compared with 
observations, is determined by integrating the distributions over the entire 
ejecta regions (both radial and $\theta$ directions). It is the mass ratio of 
each ejected element. Tables~\ref{tab:EjectStellarH} and ~\ref{tab:EjectStellarL} show the ejected masses of some isotopes excluding 
the jet materials, and Table~\ref{tab:Yields1} shows the masses of all the isotopes including 
the jet materials.

\subsection{Abundance Patterns without Jet Materials}

Tables~\ref{tab:EjectStellarH} and \ref{tab:EjectStellarL} show the ejected mass of $^{56}{\rm Ni}$ (which decays
into $^{56}{\rm Fe}$), excluding the jet material. Masses of
$^{16}{\rm O}$ and $^{28}{\rm Si}$ are also shown as
representative $\alpha$-elements to see the abundance and ratios
of these elements. In models for Case A (except Model A-4), the
ratios of the ejected masses of these elements to their progenitor
mass are rather small, compared with those ratios in ordinary
massive stars such as a 25$M_{\odot}$ star. The typical ejected
$^{56}{\rm Ni}$ mass in the 25$M_{\odot}$ star is $\sim$ 0.1
$M_{\odot}$ (Maeda \& Nomoto 2003). In the models with
$\theta_{\rm jet} ={15}^o$, asphericity is so strong that it is
only toward small $\theta$ direction where $^{56}{\rm Ni}$ and
Fe-group elements are synthesized and ejected. On the other hand,
in models for Case B and Model A-4, these masses are much larger
than the other models. The ejected $^{56}{\rm Ni}$ mass is about 5 -
10$M_{\odot}$. If this kind of supernova occurs, it is very bright in its tail 
because the heating source of a supernova is $\gamma$-rays from
radioactive decays of $^{56}{\rm Ni} \to ^{56}{\rm Co} \to ^{56}{\rm Fe}$. However, it is very difficult
to observe directly the explosions of first generation stars by
present observational devices since they are very distant ($z \gsim 20$).

\subsection{Total Abundance Pattern and Comparison with Observational Data}

\subsubsection{Intracluster Matter and Hot Gas in M82}
\begin{figure}
\epsscale{0.9}
\plottwo{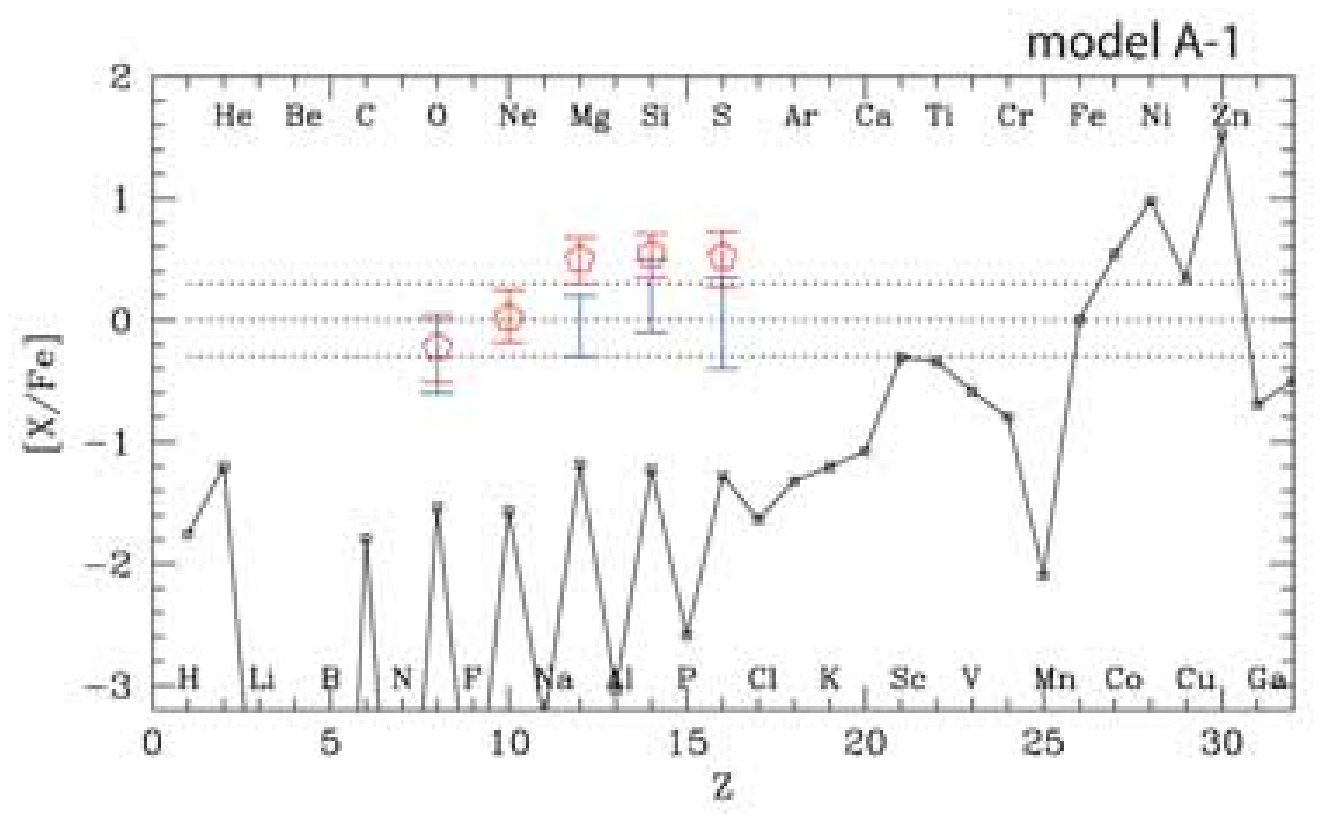}{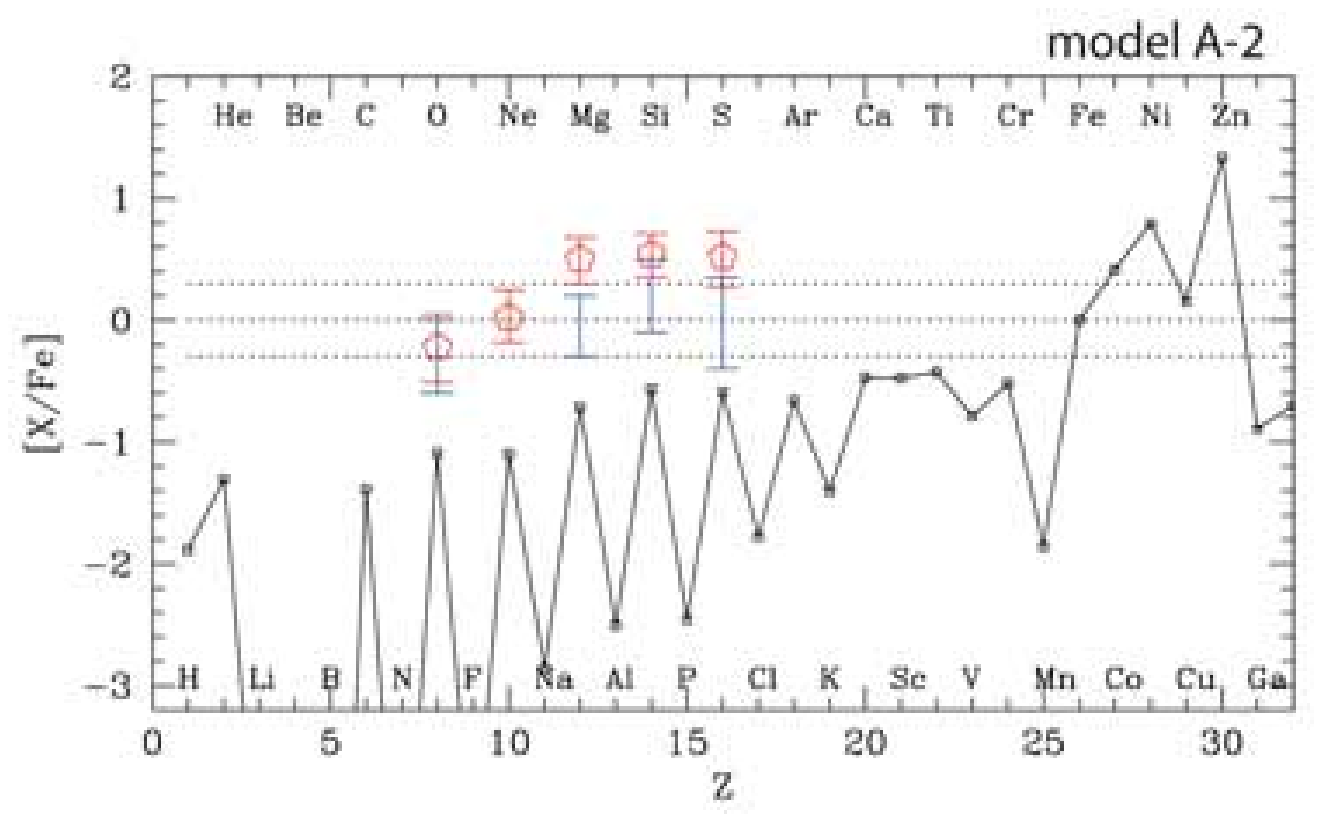}
\plottwo{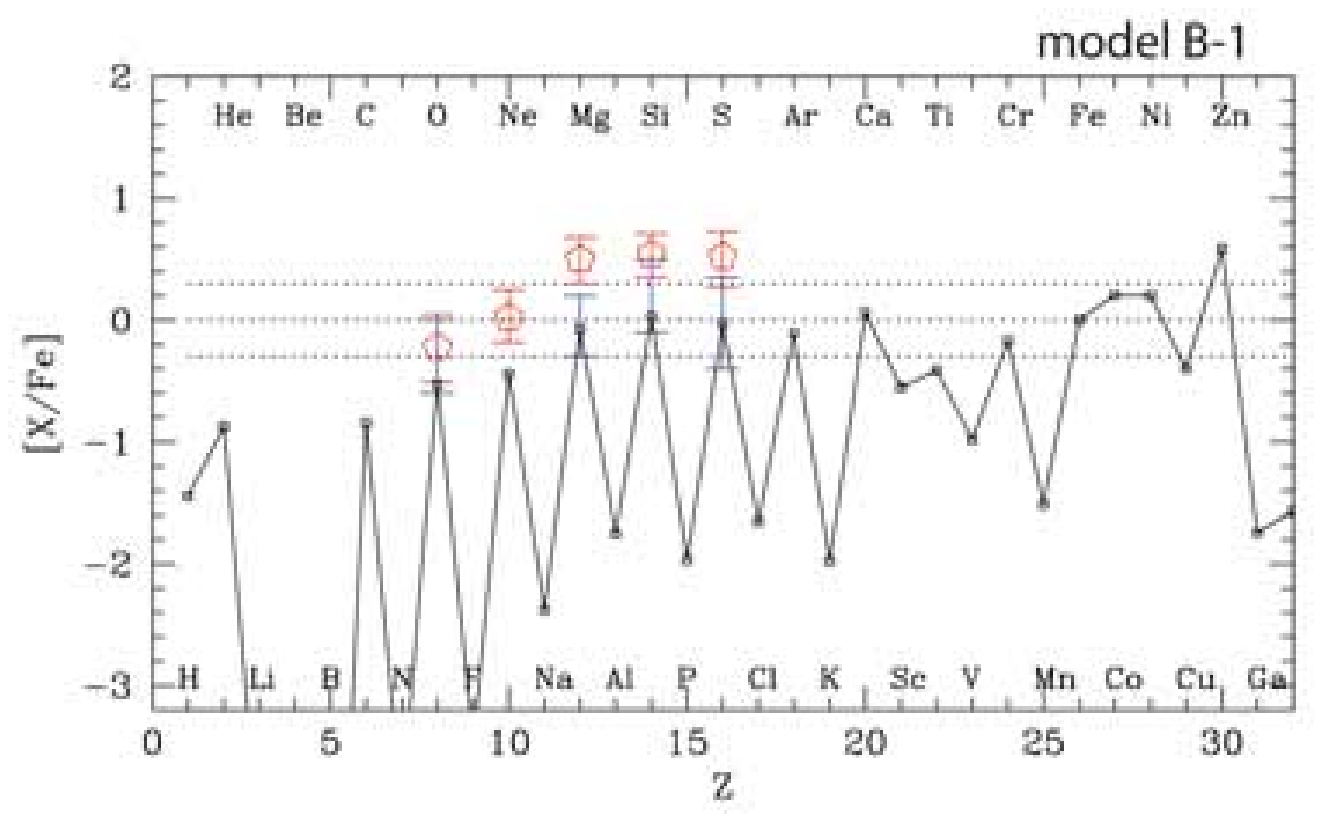}{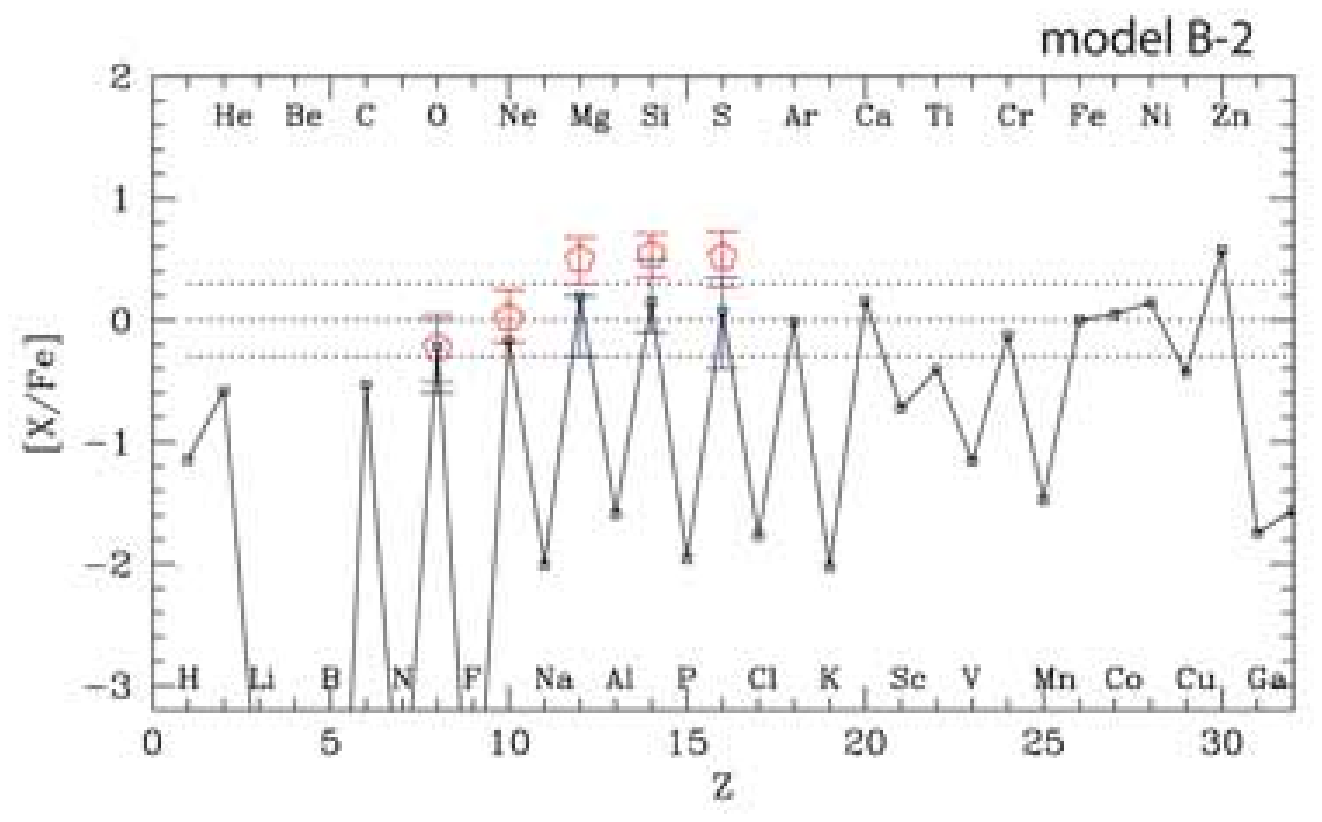}
\epsscale{0.45}
\plotone{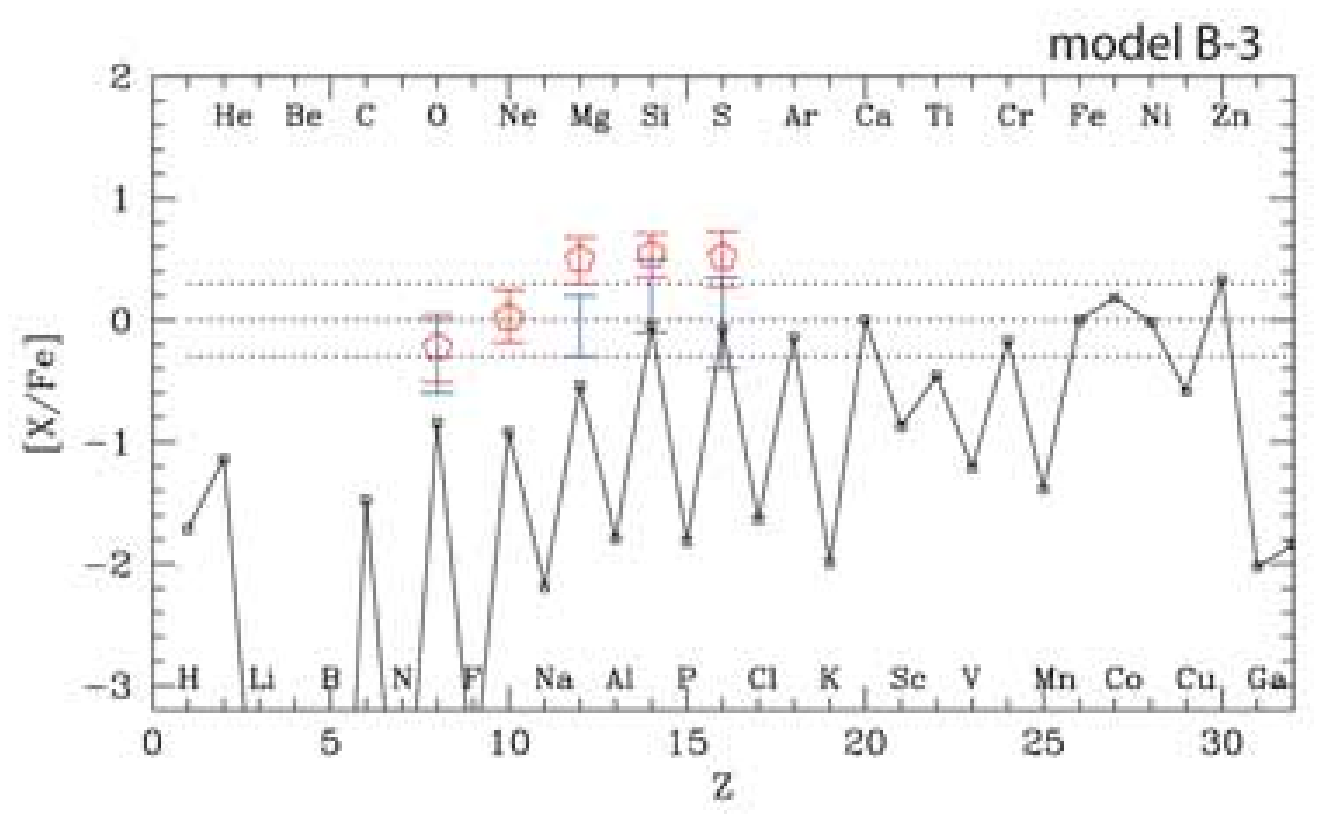}
\caption{Total abundance patterns including jet contribution for higher resolution models.
The open pentagons show the abundance ratios of gas of the central region in M82 (Ranalli et al. 2005).
The bars show the range of abundance ratios observed in ICM (Baumgartner et al. 2004; Peterson et al. 2003).
\label{fig:ABPtotal1}}
\end{figure}

\begin{figure}
\epsscale{0.9}
\plottwo{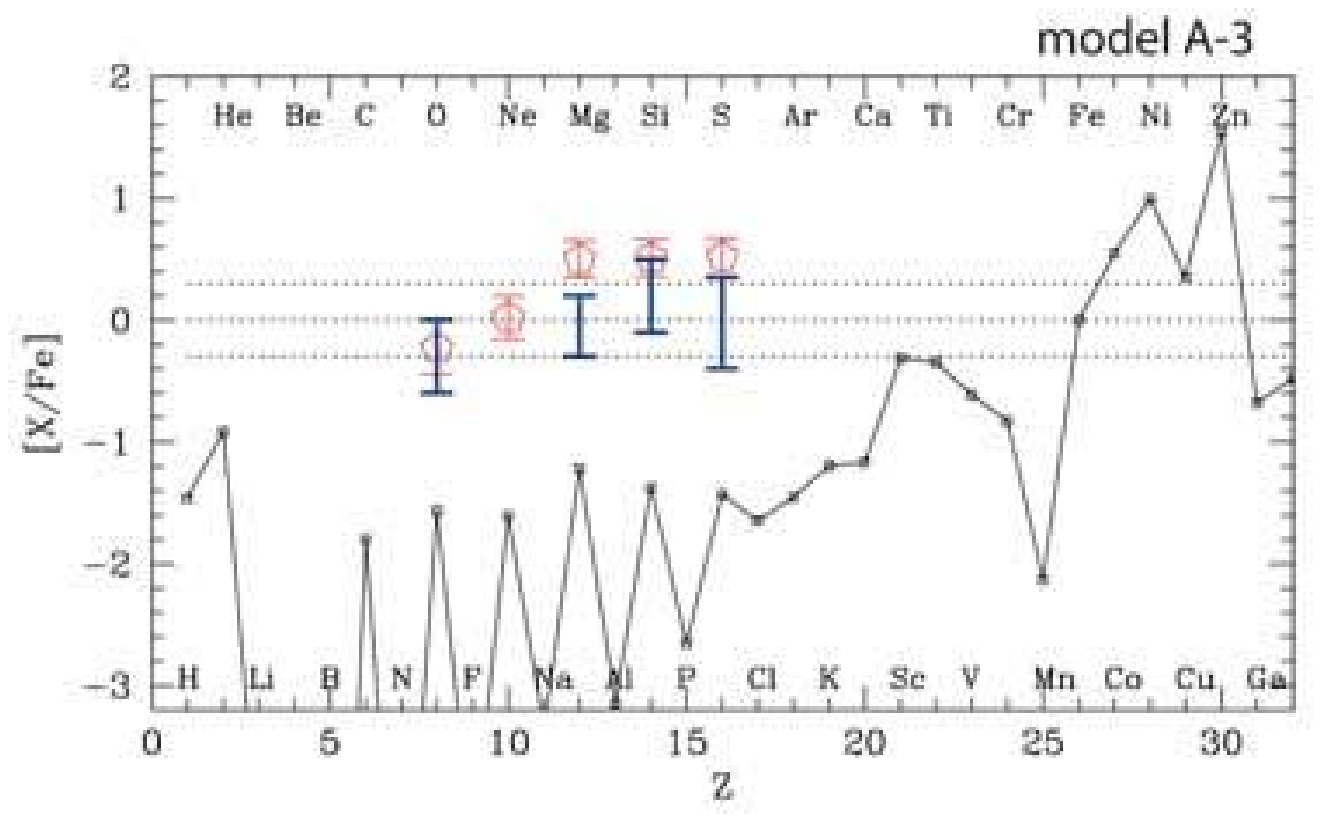}{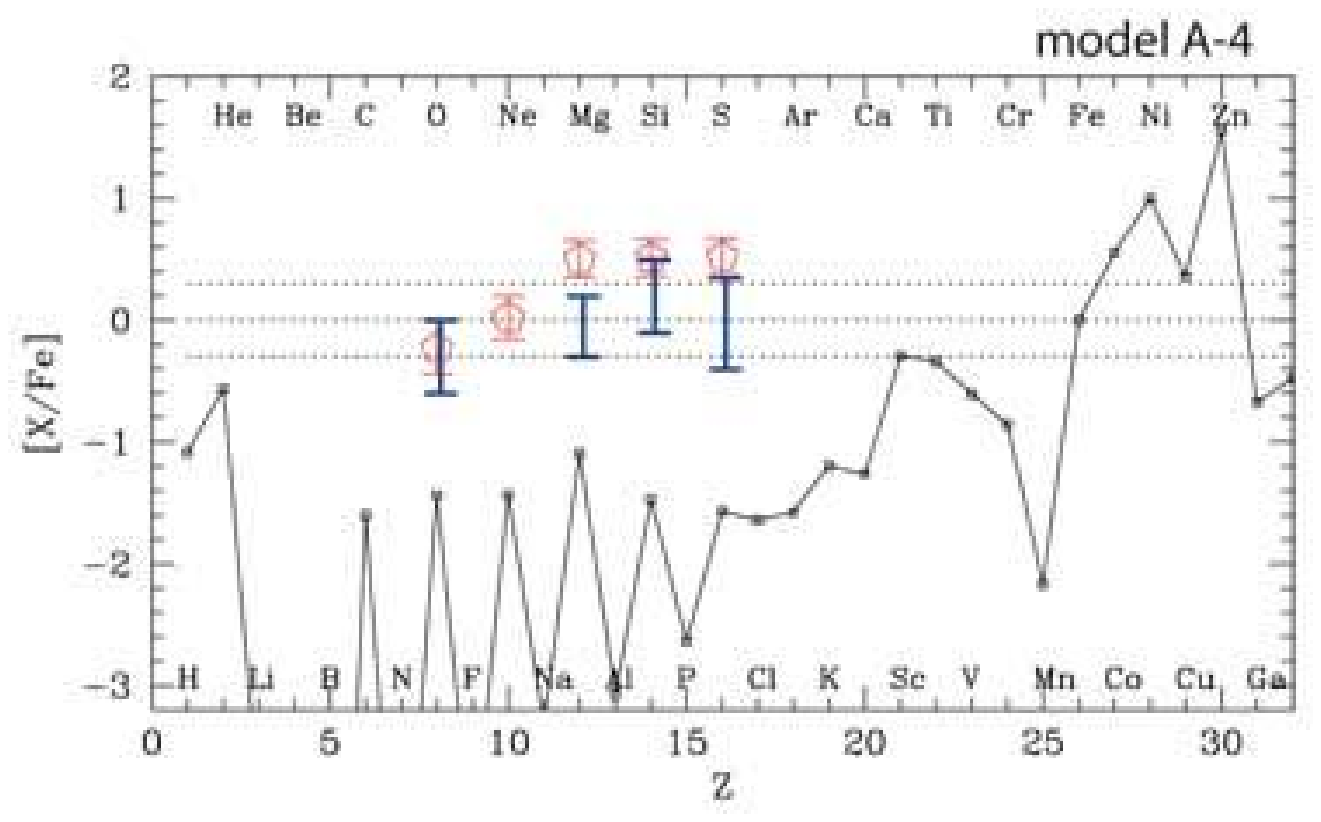}
\plottwo{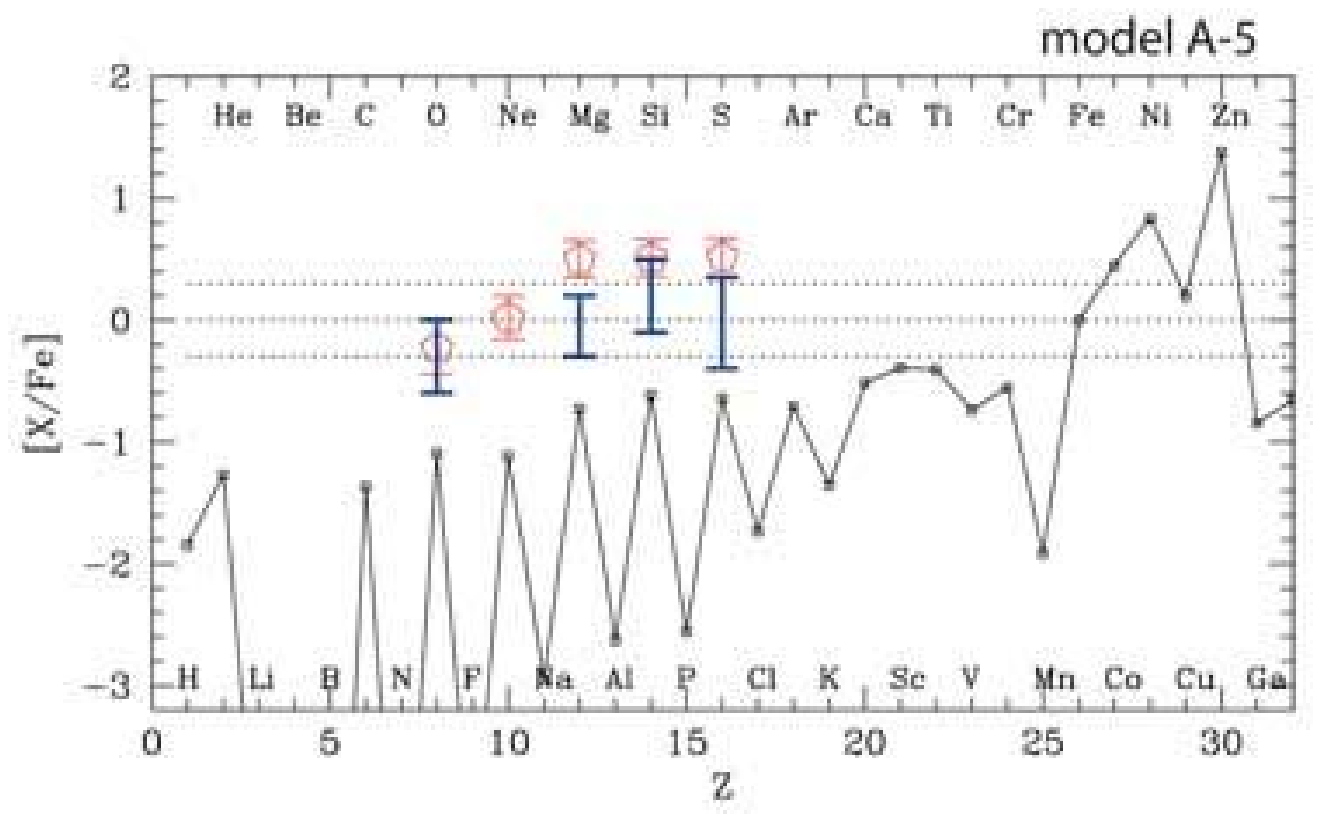}{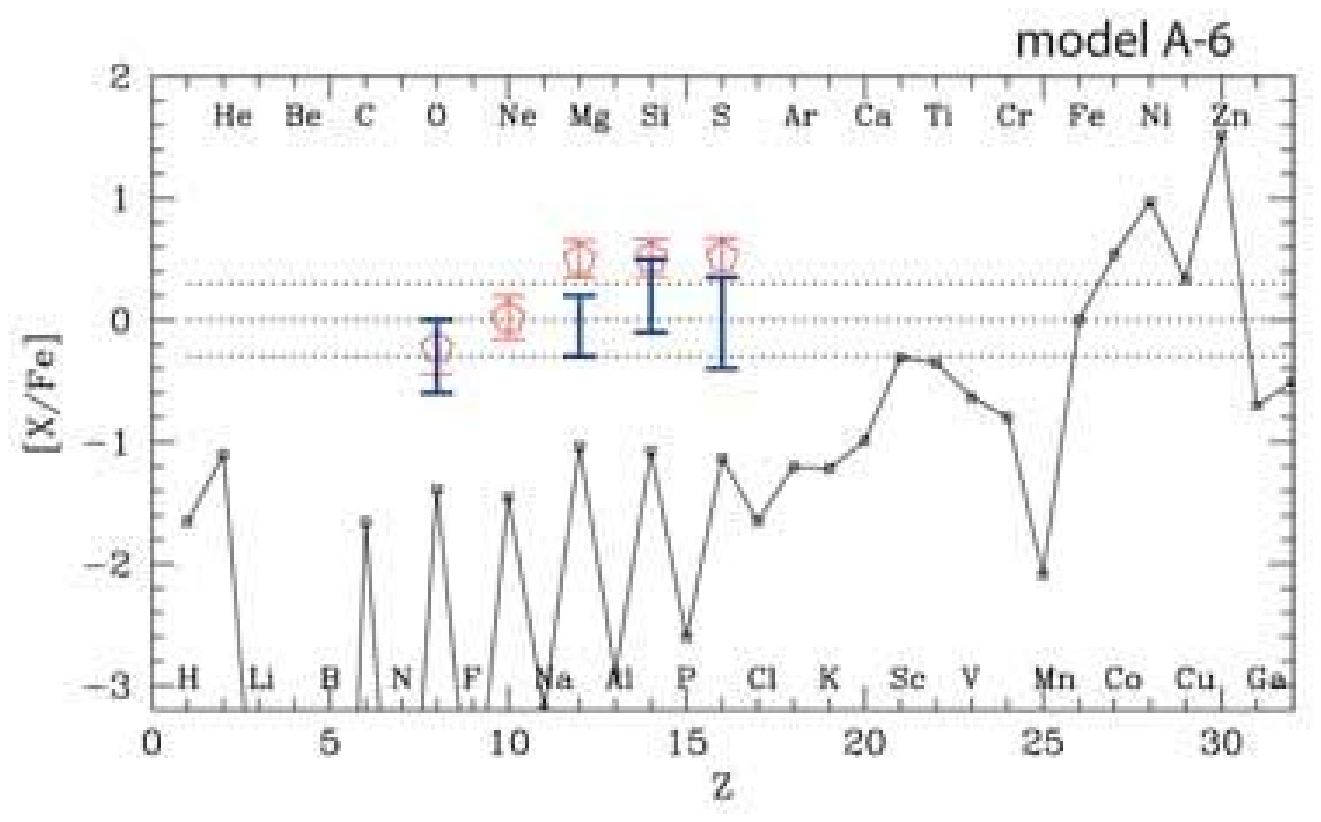}
\plottwo{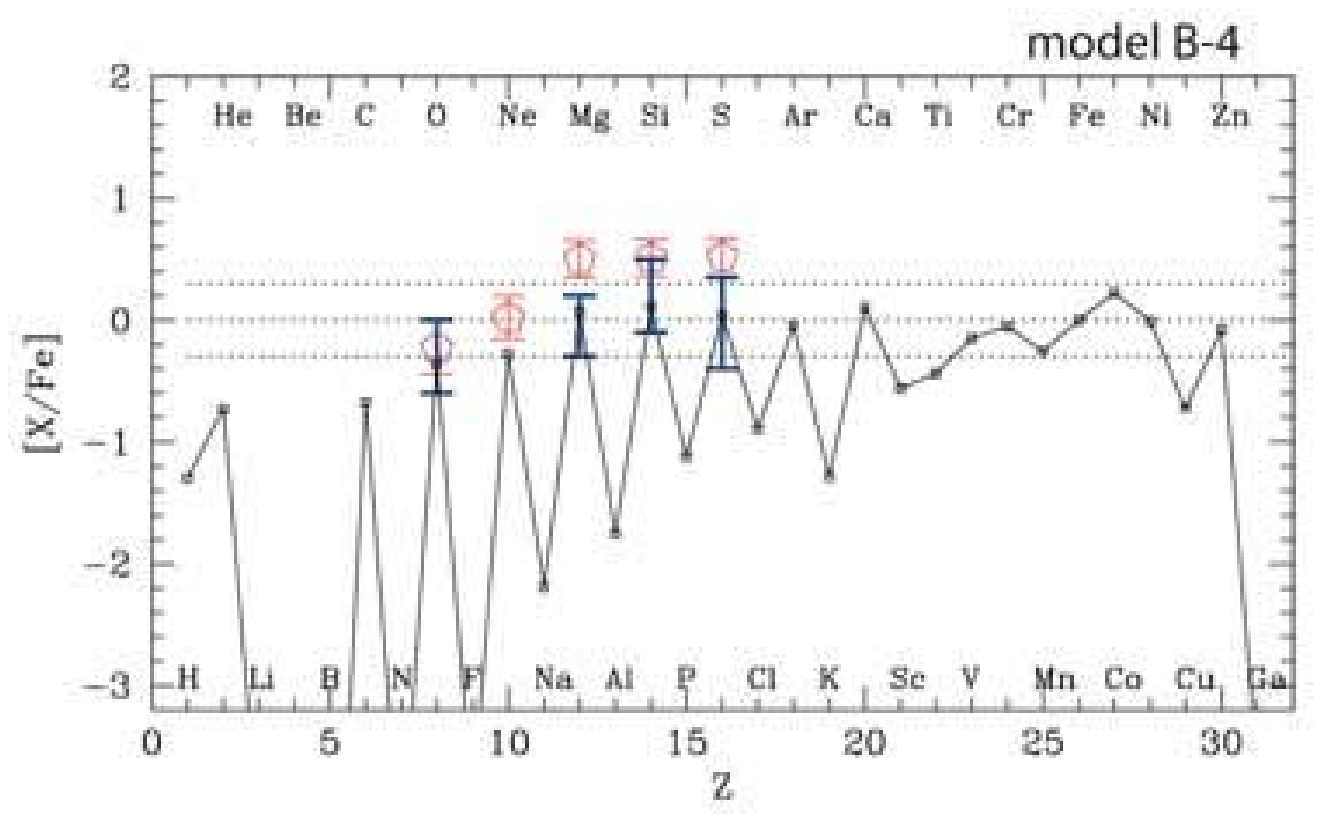}{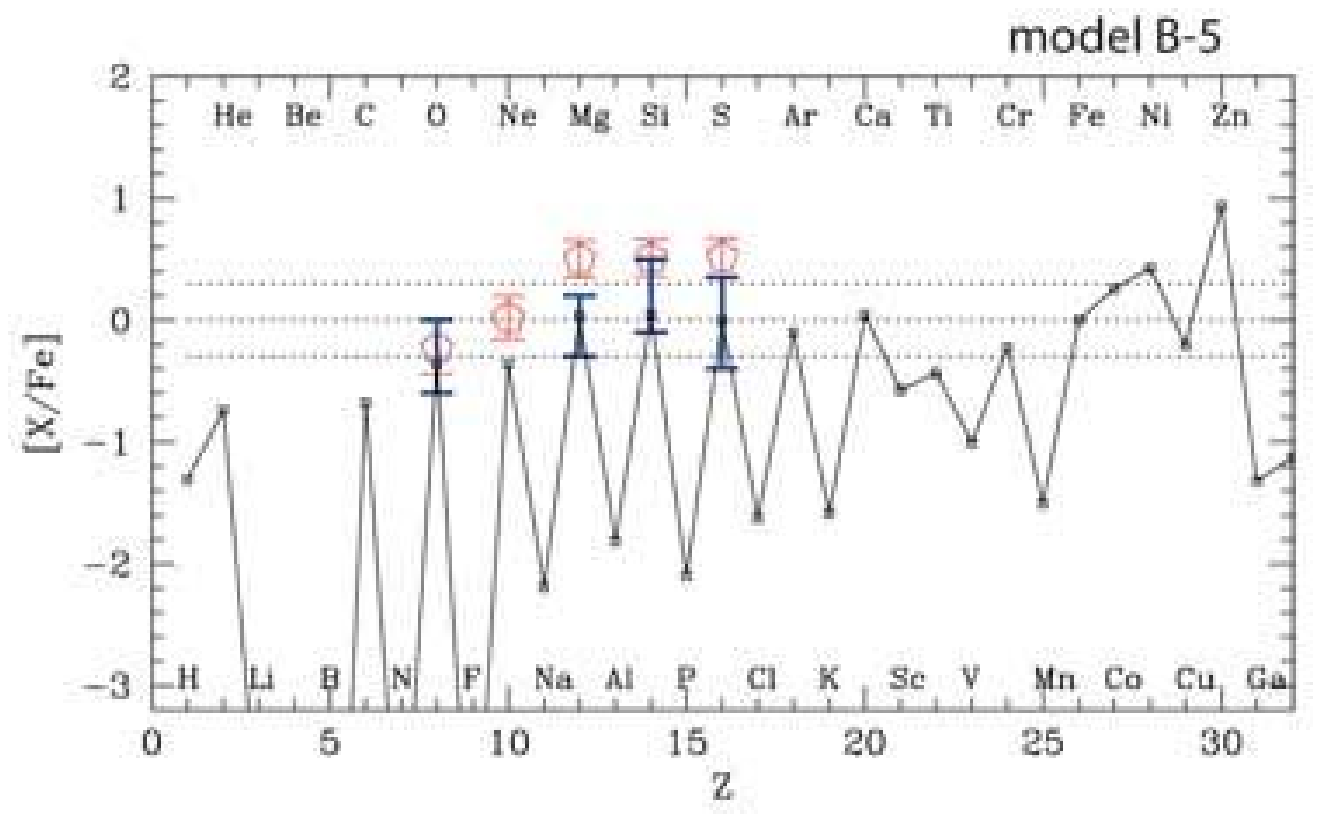}
\caption{Same as Figure~\ref{fig:ABPtotal1}, but for lower resolutoin 
models. 
\label{fig:ABPtotal2}}
\end{figure}

Figure~\ref{fig:ABPtotal1} and Figure~\ref{fig:ABPtotal2} show the
total abundance pattern for each model of higher resolution and lower 
resolution models, respectively, which is compared with the
observational data of intracluster medium (ICM) and M82. Note that the 
following discussions do not depend on the resolution. In these
figures the abundance data for the ICM gas are shown with the 
bars (Baumgartner et al. 2005; Peterson et al. 2003), while the
pentagons show the data for the gas of the central region of M82 (Ranalli et al.
2005; see also Tsuru et al. 1997). These data show that (1) the ratio [${\rm O}/{\rm Fe}$]
is smaller than the solar value, (2) [${\rm Ne}/{\rm Fe}$] is
about the solar value, and (3) the intermediate-mass
$\alpha$-elements such as Mg, Si, S exhibit oversolar abundance -
that is, [Mg/Fe], [Si/Fe] and [S/Fe] $\sim 0.5$ 
(Origlia et al. 2004; Ranalli et al. 2005). 

Note that these data are not explained by standard Type II SN
nucleosynthesis models. Also, if underabundance of [O/Fe] is due
to the contribution of Type Ia SNe, other $\alpha$-elements such
as Si and S should be also underabundant. Loewenstein (2001)
suggested the contribution of Pop III hypernovae (supernovae of ordinary massive 
stars such as 25$M_\odot$, with the explosion energy of at least $\sim 10$ times larger 
than normal supernovae; e.g., Nomoto et al. 2003) to the enrichment of ICM, in
order to explain low [O/Fe] and high [Si/Fe], using the hypernovae
models by Nakamura et al. (2001) and Heger et al. (2001). Umeda et
al. (2002) also discussed this feature, but they predicted smaller
[Ne/Fe] and [Mg/Fe] than the data given by Ranalli et al.
(2005).

Here we compare nucleosynthesis calculations of our CVMS models
with these observational data. The results are summarized as
follows. For our Case B models we obtain the abundance pattern
generally close to the observations of both ICM and M82 - for
instance, the underabundance of [O/Fe]and [Ne/Fe], and the over
solar values of [Mg/Fe], [Si/Fe] and [S/Fe]. On the other hand,
all Case A models result in very underabundant values of
[$\alpha$/Fe] $\lsim -1$ because the mass fraction of $^{56}{\rm
Ni}$ synthesized in the jet matter is much larger than that
synthesized in the matter which does not accrete. What is more,
contribution by the jet material is dominant in such models and
the uncertainty is very large. The yields 
of PISNe (Heger \& Woosley 2002; Umeda \& Nomoto 2002) are [${\rm O}/{\rm Fe}$] $\sim$ [Mg/Fe], [${\rm Ne}/{\rm Fe}$] $\sim 0$, and [Si/Fe] and [S/Fe] $\sim 1.0$, not consistent with these data.  
Therefore the yields of our Case B models for CVMSs can explain these abundance patterns better than those of PISNe. 

\subsubsection{Intergalactic Medium}
The abundances in the intergalactic medium (IGM) at high redshift also provide 
important information on the early chemical evolution 
of the universe. Many researchers have attempted to measure 
the metallicity of IGM at high redshift (Songaila \& Cowie 1996; Songaila 2001; Schaye et al. 2003). Aguirre et al. (2004) observed the abundances of C and Si 
in IGM at redshift $1.5 \lsim z \lsim 4.5$, argued that Si and C have the same origin, and obtained [C/Si] $\sim -0.77$. This value is 
considerably lower than the yields by Population III ordinary massive 
stars ($M \lsim 40M_{\odot}$) (see also Heger \& Woosley
 2002; Chieffi \& Limongi 2002; Umeda \& Nomoto 2002).

Matteucci \& Calura (2005) discussed whether this ratio could be reproduced with their chemical evolution models 
and obtained [C/Si] $= -2.0$ - $-1.7$ by including the contribution of Pop III stars over 100$M_{\odot}$. This is too small
to be compatible with the observed value. Thus they concluded that the contribution of VMSs could not be large. However, they adopted the
yields of PISNe ($130-300M_{\odot}$) only (Umeda \& Nomoto 2002; Heger \& Woosley 2002). PISNe enrich much more Si than C. 

Although there 
is a similar feature between PISNe and CVMSs in that [C/Fe] $<$ 0, [Si/Fe] 
$>$ 0 and thus [C/Si] $<$ 0,
[C/Si] from CVMSs is not so extreme as that by PISNe. With our yields of 1000$M_{\odot}$ CVMS,
[C/Si] $\sim -0.86$ - $-0.68$ (including both high and low resolution models of $1000M_{\odot}$), which is more than 10 times larger 
than the results by Matteucci \& Calura (2005). Our values are 
compatible with the observed value (Aguirre et al. 2004), and they 
show that the contribution of CVMS to the IGM enrichment 
is significant. 

\subsubsection{Extremely Metal-Poor Stars}
\begin{figure}
\epsscale{1.1}
\plottwo{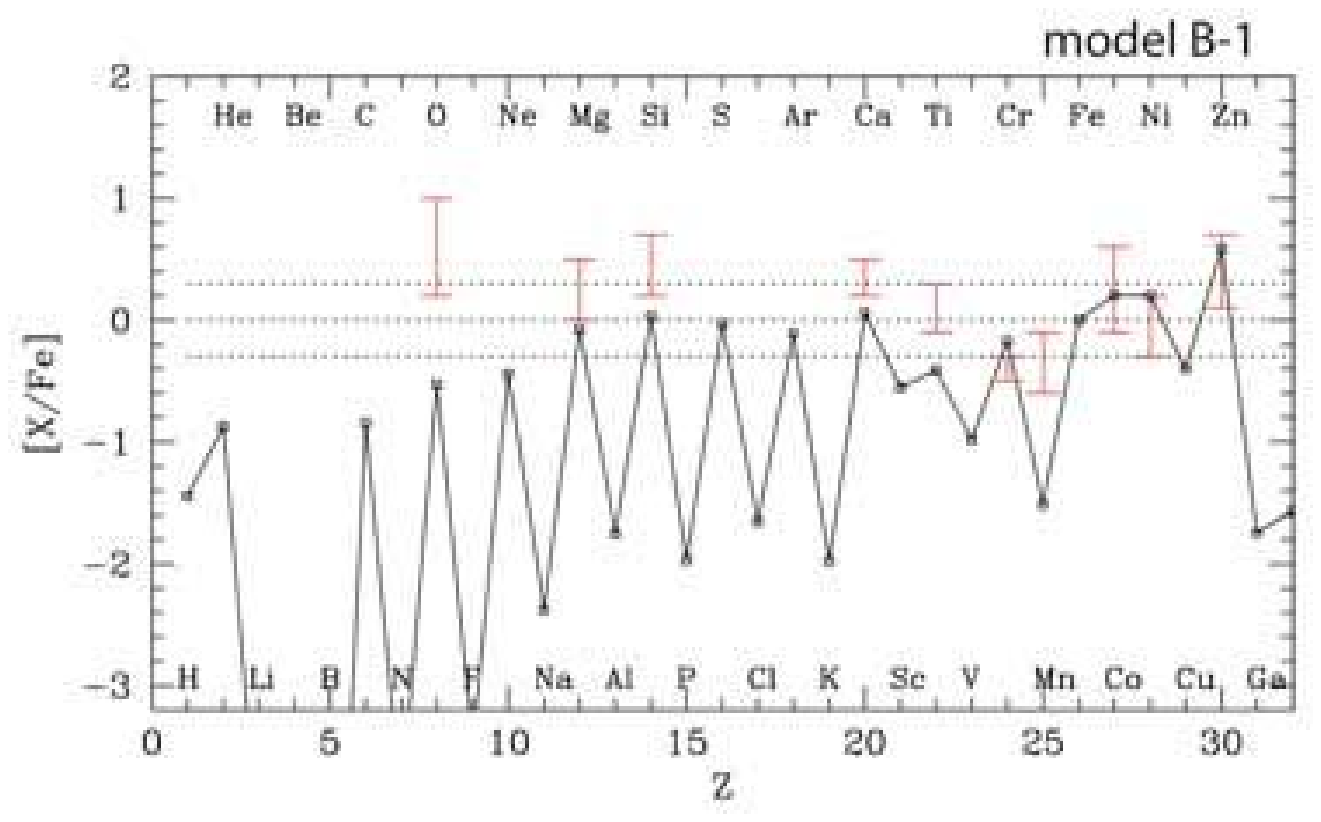}{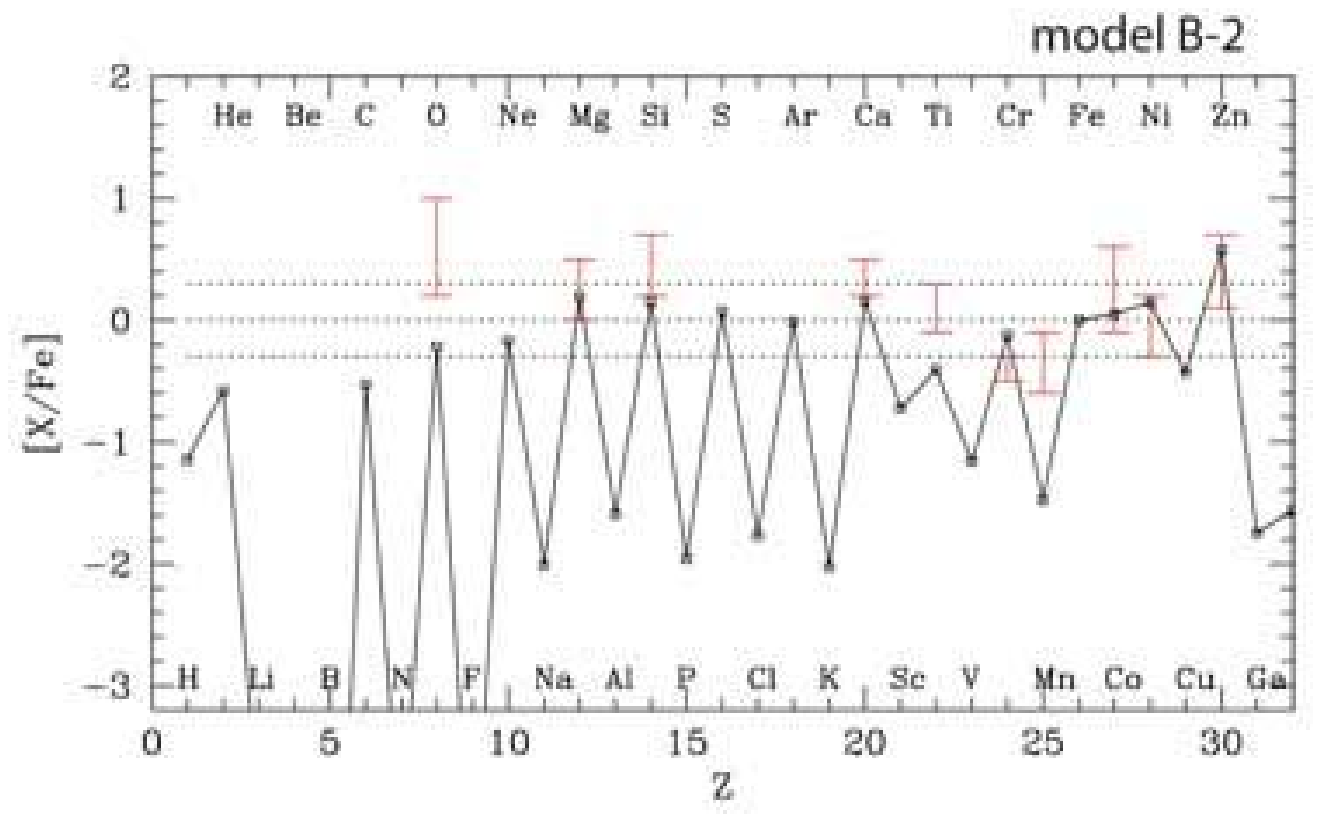}
\epsscale{0.55}
\plotone{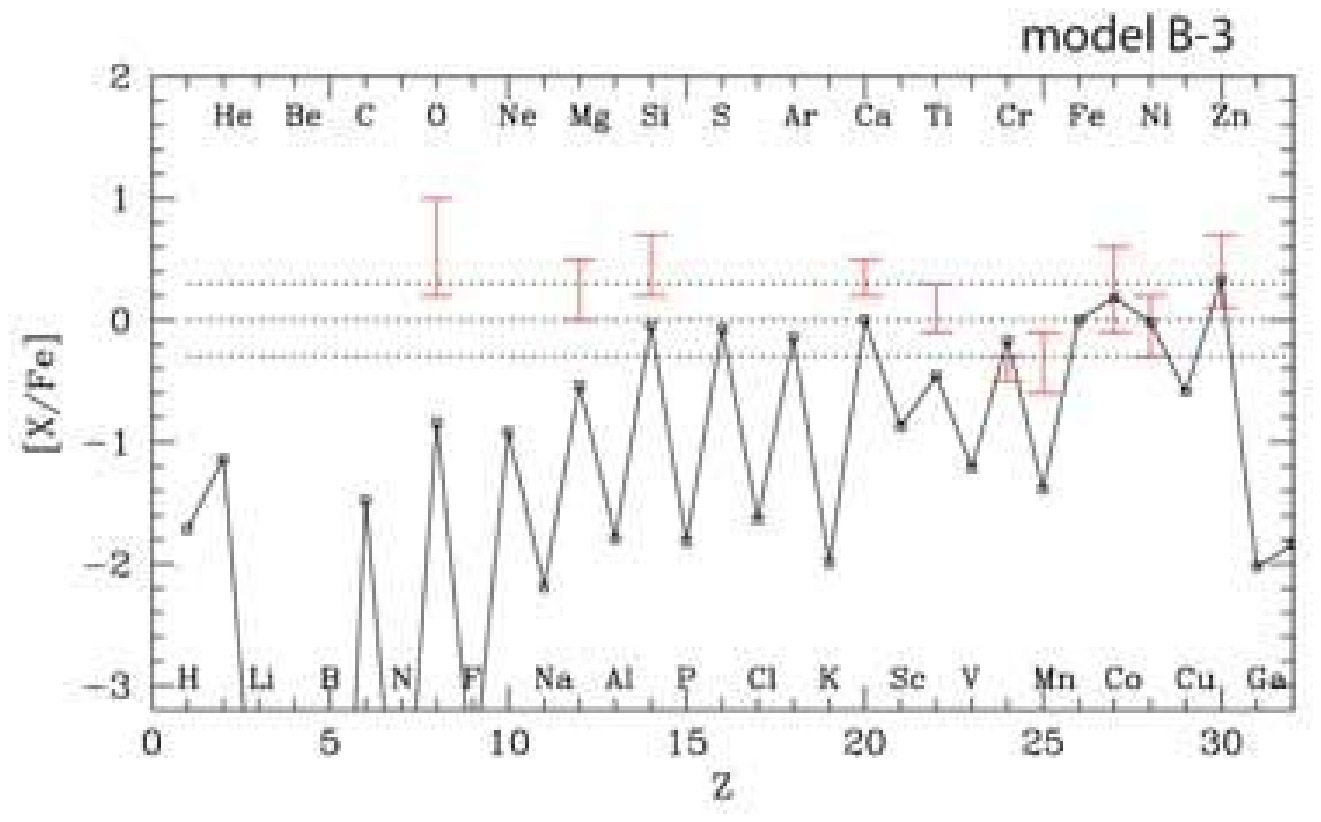}
\caption{Total abundance patterns including jet contribution for Case B.
The bars show the observational ranges of EMP stars' abundances in Galactic halo (Cayrel et al. 2004).
\label{fig:ABPtotalHALO}}
\end{figure}

Figure~\ref{fig:ABPtotalHALO} compares the yields of our models with extremely
metal-poor (EMP) stars in the Galactic halo (the data by Cayrel et
al. 2004). The result is that our Case B CVMS models mostly agree
with these Galactic halo star data for both $\alpha$-elements and
iron-peak elements. [Mg/Fe] in the EMP stars is oversolar for a wide
range of metallicity. [Cr/Fe] and [Mn/Fe] are small while [Co/Fe]
and [Zn/Fe] are large. Cr (produced as $^{48}{\rm Ti}$) and Mn
(produced as $^{55}{\rm Co}$) are mainly produced in the
incomplete silicon burning region, while Co (produced as
$^{59}{\rm Cu}$) and Zn (produced as $^{64}{\rm Ge}$) are mainly
produced in the complete silicon burning region. We note that the
aspherical models for ordinary massive stars with 25$M_{\odot}$
and 40$M_{\odot}$ in Maeda \& Nomoto (2003) are also consistent
with EMP star's abundance patterns. Umeda \& Nomoto (2003) obtained
similar results with spherical models by introducing a mixing and
fallback scenario.

It has been reported that [O/Fe] is generally oversolar for EMP stars,
which does not agree with our models. However, there are little
data for [O/Fe] at [Fe/H] $\lsim -3$ and the uncertainties involved
in the non-local thermal equilibrium (NLTE) effects and 3D effects may be too large to make conclusive statements.
Therefore, to answer the question of whether metal-free CVMSs could
contribute to the enrichment at [Fe/H] $< -3$, we will need more
accurate observational data of [O/Fe].

\section{Summary \& Discussion}

\subsection{Summary}
We first calculated the evolution of Pop III CVMSs with
$M=500M_{\odot}$ and 1000$M_{\odot}$ from the pre-main-sequence through the 
collapse with spherical symmetry. These
CVMSs are thought not to explode if they undergo spherical collapse (Fryer et al. 2001). We
assumed that these stars explode in a form of bipolar jets, and explored the
required constraints.  The results of our nucleosynthesis
calculations were used to examine their contribution to the
chemical evolution of galaxies. Our major findings are:
\begin{itemize}
   \item[1] The region which experiences explosive silicon burning to produce
iron-peak elements is more than 20\% of the total mass, much
larger than those of ordinary massive stars such as a
25$M_{\odot}$ star.  Note that for the metal-free 25$M_{\odot}$
star model, this fraction is less than 10\% (Umeda \& Nomoto
2002). This is because for the 500$M_{\odot}$ and 1000$M_{\odot}$
models the density and temperature distributions are much flatter
than those of 25$M_{\odot}$ stars.
   \item[2] Typical explosion energy is of the order of ${10}^{54}$ erg for
1000$M_{\odot}$ models for the parameter ranges in this study.
   \item[3] Black hole masses are $\sim$ 500$M_{\odot}$ for the
1000$M_{\odot}$ star models. Note that such a black hole mass is
very similar to those of IMBHs, e.g., a claimed $\sim$
700$M_{\odot}$ black hole in M82. It is quite possible that CVMSs
could be the progenitors of IMBHs.
   \item[4] Nucleosynthesis yields of CVMS have similar patterns of
[$\alpha$/Fe] to the observed abundance patterns of both ICM and gases
of the central region of M82 if the contribution of the jet is small (Case B).
Specifically, for Case B small ratios of [O/Fe] and [Ne/Fe]
combined with large [Mg/Fe], [Si/Fe] and [S/Fe] (i.e. large [(Mg, Si, S)/O]) 
are generally 
more consistent with these observational data than those of hypernovae and PISNe. 
   \item[5] For IGM, [C/Si] of our CVMS models is compatible with that of IGM at high redshift ($z=5$), which is
sufficiently higher than those of PISNe.  
   \item[6] For Fe-peak elements, the main feature of the yields of our
Case B CVMSs is that [Cr/Fe] and [Mn/Fe] are small while [Co/Fe]
and [Zn/Fe] are large. This is consistent with the observed ratios
in the extremely metal-poor (EMP) stars.  The oversolar ratios
of some $\alpha$-elements, such as [Mg/Fe] and [Si/Fe], are also
consistent with EMP stars. Our CVMS models do not agree with the
oversolar [O/Fe] of EMP stars. However, more data of [O/Fe] in EMP stars
will be needed in order to see whether CVMSs can contribute to the
early galactic chemical evolution. In this sense [O/Fe] would be
important to discriminate between different models.
\end{itemize}

\subsection{Discussion}
\subsubsection{Mass Accretion and Mass Loss}
It was pointed out (Omukai \& Palla 2003; Tan \& McKee 2004) that after a protostar
starts shining as a main-sequence star the accretion still
continues. In our current study, as a starting point the effect of
accretion on mass growth during the presupernova evolution is not
included. In our next more realistic models
such accretion will be included in the evolutionary calculations
also. However, Omukai \& Palla (2003) find that when the protostar
simulation of very massive stars is carried out properly with
time-dependent accretion rates, the rates generally decrease
toward the end of the protostar era and after the onset of the
main-sequence stellar phase. 

Also, it was pointed out 
(Maeder \& Meynet 2004) that mass loss will not be negligible even
for zero-metallicity stars when they are rotating, and hence mass
loss also will be included in the next step of our models.
However, we expect that our major conclusions as summarized above
are still valid, at least qualitatively. Somewhat more massive
CVMSs, however, may be needed to obtain the same mass black holes
if mass loss is significant.

\subsubsection{Reionization and Chemical Enrichment}
For our CVMSs, the
timescale of evolution from the zero-age main sequence to
core-collapse is $\sim$ $2 \times {10}^6$ years --
only 1/3 - 1/10 as long as for ordinary massive stars (13 - 25
$M_\odot$). So if these CVMSs were formed as
the first generation stars they would be the first contributor to
reionize and enrich the universe (Omukai \& Palla 2003; Tumlinson
\& Shull 2000; Bromm et al. 2001; Schaerer 2002).

Concerning the existence of VMSs, it was proposed (e.g., see
Wasserburg, \& Qian 2000; Qian et al. 2002; Qian \& Wasserburg
2002) that the prompt inventory involving VMSs produced the
elements from C to the Fe group, in order to explain the observed
jump in the abundances of heavy r-process elements at [Fe/H] $\sim
-3$, and also that while VMSs themselves produced no heavy
r-elements, these stars dominated chemical evolution earlier at
[Fe/H] $< -3$. Some others (e.g., Venkatesan \& Truran 2003;
Tumlinson, Venkatesan, \& Shull 2004) argue that various
observational data on reionization, the microwave background, the
metal enrichment of the high redshift IGM, etc., indicate that the
IMF of the first stars need not necessarily have been biased
toward high masses. In what follows, we revisit this issue on the
basis of our present models.

In section \S{3.4} we estimated the ionization efficiency of our
CVMSs. It was found that the number of ionizing photons per baryon
in the universe, generated in association with the IGM metallicity
$Z_{\rm IGM} \sim 10^{-4}$, is $N_{\rm Lyc}/N_{\rm b} \sim 150$,
and so CVMSs can contribute significantly to reionization of IGM in
the early epoch. Here we emphasize that our current result for CVMS is 
contributed from the mass range with $ \sim 300 - 1000 M_{\odot}$, and hence the
PISN (pair instability supernova) range is not included. On the other
hand, Venkatesan \& Truran (2003) give $N_{\rm Lyc}/N_{\rm b} \sim
10$ for $Z_{\rm IGM} \sim 10^{-4}$ for the mass range $\sim 100 -
1000 M_{\odot}$ which reflects the large contribution of PISNe to
metal enrichment. Daigne et al. (2004) also considered  
reionization and chemical enrichment of IGM simultaneously and 
reached similar conclusion that VMSs are not necessary. However, 
in their models CVMSs do not explode. Here we may note that less massive Pop III
stars ($\lsim 100 M_{\odot}$) can also produce the amount of
ionizing photons per metal similar to CVMS (Venkatesan \& Truran 2003;
Tumlinson et al. 2004).

The relation between reionization and metal enrichment of IGM
becomes clearer if we solve the equation for $N_{\rm Lyc}/N_{\rm
b}$ (Eqn. 2 of Venkatesan \& Truran 2003) for a given value of
$Z_{\rm IGM}$. For a 1000 $M_{\odot}$ star, $Z/Z_{\odot} \sim
10^{-3.4}$ and $10^{-4.4}$ for the required number of ionizing
photons per baryon 10 and 1, respectively. This is about one order
of magnitude smaller than the case for the mass range $100 - 1000
M_{\odot}$ (mainly contributed by PISNe). The difference between
CVMSs and PISNe is larger if we consider the enrichment of iron.
The 260 $M_{\odot}$ PISN of Heger \& Woosley (2002) gives
$Z_{\rm Fe}/Z_{\rm Fe, \odot} \sim 10^{-2} - 10^{-3}$, while our 500$M_{\odot}$ and 1000$M_{\odot}$ star gives $\sim 10^{-3.2} - 10^{-4.2}$.

A main critique against the existence of PISNe comes from the fact
that we do not see the abundance patterns of PISNe in EMP stars (Umeda \& Nomoto 2002; Tumlinson et al. 2004). The EMP abundances are
indeed suggested to be accounted for by hypernovae or faint
supernovae of less massive stars of $\lsim 100 M_{\odot}$ (Umeda
\& Nomoto 2003). However, the apparent
lack of evidence of VMSs by no means contradicts the existence of CVMSs at 
earlier epochs, if the majority of first stars in the
earlier epoch has masses $\gsim 300 M_{\odot}$. 
First, PISNe from stars of $\lsim 300 M_{\odot}$
will be just a minor fraction in such a case, explaining the lack of
the signature of PISNe. Second, $Z/Z_{\odot}$ expected from our
CVMSs is smaller than PISNe. Namely, the metal enrichment by CVMSs
might be finished before ordinary core-collapse SNe become dominant. 
Note that the abundance, especially of oxygen, in EMP stars and IGM is different. Here we have 
shown that the yields of our CVMSs can reproduce the abundance of IGM (Section 4). Therefore, 
it would be worthwhile studying a scenario where CVMSs are first formed in pregalactic mini halos, 
and then subsequently ordinary core-collapse SNe took place in the galactic halo.

We mentioned that the early universe would have been contaminated too 
much if there were many PISNe. This conclusion is not affected when 
the calculation of PISNe considers the effect of rotation and asymmetric 
explosion (see Stringfellow \& Woosley 1988). In PISNe with rotation, all matter is ejected with nothing left as in non-rotating models, and
the total amount of matter ejected is almost independent of the 
geometry of the explosion, whether spherical or not. The explosive 
nucleosynthesis itself and hence the resulting exact composition of the 
ejected matter change with rotation and under the consequent asymmetric 
environment, but that does not change our conclusion that too many of 
PISNe result in the overabundance of heavy elements in the early 
universe which contradicts with the observation.

\subsubsection{Initial Mass Function}
Tumlinson et al. (2004) raised two problems associated with top
heavy IMF of the first stars.  (a) If stars are all $M \gsim 300
M_{\odot}$, no metals are released from Pop III stars to trigger
the transition from the first stars to present IMF star formation,
and (b) no mechanism has been proposed for forming stars more
massive than $\sim$ 300$M_{\odot}$ without forming PISNs.

Here we discuss how our CVMS models could resolve these apparent
problems. Concerning (a), our present CVMS models do eject metals
(though less than PISNe), leading to metal enrichment of IGM. In
this connection, note that the existing literature concerning the
effects of VMSs (in Pop III IMF) on reionization, etc., includes
only contribution by PISNs but not those heavier because they assumed that 
heavier stars do not explode, and hence make no contribution.
However, we emphasize the importance of the explosion of these heavier stars
($\gsim$ 500$M_{\odot}$). 

Concerning (b), collisions and merging of ordinary massive stars
in very dense clusters are expected to lead to the formation of
more massive stars (Ebisuzaki et al. 2001; Portegies Zwart et
al. 1999, 2004a, 2004b) which can easily lead to CVMSs ($\gsim
300 M_{\odot}$) without or with only minor fraction of stars
responsible for PISNe ($\lsim 300 M_{\odot}$), and hence the
problem in question can disappear. Specifically, in this scenario
PISN stars will have no time to explode before merging into
heavier stars when the timescale of the PISN star evolution is
longer than merging timescale. Also, even in the case of single
star formation (no merging) there is yet no reason to exclude a
possibility of the first star IMF with the minimum mass of $\sim
300 M_{\odot}$.

As to the question of how CVMSs are formed, the first generation
stars are generally thought to have formed in low-mass halos with
the virial temperature $T_{\rm vir} <$ 10$^4$ K.  Then the upper
limit to the mass of first stars may be $\sim$ 300$M_{\odot}$
(e.g., Bromm \& Loeb 2004). However, Oh \& Haiman (2002)
investigated halos with higher mass, with $T_{\rm vir} >$ 10$^4$
K. The evolution of these high mass halos, e.g., of $\sim
10^9M_{\odot}$, is found to be quite different from the low mass
case, and the degree of fragmentation of the gas is still highly
uncertain. Therefore, it appears that whether more massive stars
can be directly formed is still an open question. However,
regardless of the feasibility of direct formation, it has been
emphasized by, e.g., Ebisuzaki et al. (2001), and Portegies Zwart
et al. (1999, 2004a, 2004b), that CVMSs will be formed easily by
merging of less massive stars in very dense star clusters, and
hence there appears to be essentially no problem for CVMS
formation.

\subsubsection{Black Hole Mass}
As to the existence of IMBHs, currently extensive effort is under
way to try to detect them in nearby galaxies (see, e.g., Barth et al. 
2005 for recent review). Already several of these objects have been
identified, mostly in dwarf elliptical galaxies, but some in
spiral galaxies, e.g., M33, IC 342, Pox 52, NGC 4395 and several galaxies 
in the Sloan Digital Sky Survey (SDSS) sample. For example, the black hole mass obtained for M33
is less than $\sim$ 3000$M_{\odot}$ but larger than the mass of a
stellar mass black hole.  The black hole mass obtained for NGC
4395 is $\sim$ 10$^4$ - 10$^5 M_{\odot}$, but its spectra are
unlike NLSI (narrow line Seyfert I) - a class of AGN which tends
to have small mass. We expect more of these IMBH candidates, with
better mass measurement, to be identified in the very near future.

As an example of possibly more recently formed IMBHs, Matsumoto et
al. (2001) reported the possible discovery of a $\gsim$
700$M_{\odot}$ black hole in M82 as an ULX (ultra-luminous X-ray
source).  Since an ULX was first detected in 1989 by the {\sl
Einstein} Observatory (Fabbiano 1989), many of these objects have
been discovered. Possible scenarios for formation of IMBHs
associated with ULXs are speculated in a recent article by Krolik
(2004). Colbert \& Mushotzky (1999) first suggested that these
luminous objects are indeed IMBHs, because their luminosity is
super Eddington for stellar mass black holes if spherical
accretion is adopted.  That may not be necessary if beaming, etc.,
is assumed. However, ULXs may be heterogeneous, and at least some
of these objects may well prove to be IMBHs. The prospect is
bright because various observations in multifreuency bands can
distinguish between different interpretations.

\acknowledgements

This work has been supported in part by the Grant-in-Aid for Scientific Research (15204010, 16540229, 17030005, 17033002) and the 21st Century COE Program (QUEST) from the JSPS and MEXT of Japan.

We thank the referee for useful comments.

%\end{document}

\clearpage

%tables

\begin{deluxetable}{ccccccccc}
\tablecaption{The higher resoluton models.} 
\tablewidth{0pt}
\tablehead{
\colhead{Models} &
\colhead{progenitor($M_{\odot}$)} &
\colhead{$M_{\rm BH0}$($M_{\odot}$)} &
\colhead{$R_0$(km)} &
\colhead{${\theta_{\rm jet}}^o$} &
\colhead{$\epsilon$} &
\colhead{$\mu$} &
\colhead{$F_{\rm thermal}$} &
\colhead{$v_{\rm jet}/$c}
}
\startdata 
 {\bf A-1} &  1000 & 100 & $1.5 \times {10}^4$ & 15 & 0.01  & 0.1  & 0.01 & 0.45 \\
 {\bf A-2} &  1000 & 100 & $1.5 \times {10}^4$ & 30 & 0.01  & 0.1  & 0.01 & 0.45 \\
\hline 
 {\bf B-1} &  1000 & 100 & $1.5 \times {10}^4$ & 15 & 0.01  & 0.005  & 0.95 & 0.45 \\  
 {\bf B-2} &  1000 & 100 & $1.5 \times {10}^4$ & 15 & 0.005 & 0.0025 & 0.95 & 0.45 \\
 {\bf B-3} &   500 &  50 & $1.1 \times {10}^4$ & 15 & 0.01  & 0.005  & 0.95 & 0.45 \\ 
\enddata
\tablecomments{initial black hole mass $M_{\rm BH0}$, the radius at the inner boundary of the simulations $R_0$, jet injected angle $\theta_{\rm jet}$, energy tranformation efficiency $\epsilon$, the mass fraction in which accreted matter is ejected as jet $\mu$, and the jet velocity normalized by the speed of light $v_{\rm jet}$.} 
\label{tab:ModelsH}
\end{deluxetable}

\begin{deluxetable}{ccccccccc}
\tablecaption{The lower resolution models.}
\tablewidth{0pt}
\tablehead{
\colhead{Models} &
\colhead{progenitor($M_{\odot}$)} &
\colhead{$M_{\rm BH0}$($M_{\odot}$)} &
\colhead{$R_0$(km)} &
\colhead{${\theta_{\rm jet}}^o$} &
\colhead{$\epsilon$} &
\colhead{$\mu$} &
\colhead{$F_{\rm thermal}$} &
\colhead{$v_{\rm jet}/$c}
}
\startdata
 {\bf A-3} &  1000 & 100 & $1.5 \times {10}^4$ & 15 & 0.005 & 0.05 & 0.01 & 0.45 \\
 {\bf A-4} &  1000 & 100 & $1.5 \times {10}^4$ & 15 & 0.002 & 0.02 & 0.01 & 0.45 \\  
 {\bf A-5} &  1000 &  50 & $8.4 \times {10}^3$ & 15 & 0.005 & 0.05 & 0.01 & 0.45 \\
 {\bf A-6} &  1000 & 200 & $2.7 \times {10}^4$ & 15 & 0.01  & 0.1  & 0.01 & 0.45 \\
 {\bf A-7} &  500  &  50 & $8.4 \times {10}^3$ & 15 & 0.01  & 0.1  & 0.01 & 0.45 \\
\hline
 {\bf B-2} &  1000 & 100 & $1.5 \times {10}^4$ & 15 & 0.005 & 0.0025 & 0.95 & 0.45 \\
 {\bf B-4} &  1000 & 100 & $1.5 \times {10}^4$ & 15 & 0.01  & 0.02   & 0.9  & 0.32 \\
 {\bf B-5} &  1000 & 100 & $1.5 \times {10}^4$ & 15 & 0.005 & 0.01   & 0.9  & 0.32 \\
\enddata 
\label{tab:ModelsL}

\end{deluxetable}

\begin{deluxetable}{cccccc}
\tablecaption{Caracteristics of explosion models in Table~\ref{tab:ModelsH}.}  
\tablewidth{0pt}
\tablehead{
\colhead{Models} &
\colhead{$E$ (erg)} &
\colhead{$M_{\rm BH}$($M_{\odot}$)} &
\colhead{$M_{\rm e0}$($M_{\odot}$)} &
\colhead{$M_{\rm jet}$($M_{\odot}$)} &
\colhead{$M_{\rm e}$($M_{\odot}$)} 
}
\startdata
 {\bf A-1} &  $6.7 \times {10}^{54}$ & $5.0 \times {10}^{2}$ & $4.6 \times {10}^{2}$ & $44$  & $5.0 \times {10}^{2}$   \\
 {\bf A-2} &  $2.2 \times {10}^{54}$ & $4.4 \times {10}^{2}$ & $5.2 \times {10}^{2}$ & $38$  & $5.6 \times {10}^{2}$   \\
 \hline
 {\bf B-1} &  $4.9 \times {10}^{54}$ & $4.6 \times {10}^{2}$ & $5.4 \times {10}^{2}$ & $1.8$ & $5.4 \times {10}^{2}$   \\
 {\bf B-2} &  $1.6 \times {10}^{54}$ & $4.8 \times {10}^{2}$ & $5.2 \times {10}^{2}$ & $0.94$ & $5.2 \times {10}^{2}$   \\
 {\bf B-3} &  $2.9 \times {10}^{54}$ & $2.3 \times {10}^{2}$ & $2.7 \times {10}^{2}$ & $0.90$ & $2.7 \times {10}^{2}$   \\
\enddata   
\tablecomments{Explosion energy $E$, final black hole mass $M_{\rm BH}$, ejected mass excluding jet material $M_{\rm e0}$, mass of jet $M_{\rm jet}$, and total ejected mass $M_{\rm e}$.}
\label{tab:EMtableH}
\end{deluxetable}

\begin{deluxetable}{cccccc}
\tablecaption{Same as Table~\ref{tab:EMtableH}, but for the models in Table~\ref{tab:ModelsL}.}
\tablewidth{0pt}
\tablehead{
\colhead{Models} &
\colhead{$E$ (erg)} &
\colhead{$M_{\rm BH}$($M_{\odot}$)} &
\colhead{$M_{\rm e0}$($M_{\odot}$)} &
\colhead{$M_{\rm jet}$($M_{\odot}$)} &
\colhead{$M_{\rm e}$($M_{\odot}$)} 
}
\startdata
 {\bf A-3} &  $4.0 \times {10}^{54}$ & $5.3 \times {10}^{2}$ & $4.5 \times {10}^{2}$ & $23$  & $4.7 \times {10}^{2}$   \\
 {\bf A-4} &  $5.2 \times {10}^{53}$ & $5.6 \times {10}^{2}$ & $4.3 \times {10}^{2}$ & $9.5$ & $4.4 \times {10}^{2}$   \\
 {\bf A-5} &  $3.0 \times {10}^{54}$ & $5.1 \times {10}^{2}$ & $4.5 \times {10}^{2}$ & $45$  & $4.9 \times {10}^{2}$   \\
 {\bf A-6} &  $8.2 \times {10}^{54}$ & $5.1 \times {10}^{2}$ & $4.6 \times {10}^{2}$ & $34$  & $4.9 \times {10}^{2}$   \\
 {\bf A-7} &  $6.1 \times {10}^{54}$ & $2.4 \times {10}^{2}$ & $2.4 \times {10}^{2}$ & $21$  & $2.6 \times {10}^{2}$   \\
 \hline
 {\bf B-2} &  $1.9 \times {10}^{54}$ & $4.8 \times {10}^{2}$ & $5.2 \times {10}^{2}$ & $0.95$ & $5.2 \times {10}^{2}$  \\
 {\bf B-4} &  $6.4 \times {10}^{54}$ & $4.6 \times {10}^{2}$ & $5.3 \times {10}^{2}$ & $7.4$ & $5.4 \times {10}^{2}$   \\
 {\bf B-5} &  $1.8 \times {10}^{54}$ & $4.7 \times {10}^{2}$ & $5.3 \times {10}^{2}$ & $3.7$ & $5.3 \times {10}^{2}$   \\
\enddata  
\label{tab:EMtableL} 
\end{deluxetable}

\begin{deluxetable}{cccccccc}
\tablecaption{The models in which explosion does not occur.}
\tablewidth{0pt}
\tablehead{
\colhead{Models} &
\colhead{progenitor($M_{\odot}$)} &
\colhead{$M_{\rm BH0}$($M_{\odot}$)} &
\colhead{${\theta_{\rm jet}}^o$} &
\colhead{$\epsilon$} &
\colhead{$\mu$} &
\colhead{$F_{\rm thermal}$} &
\colhead{$v_{\rm jet}/$c}
}
\startdata
 {\bf F-1} &  1000 & 100 & 15 & 0.001 & 0.01   & 0.01 & 0.45 \\
 {\bf F-2} &  1000 & 100 & 30 & 0.005 & 0.05   & 0.01 & 0.45 \\
 {\bf F-3} &  1000 & 100 & 45 & 0.01  & 0.1    & 0.01 & 0.45 \\
 {\bf F-4} &  1000 & 100 & 15 & 0.003 & 0.0015 & 0.95 & 0.45 \\  
\enddata  
\label{tab:Failure}
\end{deluxetable}

\begin{deluxetable}{cccc}
\tablecaption{Ejected mass of $^{56}{\rm Ni}$, $^{16}{\rm O}$, $^{28}{\rm Si}$, excluding jet material for higher resolution models.}
\tablewidth{0pt}
\tablehead{
\colhead{Models} &
\colhead{$M(^{56}{\rm Ni}$) ($M_{\odot}$)} &
\colhead{$M(^{16}{\rm O}$) ($M_{\odot}$)} &
\colhead{$M(^{28}{\rm Si}$) ($M_{\odot}$)} 
}
\startdata
 {\bf A-1} &  $1.5 $ & $4.5 $ & $0.69$ \\ 
 {\bf A-2} &  $12 $ & $18  $ & $4.3 $ \\  
\hline  
 {\bf B-1} &  $9.3 $ & $23 $ & $6.1 $ \\  
 {\bf B-2} &  $4.7 $ & $24 $ & $4.0 $ \\ 
 {\bf B-3} &  $8.9 $ & $10 $ & $4.8 $ \\  
\enddata 
\label{tab:EjectStellarH}
\end{deluxetable}

\begin{deluxetable}{cccc}
\tablecaption{Same as Table~\ref{tab:EjectStellarH}, but for lower resolution models.}
\tablewidth{0pt}
\tablehead{
\colhead{Models} &
\colhead{$M(^{56}{\rm Ni}$) ($M_{\odot}$)} &
\colhead{$M(^{16}{\rm O}$) ($M_{\odot}$)} &
\colhead{$M(^{28}{\rm Si}$) ($M_{\odot}$)} 
}
\startdata 
 {\bf A-3} &  $0.4 $ & $2.1 $ & $0.24$ \\  
 {\bf A-4} &  $0.12$ & $1.2 $ & $0.080$ \\  
 {\bf A-5} &  $0.36$ & $2.4 $ & $0.23$ \\  
 {\bf A-6} &  $1.4 $ & $5.1 $ & $0.77$ \\  
 {\bf A-7} &  $1.1 $ & $1.8 $ & $0.41$ \\  
\hline  
 {\bf B-2} &  $4.4 $ & $22 $ & $3.8 $ \\ 
 {\bf B-4} &  $7.0 $ & $24 $ & $5.3 $ \\  
 {\bf B-5} &  $5.8 $ & $25 $ & $4.6 $ \\   
\enddata 
\label{tab:EjectStellarL} 
\end{deluxetable}

\clearpage

% \begin{deluxetable}
% \begin{center}
\begin{deluxetable}{rrrrr|rrrrr}
\tabletypesize{\scriptsize}
\tablecaption{Nucleosynthesis products ($M_{\odot}$)
 of models A-1, B-1, B-2, and B-3.}
%\label{tab:Yields1}
\tablewidth{0pt}
\tablehead{
\colhead{Models} &
\colhead{A-1} &
\colhead{B-1} &
\colhead{B-2} &
\colhead{B-3} &
\colhead{} &
\colhead{A-1} &
\colhead{B-1} &
\colhead{B-2} &
\colhead{B-3} 
}
\startdata

$^{  }$ n& 1.68E-13& 6.94E-12& 5.19E-13& 5.65E-14& $^{50}$Cr& 1.04E-03& 1.75E-04& 9.71E-05& 1.29E-04\\
$^{  }$ p& 1.90E+02& 1.94E+02& 1.94E+02& 9.59E+01& $^{51}$Cr& 1.75E-05& 8.55E-07& 7.88E-07& 6.45E-07\\
$^{  }$ d& 2.78E-15& 3.65E-15& 2.84E-15& 4.00E-15& $^{52}$Cr& 3.81E-05& 1.61E-06& 8.16E-07& 9.77E-07\\
$^{ 3}$He& 2.12E-04& 2.12E-04& 2.12E-04& 2.02E-05& $^{53}$Cr& 6.53E-09& 4.55E-10& 2.15E-10& 2.28E-10\\
$^{ 4}$He& 2.72E+02& 2.85E+02& 2.83E+02& 1.43E+02& $^{54}$Cr& 4.17E-10& 1.30E-10& 5.06E-11& 8.69E-11\\
$^{ 6}$Li& 1.09E-18& 9.64E-18& 1.20E-17& 6.99E-18& $^{55}$Cr& 7.20E-12& 1.41E-10& 4.98E-11& 5.59E-11\\
$^{ 7}$Li& 2.77E-14& 1.36E-13& 1.52E-13& 3.47E-14& $^{48}$Mn& 6.84E-05& 2.99E-06& 1.37E-06& 1.40E-06\\
$^{ 7}$Be& 3.77E-09& 3.86E-09& 3.89E-09& 1.86E-09& $^{49}$Mn& 3.54E-04& 1.12E-04& 2.88E-05& 6.88E-05\\
$^{ 9}$Be& 6.40E-21& 3.83E-19& 3.10E-20& 8.71E-21& $^{50}$Mn& 2.30E-04& 6.94E-05& 2.08E-05& 3.61E-05\\
$^{ 8}$B & 2.93E-14& 3.05E-14& 3.05E-14& 4.54E-13& $^{51}$Mn& 1.16E-03& 2.68E-04& 9.65E-05& 1.56E-04\\
$^{10}$B & 7.04E-15& 8.02E-14& 9.11E-14& 2.96E-14& $^{52}$Mn& 1.09E-04& 9.84E-06& 7.71E-06& 9.95E-06\\
$^{11}$B & 1.08E-15& 1.22E-14& 1.20E-14& 5.68E-15& $^{53}$Mn& 1.66E-04& 2.00E-05& 1.37E-05& 2.87E-05\\
$^{11}$C & 1.74E-12& 1.05E-11& 4.28E-12& 3.14E-12& $^{54}$Mn& 3.20E-07& 1.38E-08& 6.94E-09& 7.63E-09\\
$^{12}$C & 8.11E-01& 3.56E+00& 3.68E+00& 7.61E-01& $^{55}$Mn& 1.73E-08& 9.05E-10& 4.36E-10& 5.97E-10\\
$^{13}$C & 6.07E-09& 2.23E-08& 2.30E-08& 4.58E-08& $^{56}$Mn& 8.12E-12& 1.03E-10& 4.28E-11& 8.74E-11\\
$^{13}$N & 1.16E-10& 3.20E-10& 6.14E-10& 2.31E-08& $^{57}$Mn& 7.48E-12& 1.08E-10& 4.22E-11& 7.60E-11\\
$^{14}$N & 1.16E-05& 3.49E-05& 2.77E-05& 7.15E-06& $^{50}$Fe& 2.01E-04& 1.09E-05& 3.63E-06& 4.36E-06\\
$^{15}$N & 3.13E-09& 1.30E-07& 3.93E-08& 1.40E-07& $^{51}$Fe& 4.36E-04& 5.11E-05& 9.14E-06& 1.84E-05\\
$^{14}$O & 2.54E-05& 9.14E-04& 1.86E-04& 1.41E-04& $^{52}$Fe& 4.31E-02& 9.34E-02& 5.22E-02& 8.61E-02\\
$^{15}$O & 2.18E-06& 7.72E-06& 4.25E-06& 5.12E-06& $^{53}$Fe& 1.36E-03& 3.34E-03& 2.00E-03& 3.61E-03\\
$^{16}$O & 4.55E+00& 2.31E+01& 2.37E+01& 1.03E+01& $^{54}$Fe& 2.26E-03& 5.19E-03& 4.99E-03& 1.11E-02\\
$^{17}$O & 1.74E-08& 1.29E-07& 5.67E-08& 3.57E-08& $^{55}$Fe& 3.96E-05& 3.52E-06& 2.16E-06& 1.06E-05\\
$^{18}$O & 2.50E-07& 5.92E-07& 9.19E-07& 6.93E-11& $^{56}$Fe& 5.40E-05& 2.73E-06& 1.33E-06& 6.56E-06\\
$^{17}$F & 1.24E-08& 2.86E-09& 1.58E-09& 7.01E-10& $^{57}$Fe& 6.49E-08& 2.94E-09& 1.45E-09& 1.63E-09\\
$^{18}$F & 8.20E-08& 8.77E-06& 1.25E-06& 7.24E-07& $^{58}$Fe& 6.00E-09& 4.07E-10& 1.87E-10& 2.09E-10\\
$^{19}$F & 6.80E-11& 5.56E-08& 1.37E-09& 1.43E-08& $^{59}$Fe& 6.49E-12& 1.16E-10& 5.11E-11& 6.46E-11\\
$^{18}$Ne& 6.92E-06& 3.10E-07& 1.48E-07& 4.68E-07& $^{60}$Fe& 7.62E-12& 1.31E-10& 4.37E-11& 7.54E-11\\
$^{19}$Ne& 2.91E-07& 1.07E-06& 2.77E-08& 5.09E-07& $^{61}$Fe& 3.56E-12& 1.20E-10& 2.07E-11& 7.08E-11\\
$^{20}$Ne& 7.83E-01& 5.18E+00& 4.82E+00& 1.58E+00& $^{51}$Co& 2.89E-10& 1.19E-11& 5.95E-12& 5.71E-12\\
$^{21}$Ne& 1.23E-05& 2.56E-05& 4.48E-05& 2.35E-05& $^{52}$Co& 1.22E-05& 4.82E-07& 2.27E-07& 2.35E-07\\
$^{22}$Ne& 1.34E-05& 2.27E-05& 4.64E-05& 1.50E-06& $^{53}$Co& 1.56E-04& 1.39E-05& 4.15E-06& 6.23E-06\\
$^{21}$Na& 4.98E-07& 2.84E-06& 6.22E-07& 1.44E-06& $^{54}$Co& 3.07E-04& 6.25E-05& 1.14E-05& 3.25E-05\\
$^{22}$Na& 1.64E-06& 1.13E-05& 1.53E-05& 3.55E-05& $^{55}$Co& 1.37E-03& 3.30E-03& 1.79E-03& 4.02E-03\\
$^{23}$Na& 4.18E-05& 3.46E-04& 4.14E-04& 5.56E-04& $^{56}$Co& 1.72E-04& 1.89E-05& 1.32E-05& 1.99E-05\\
$^{22}$Mg& 2.37E-04& 1.58E-05& 6.22E-06& 1.26E-05& $^{57}$Co& 5.33E-04& 2.71E-05& 1.60E-05& 1.43E-05\\
$^{23}$Mg& 2.14E-04& 7.84E-04& 9.12E-04& 1.04E-03& $^{58}$Co& 7.77E-06& 3.29E-07& 1.66E-07& 1.60E-07\\
$^{24}$Mg& 7.05E-01& 4.55E+00& 4.14E+00& 1.43E+00& $^{59}$Co& 3.23E-05& 1.37E-06& 6.89E-07& 6.58E-07\\
$^{25}$Mg& 1.47E-04& 6.76E-04& 7.04E-04& 2.88E-04& $^{60}$Co& 4.68E-09& 4.20E-10& 1.80E-10& 2.39E-10\\
$^{26}$Mg& 3.65E-05& 1.70E-04& 1.50E-04& 4.11E-05& $^{61}$Co& 1.82E-11& 2.24E-10& 8.42E-11& 1.02E-10\\
$^{27}$Mg& 1.54E-11& 9.35E-10& 1.85E-10& 3.38E-11& $^{62}$Co& 6.29E-12& 1.32E-10& 5.11E-11& 8.78E-11\\
$^{25}$Al& 2.97E-05& 1.10E-04& 1.25E-05& 2.79E-04& $^{54}$Ni& 5.26E-05& 1.42E-06& 5.54E-07& 6.72E-07\\
$^{26}$Al& 3.88E-06& 1.72E-05& 1.42E-05& 2.54E-05& $^{55}$Ni& 3.16E-04& 1.57E-05& 4.70E-06& 5.77E-06\\
$^{27}$Al& 5.97E-04& 7.06E-03& 4.95E-03& 5.78E-03& $^{56}$Ni& 2.04E+01& 1.01E+01& 5.07E+00& 9.32E+00\\
$^{28}$Al& 1.98E-08& 5.58E-07& 1.88E-07& 9.17E-08& $^{57}$Ni& 3.80E-01& 1.68E-01& 8.32E-02& 1.30E-01\\
$^{29}$Al& 1.42E-12& 7.85E-11& 1.31E-11& 1.24E-11& $^{58}$Ni& 7.66E+00& 3.83E-01& 1.93E-01& 2.03E-01\\
$^{26}$Si& 5.63E-04& 7.80E-05& 2.16E-05& 5.95E-05& $^{59}$Ni& 1.47E-01& 6.79E-03& 3.76E-03& 3.47E-03\\
$^{27}$Si& 2.62E-04& 1.21E-03& 1.06E-03& 1.14E-03& $^{60}$Ni& 2.31E+00& 9.74E-02& 5.21E-02& 4.69E-02\\
$^{28}$Si& 6.92E-01& 6.19E+00& 4.04E+00& 4.81E+00& $^{61}$Ni& 2.29E-03& 9.65E-05& 4.97E-05& 4.65E-05\\
$^{29}$Si& 3.07E-04& 2.26E-03& 1.94E-03& 3.03E-03& $^{62}$Ni& 2.13E-04& 9.00E-06& 4.64E-06& 4.33E-06\\
$^{30}$Si& 1.66E-05& 2.38E-04& 7.63E-05& 5.81E-04& $^{63}$Ni& 3.05E-08& 1.40E-09& 6.71E-10& 6.68E-10\\
$^{31}$Si& 4.78E-11& 8.74E-09& 6.55E-10& 3.94E-09& $^{64}$Ni& 6.81E-10& 2.02E-10& 9.08E-11& 1.11E-10\\
$^{32}$Si& 5.46E-13& 1.14E-11& 5.42E-12& 3.31E-12& $^{65}$Ni& 6.02E-12& 1.32E-10& 4.23E-11& 7.07E-11\\
$^{27}$P & 1.93E-06& 8.17E-08& 4.11E-08& 3.99E-08& $^{66}$Ni& 7.20E-12& 2.09E-10& 5.35E-11& 8.03E-11\\
$^{28}$P & 7.79E-05& 8.57E-06& 2.43E-06& 3.40E-06& $^{56}$Cu& 1.23E-09& 5.07E-11& 2.53E-11& 2.45E-11\\
$^{29}$P & 1.04E-04& 4.08E-04& 6.05E-05& 1.35E-03& $^{57}$Cu& 1.29E-03& 5.67E-05& 2.78E-05& 2.87E-05\\
$^{30}$P & 1.26E-04& 3.25E-04& 1.75E-04& 2.79E-04& $^{58}$Cu& 3.60E-02& 4.92E-02& 1.41E-02& 4.46E-02\\
$^{31}$P & 8.43E-05& 4.22E-04& 2.66E-04& 7.31E-04& $^{59}$Cu& 4.34E-02& 3.67E-02& 1.15E-02& 3.46E-02\\
$^{32}$P & 1.37E-09& 5.33E-09& 3.76E-09& 1.15E-08& $^{60}$Cu& 6.17E-03& 3.54E-03& 2.70E-03& 1.71E-03\\
$^{33}$P & 4.73E-10& 1.16E-09& 1.25E-09& 4.37E-09& $^{61}$Cu& 6.53E-02& 2.80E-03& 1.55E-03& 1.35E-03\\
$^{34}$P & 1.76E-12& 2.02E-11& 1.41E-11& 2.85E-12& $^{62}$Cu& 2.45E-03& 1.04E-04& 5.26E-05& 5.00E-05\\
$^{30}$S & 1.05E-03& 1.58E-04& 3.32E-05& 7.83E-05& $^{63}$Cu& 1.81E-03& 7.66E-05& 4.00E-05& 3.69E-05\\
$^{31}$S & 2.60E-04& 2.94E-04& 9.71E-05& 1.88E-04& $^{64}$Cu& 1.85E-06& 7.84E-08& 3.96E-08& 3.78E-08\\
$^{32}$S & 3.55E-01& 3.17E+00& 2.00E+00& 2.68E+00& $^{65}$Cu& 7.68E-07& 3.26E-08& 1.64E-08& 1.57E-08\\
$^{33}$S & 1.71E-04& 1.11E-03& 8.41E-04& 1.31E-03& $^{66}$Cu& 1.23E-10& 1.65E-10& 7.10E-11& 8.85E-11\\
$^{34}$S & 1.45E-04& 3.39E-05& 3.47E-05& 3.59E-04& $^{67}$Cu& 1.21E-11& 2.06E-10& 6.61E-11& 7.34E-11\\
$^{35}$S & 3.83E-10& 3.96E-10& 4.72E-10& 1.77E-09& $^{68}$Cu& 9.76E-12& 1.65E-10& 5.32E-11& 6.91E-11\\
$^{36}$S & 1.45E-09& 3.52E-10& 3.41E-10& 2.03E-10& $^{59}$Zn& 1.07E-03& 3.81E-04& 6.15E-05& 6.95E-04\\
$^{37}$S & 1.87E-12& 2.00E-11& 1.48E-11& 3.27E-12& $^{60}$Zn& 6.55E-01& 3.70E-01& 1.29E-01& 2.08E-01\\
$^{32}$Cl& 4.80E-04& 2.21E-05& 1.02E-05& 1.04E-05& $^{61}$Zn& 1.35E-02& 2.74E-03& 1.51E-03& 1.78E-03\\
$^{33}$Cl& 9.07E-05& 6.11E-05& 9.32E-06& 7.32E-05& $^{62}$Zn& 7.68E-01& 3.62E-02& 1.80E-02& 1.93E-02\\
$^{34}$Cl& 2.31E-04& 7.34E-05& 1.68E-05& 3.23E-05& $^{63}$Zn& 2.37E-02& 1.06E-03& 6.10E-04& 5.23E-04\\
$^{35}$Cl& 3.10E-04& 1.45E-04& 5.12E-05& 2.93E-04& $^{64}$Zn& 1.06E+00& 4.49E-02& 2.30E-02& 2.16E-02\\
$^{36}$Cl& 2.45E-07& 3.80E-08& 2.64E-08& 7.18E-08& $^{65}$Zn& 2.37E-03& 1.00E-04& 5.13E-05& 4.82E-05\\
$^{37}$Cl& 1.46E-07& 1.60E-08& 1.35E-08& 3.73E-08& $^{66}$Zn& 2.14E-03& 9.02E-05& 4.55E-05& 4.34E-05\\
$^{38}$Cl& 3.17E-12& 3.17E-11& 2.95E-11& 8.79E-12& $^{67}$Zn& 6.02E-07& 2.57E-08& 1.30E-08& 1.24E-08\\
$^{34}$Ar& 1.84E-03& 1.69E-04& 4.99E-05& 7.78E-05& $^{68}$Zn& 1.09E-08& 6.34E-10& 3.16E-10& 3.11E-10\\
$^{35}$Ar& 8.13E-04& 2.26E-04& 5.11E-05& 9.33E-05& $^{69}$Zn& 9.27E-12& 2.04E-10& 6.15E-11& 7.40E-11\\
$^{36}$Ar& 6.96E-02& 5.88E-01& 3.65E-01& 4.97E-01& $^{70}$Zn& 1.00E-11& 2.42E-10& 6.58E-11& 8.77E-11\\
$^{37}$Ar& 5.31E-05& 1.68E-04& 1.16E-04& 1.79E-04& $^{71}$Zn& 6.19E-12& 1.08E-10& 4.16E-11& 5.50E-11\\
$^{38}$Ar& 3.94E-04& 4.02E-05& 3.34E-05& 3.25E-04& $^{61}$Ga& 8.83E-05& 3.80E-06& 1.89E-06& 1.84E-06\\
$^{39}$Ar& 7.03E-08& 3.00E-09& 1.53E-09& 1.54E-09& $^{62}$Ga& 4.83E-05& 5.15E-06& 1.67E-06& 4.16E-06\\
$^{40}$Ar& 8.68E-08& 3.75E-09& 1.89E-09& 1.79E-09& $^{63}$Ga& 1.37E-03& 1.39E-03& 4.86E-04& 9.56E-04\\
$^{41}$Ar& 2.41E-12& 3.70E-11& 2.57E-11& 2.06E-11& $^{64}$Ga& 9.88E-04& 4.41E-04& 3.13E-04& 2.06E-04\\
$^{42}$Ar& 2.80E-12& 4.55E-11& 2.62E-11& 1.35E-11& $^{65}$Ga& 7.21E-04& 3.52E-05& 2.43E-05& 1.63E-05\\
$^{43}$Ar& 4.53E-12& 5.33E-11& 3.00E-11& 1.73E-11& $^{66}$Ga& 8.58E-05& 3.66E-06& 2.18E-06& 1.77E-06\\
$^{36}$K & 6.58E-04& 3.03E-05& 1.39E-05& 1.39E-05& $^{67}$Ga& 5.36E-04& 2.26E-05& 1.15E-05& 1.09E-05\\
$^{37}$K & 1.11E-04& 6.81E-05& 1.65E-05& 2.06E-05& $^{68}$Ga& 1.44E-06& 6.12E-08& 3.08E-08& 2.95E-08\\
$^{38}$K & 7.55E-05& 3.07E-05& 1.31E-05& 9.38E-06& $^{69}$Ga& 1.02E-06& 4.32E-08& 2.18E-08& 2.08E-08\\
$^{39}$K & 6.15E-04& 6.88E-05& 3.66E-05& 1.59E-04& $^{70}$Ga& 6.56E-10& 2.11E-10& 9.27E-11& 1.24E-10\\
$^{40}$K & 5.99E-07& 2.73E-08& 1.35E-08& 1.65E-08& $^{71}$Ga& 1.99E-11& 1.74E-10& 7.51E-11& 1.13E-10\\
$^{41}$K & 9.58E-07& 4.06E-08& 2.05E-08& 1.99E-08& $^{72}$Ga& 9.93E-12& 1.38E-10& 5.53E-11& 7.30E-11\\
$^{42}$K & 1.79E-11& 6.95E-11& 3.91E-11& 3.79E-11& $^{73}$Ga& 1.10E-11& 1.52E-10& 5.98E-11& 8.06E-11\\
$^{43}$K & 7.21E-12& 8.11E-11& 4.19E-11& 3.41E-11& $^{63}$Ge& 2.79E-05& 1.48E-06& 4.88E-07& 1.60E-06\\
$^{44}$K & 4.45E-12& 7.90E-11& 3.84E-11& 3.77E-11& $^{64}$Ge& 4.06E-02& 1.69E-02& 7.24E-03& 1.05E-02\\
$^{45}$K & 4.16E-12& 1.00E-10& 3.78E-11& 4.05E-11& $^{65}$Ge& 1.50E-04& 7.10E-05& 5.23E-05& 4.21E-05\\
$^{38}$Ca& 1.42E-03& 9.17E-05& 3.61E-05& 3.80E-05& $^{66}$Ge& 8.48E-03& 5.55E-04& 2.80E-04& 3.00E-04\\
$^{39}$Ca& 3.00E-03& 2.06E-04& 7.89E-05& 8.39E-05& $^{67}$Ge& 2.61E-04& 1.18E-05& 6.26E-06& 5.58E-06\\
$^{40}$Ca& 7.03E-02& 5.71E-01& 3.59E-01& 4.63E-01& $^{68}$Ge& 4.68E-02& 1.98E-03& 9.97E-04& 9.52E-04\\
$^{41}$Ca& 4.81E-05& 3.77E-05& 2.35E-05& 3.64E-05& $^{69}$Ge& 2.27E-04& 9.59E-06& 4.84E-06& 4.62E-06\\
$^{42}$Ca& 8.38E-04& 3.65E-05& 1.84E-05& 2.47E-05& $^{70}$Ge& 1.04E-03& 4.38E-05& 2.21E-05& 2.11E-05\\
$^{43}$Ca& 1.55E-06& 6.57E-08& 3.37E-08& 3.32E-08& $^{71}$Ge& 8.96E-07& 3.81E-08& 1.92E-08& 1.84E-08\\
$^{44}$Ca& 1.14E-06& 4.81E-08& 2.42E-08& 2.33E-08& $^{72}$Ge& 2.78E-08& 1.41E-09& 7.61E-10& 6.81E-10\\
$^{45}$Ca& 6.55E-11& 1.45E-10& 5.04E-11& 7.25E-11& $^{73}$Ge& 2.33E-11& 1.63E-10& 9.63E-11& 8.29E-11\\
$^{46}$Ca& 5.62E-12& 9.28E-11& 5.45E-11& 4.61E-11& $^{74}$Ge& 1.26E-11& 1.53E-10& 8.87E-11& 8.11E-11\\
$^{47}$Ca& 4.46E-12& 6.63E-11& 3.61E-11& 4.57E-11& $^{75}$Ge& 6.60E-12& 1.63E-10& 8.86E-11& 5.69E-11\\
$^{48}$Ca& 4.01E-12& 8.17E-11& 4.34E-11& 3.33E-11& $^{65}$As& 3.70E-09& 1.69E-10& 8.19E-11& 8.45E-11\\
$^{40}$Sc& 8.98E-08& 3.80E-09& 1.91E-09& 1.83E-09& $^{66}$As& 5.56E-07& 1.83E-07& 3.83E-08& 5.49E-08\\
$^{41}$Sc& 1.51E-04& 6.64E-06& 3.22E-06& 3.16E-06& $^{67}$As& 3.99E-05& 1.63E-05& 5.84E-06& 7.92E-06\\
$^{42}$Sc& 2.13E-05& 1.05E-06& 4.67E-07& 4.56E-07& $^{68}$As& 4.13E-06& 9.82E-07& 4.16E-07& 4.04E-07\\
$^{43}$Sc& 1.43E-04& 7.64E-06& 3.29E-06& 3.55E-06& $^{69}$As& 2.77E-06& 1.80E-07& 7.38E-08& 7.03E-08\\
$^{44}$Sc& 1.56E-06& 6.62E-08& 3.34E-08& 3.25E-08& $^{70}$As& 7.92E-07& 3.38E-08& 1.76E-08& 1.63E-08\\
$^{45}$Sc& 8.43E-05& 3.56E-06& 1.87E-06& 1.72E-06& $^{71}$As& 1.87E-05& 7.90E-07& 3.99E-07& 3.80E-07\\
$^{46}$Sc& 3.40E-08& 1.53E-09& 7.62E-10& 7.26E-10& $^{72}$As& 1.55E-07& 6.74E-09& 3.38E-09& 3.25E-09\\
$^{47}$Sc& 1.45E-09& 1.89E-10& 7.61E-11& 1.21E-10& $^{73}$As& 1.09E-07& 4.92E-09& 2.52E-09& 2.37E-09\\
$^{48}$Sc& 5.61E-12& 1.23E-10& 4.45E-11& 6.76E-11& $^{74}$As& 1.50E-10& 1.38E-10& 8.36E-11& 9.52E-11\\
$^{49}$Sc& 5.77E-12& 7.81E-11& 5.26E-11& 4.27E-11& $^{75}$As& 8.20E-11& 2.17E-10& 1.15E-10& 1.39E-10\\
$^{42}$Ti& 5.85E-04& 2.46E-05& 1.22E-05& 1.18E-05& $^{76}$As& 8.53E-12& 1.35E-10& 5.92E-11& 6.42E-11\\
$^{43}$Ti& 1.23E-04& 6.28E-06& 2.55E-06& 2.96E-06& $^{67}$Se& 2.21E-08& 9.79E-10& 2.79E-10& 3.95E-10\\
$^{44}$Ti& 1.28E-02& 3.30E-03& 1.60E-03& 2.61E-03& $^{68}$Se& 2.86E-04& 4.83E-05& 1.55E-05& 2.52E-05\\
$^{45}$Ti& 7.11E-05& 2.69E-05& 1.52E-05& 1.20E-05& $^{69}$Se& 1.26E-06& 3.68E-07& 1.39E-07& 2.36E-07\\
$^{46}$Ti& 1.16E-03& 6.18E-05& 3.25E-05& 3.07E-05& $^{70}$Se& 1.24E-05& 7.52E-07& 3.38E-07& 4.00E-07\\
$^{47}$Ti& 4.09E-06& 1.86E-07& 5.71E-07& 9.57E-08& $^{71}$Se& 1.00E-06& 4.62E-08& 2.26E-08& 2.16E-08\\
$^{48}$Ti& 1.50E-05& 6.32E-07& 3.19E-07& 3.05E-07& $^{72}$Se& 7.15E-04& 3.02E-05& 1.52E-05& 1.46E-05\\
$^{49}$Ti& 1.43E-08& 7.22E-10& 3.61E-10& 3.65E-10& $^{73}$Se& 8.71E-06& 3.68E-07& 1.86E-07& 1.77E-07\\
$^{50}$Ti& 3.54E-11& 1.17E-10& 5.98E-11& 6.37E-11& $^{74}$Se& 1.48E-04& 6.27E-06& 3.16E-06& 3.02E-06\\
$^{51}$Ti& 4.88E-12& 9.63E-11& 3.21E-11& 5.57E-11& $^{75}$Se& 9.18E-07& 3.91E-08& 1.98E-08& 1.89E-08\\
$^{44}$V & 3.86E-04& 1.65E-05& 8.17E-06& 7.89E-06& $^{76}$Se& 1.77E-06& 7.64E-08& 3.91E-08& 3.69E-08\\
$^{45}$V & 1.57E-04& 5.85E-05& 1.29E-05& 2.49E-05& $^{77}$Se& 2.21E-09& 2.86E-10& 1.28E-10& 1.08E-10\\
$^{46}$V & 7.61E-05& 1.86E-05& 4.71E-06& 6.83E-06& $^{78}$Se& 1.88E-10& 2.45E-10& 1.22E-10& 1.04E-10\\
$^{47}$V & 8.10E-04& 6.57E-05& 2.59E-05& 3.11E-05& $^{69}$Br& 4.76E-12& 2.07E-13& 1.03E-13& 1.01E-13\\
$^{48}$V & 6.82E-05& 2.94E-06& 1.54E-06& 1.47E-06& $^{70}$Br& 1.49E-09& 1.15E-10& 3.49E-11& 4.13E-11\\
$^{49}$V & 2.86E-04& 1.23E-05& 7.88E-06& 6.20E-06& $^{71}$Br& 7.23E-07& 4.28E-08& 1.86E-08& 1.93E-08\\
$^{50}$V & 9.02E-08& 3.98E-09& 2.00E-09& 1.92E-09& $^{72}$Br& 3.45E-08& 3.46E-09& 1.46E-09& 1.89E-09\\
$^{51}$V & 2.34E-07& 1.00E-08& 5.03E-09& 4.87E-09& $^{73}$Br& 5.42E-08& 1.07E-08& 3.20E-09& 5.88E-09\\
$^{52}$V & 2.79E-11& 8.86E-11& 5.14E-11& 6.41E-11& $^{74}$Br& 9.04E-09& 1.78E-09& 8.19E-10& 8.46E-10\\
$^{53}$V & 7.36E-12& 1.15E-10& 5.36E-11& 5.58E-11& $^{75}$Br& 2.73E-07& 1.54E-08& 7.13E-09& 7.77E-09\\
$^{46}$Cr& 7.13E-04& 3.69E-05& 1.53E-05& 1.68E-05& $^{76}$Br& 1.03E-07& 5.96E-09& 2.71E-09& 2.93E-09\\
$^{47}$Cr& 3.21E-04& 4.92E-05& 9.88E-06& 2.15E-05& $^{77}$Br& 1.14E-06& 4.94E-08& 2.50E-08& 2.38E-08\\
$^{48}$Cr& 1.70E-02& 8.42E-03& 4.31E-03& 7.13E-03& $^{78}$Br& 7.85E-09& 1.31E-09& 4.93E-10& 7.45E-10\\
$^{49}$Cr& 7.06E-04& 3.71E-04& 1.86E-04& 2.60E-04& $^{79}$Br& 6.43E-09& 1.66E-09& 1.12E-09& 7.53E-10\\

% \hline
% \end{tabular}
% }
% \caption{Nucleosynthesis products ($M_{\odot}$)
% of models A-1, B-1, and B-2.
% \label{tab:Yields1}}
\enddata
\label{tab:Yields1}
% \end{center}
 \end{deluxetable}

\clearpage

%figures

\clearpage

\end{document}